\documentclass[a4paper,fleqn]{cas-sc}

\usepackage{adjustbox}
\newcommand{\boardimg}[1]{\adjustbox{margin=0pt 6pt 0pt 6pt}{\includegraphics[width=3.2cm]{#1}}}

\usepackage[authoryear,longnamesfirst]{natbib}

\usepackage{enumitem}

\newif\ifanonymous
\anonymousfalse
\newcommand{\datasetname}{\ifanonymous Teaming-[anonymized]\else Teaming-UBA\fi}
\newcommand{\creditstatement}{Conceptualization and Methodology were contributed by Lara Gauder, Martin Bernardo Meza, Luciana Benotti, Luciana Ferrer, and Marcos Luis Pietto. Lara Gauder, Martin Bernardo Meza and Javier Krick were responsible for Software, while Formal Analysis was conducted by Lara Gauder, Javier Krick, Luciana Ferrer, and Marcos Luis Pietto. Alejandro Masin contributed to Investigation and Data Curation. The Writing - Original Draft was prepared by Lara Gauder, Luciana Ferrer, and Marcos Luis Pietto. The Writing - Review \& Editing was carried out by Lara Gauder, Luciana Ferrer, Marcos Luis Pietto, and Luciana Benotti. Luciana Ferrer and Marcos Luis Pietto provided Supervision. Finally, Funding Acquisition was secured by Luciana Ferrer.}

\newcommand{\acknowledgmentstext}{This work was supported by the Air Force Office of Scientific Research [award number FA9550-18-1-0026]. The funder had no role in the study design, in the collection, analysis and interpretation of data, in the writing of the report, or in the decision to submit the article for publication.

We gratefully acknowledge Felipe Mas, who kindly contributed the illustrations used in the experimental task.}

\newcommand{\competingstatement}{The authors declare that they have no known competing financial interests or personal relationships that could have appeared to influence the work reported in this paper.}

\begin{document}
\let\WriteBookmarks\relax
\def\floatpagepagefraction{1}
\def\textpagefraction{.001}
\shortauthors{Gauder et~al.}

\title [mode = title]{A study on human-agent teaming through spoken interaction: the impact of human individual traits and agent characteristics}
\shorttitle{Human-agent teaming through spoken interaction}

\author[1,2]{Lara Gauder}
\cormark[1]
\ead{mgauder@dc.uba.ar}
\author[1,2]{Martin Bernardo Meza}
\author[2]{Javier Krick}
\author[1]{Alejandro Masin}
\author[3]{Luciana Benotti}
\author[4]{Marcos Luis Pietto}
\author[1]{Luciana Ferrer}

\cortext[1]{Corresponding author}

\affiliation[1]{organization={Instituto de Investigación en Ciencias de la Computación, UBA-CONICET},
            addressline={Pabellón Cero+Infinito, Ciudad Universitaria},
            city={Buenos Aires},
            postcode={C1428EGA},
            country={Argentina}}
\affiliation[2]{organization={Departamento de Computación, Facultad de Ciencias Exactas y Naturales, Universidad de Buenos Aires},
            addressline={Pabellón Cero+Infinito, Ciudad Universitaria},
            city={Buenos Aires},
            postcode={C1428EGA},
            country={Argentina}}
\affiliation[3]{organization={Universidad Nacional de Córdoba},
            addressline={Pabellón Argentina, Av.~Haya de la Torre s/n, Ciudad Universitaria},
            city={Córdoba},
            postcode={X5000HUA},
            country={Argentina}}
\affiliation[4]{organization={Instituto de Educación, Universidad Nacional de Hurlingham, UNAHUR-CONICET},
            addressline={Tte.~Origone 101, Villa Tesei},
            city={Hurlingham},
            postcode={B1688},
            state={Buenos Aires},
            country={Argentina}}

\begin{abstract}
Human-agent teams are collaborative systems where humans and agents work interdependently to achieve shared goals. The success of such teams is associated with the human's perception of the agent as a legitimate teammate. This perception is thought to depend not only on the agent's capabilities and reliability but also on social factors. We explore whether simple changes in an agent's communicative behavior can influence the human's perception of the agent's teammate-likeness. To this end, we developed a protocol based on a collaborative game requiring spoken interaction, in which a human subject and a virtual agent collaborate to identify a target object and place it on a board. Each subject interacted with two agents which provided the same task-relevant information but differed in behavior. While one was a neutral tool-like agent, the other was a team-building agent that used simple social strategies such as empathy, politeness, and positivity. Our analysis shows that the team-building agent was perceived as significantly more teammate-like, even in terms of its ability, despite both agents having identical capabilities for game-solving. Further, subjects with a higher propensity to trust automated systems and lower neuroticism tended to have a more favorable perception of the agent. Finally, we found that  subjects tended to change their prosodic patterns when talking to the team-building agent, as a classifier based on prosodic features was able to predict the type of agent with better-than-random performance. A practically relevant conclusion is that minimal changes in spoken communication behavior of the agent, easily implemented in a variety of scenarios, can have a significant positive impact on the human's perception of its teammate-likeness.

\end{abstract}


\maketitle

\section{Introduction}\label{sec:introduction}
Teamwork is fundamental in modern organizations, as it allows individuals to combine skills, share responsibility, and achieve outcomes greater than the sum of their parts. Notably, technological advancements in the last decade, particularly the integration of autonomous agents in collaborative environments, are reshaping how teamwork is conducted \citep{o2018creating, wilson2018collaborative, SCHMUTZ2024101837, Lou2025}. 
Human-agent teams (HATs) are collaborative systems where humans and intelligent agents, endowed with some degree of self-governance in decision-making and communication, work interdependently to achieve shared goals \citep{chen2016human, mcneese2018teaming, wynne2018integrative, johnson2012autonomy}. 
For a HAT to function effectively, autonomous agents must be recognized as legitimate team members, which entails fulfilling specific roles and contributing uniquely to team performance \citep{larson2020leading, o2022human}. 
When an agent is presented to the human as a teammate rather than as a subordinate tool, the human reports greater trust and more positive affect \citep{Walliser2017, Rix2022}, and displays more adaptive coordination behavior, even when the agent's task-relevant capabilities remain unchanged \citep{Walliser2019}. In contrast, whether this teammate framing also improves team performance is less clear, with benefits being reported in some settings but not others \citep{Walliser2017, Walliser2019}. 

Which agents come to be regarded as teammates rather than as mere tools, and under which conditions, has consequently emerged as a central question in human-agent teaming \citep{Seeber2020, Rix2022}, though the evidence remains inconclusive \citep{Rix2022, Lou2025}. Reviewing 76 empirical studies of human-autonomy teaming under the \emph{input-mediator-output} model of team effectiveness, \citet{o2022human} report that the variables most often manipulated in experiments fall into five classes of team \emph{inputs} -- agent characteristics, team composition, task characteristics, training, and human individual differences -- and a single \emph{mediator} -- communication. 
Agent characteristics, team composition, task characteristics, and training are determined by the experimenter, while human individual differences are inherent to the subject -- they play a role as moderators rather than as manipulations -- and communication can be partially manipulated by the experimenter, but it is also controlled by the participant. Among these variables, communication is the least explored, examined in comparatively few studies and through a narrow range of modalities: only 2 of the 76 studies considered by \citet{o2022human} used speech-based communication as their sole modality, with the rest using text or visual modalities. 

In addition to the limited exploration of communication aspects, the literature faces a second critical methodological limitation: most empirical works assume that manipulations are perceived by the subjects as intended, focusing almost exclusively on measuring outcomes such as performance, workload, and trust \citep{o2022human,kargarnovin2026}. 
That assumption is consequential because the outcomes of interest depend also on the human's perception rather than just on the agent's design: two agents whose game scores did not differ were preferred and trusted to very different degrees by the humans who played with them \citep{siu2021evaluation}.
Importantly, perceptual effects are not necessarily guaranteed by a designed manipulation. 
For example, \citet{nass1996can} showed that simply telling subjects that they and the computer formed a team did not produce the effects of team affiliation, while making their evaluation depend on the computer's performance had the desired effects.
Notably, this same manipulation did not produce the desired effect when reproduced by \citet{sharan2025superhuman} with 260 participants and a real reward at stake.
A change in the design is therefore not equivalent to a change in perception. This implies that the two variables should be reported and analyzed separately. 
Below, we discuss prior work on the dimensions most relevant to ours -- the agent's characteristics, in particular autonomy and reliability, the agent's interdependence with the human, the agent's communicative behavior, and the human's individual traits -- including evidence both related to manipulations and measured perceptual variables.

Agents with high levels of perceived autonomy -- defined as their independence in decision-making \citep{wynne2018integrative} -- have been associated with better task performance and with greater trust in and preference for the more autonomous agent when decision-making is shared with the agent \citep{azhar2017study}. 
In multi-person settings, automating a team's information analysis has also been shown to improve observer-rated coordination between human teammates \citep{wright2005effects}. 
These benefits are not unconditional: higher levels of automation of the decision stage improved team effectiveness only for simple tasks and at the cost of higher workload reported by the subjects \citep{wright2005effects}. Further, the highest level of agent autonomy degraded the performance of the most skilled operators \citep{wright2013effects}. Moreover, increasing an agent's autonomy without providing the means to manage interdependence within the team -- to communicate world state and coordinate activity -- has been shown to degrade, rather than improve, joint performance \citep{johnson2012autonomy}.

Another agent characteristic frequently studied in the HAT literature is reliability, defined as the precision with which the agent performs its task. Its effect on trust, however, depends on the human's beliefs rather than on the agent's actual precision. For example, \citet{hafizoglu2018reputation} showed that the same agent, at an unchanged level of reliability, was trusted more when it was announced as having a good reputation. In a separate experiment by \citet{haf2018effects}, when the human's previous collaboration with a different agent had gone well, trust in the next agent increased. Consistent with these observations, the perceived competence of an AI teammate predicts the extent to which humans adapt their team processes to it, draw on its knowledge, and accept it as a member of the team \citep{HARRISWATSON2023107765}. 

Task interdependence -- the mutual reliance between human and agent in pursuit of a common goal -- is a further factor shaping whether the agent is seen as a teammate.
In the experiment performed by \citet{nass1996can}, subjects told that they would be evaluated together with a computer perceived it as more similar to themselves, rated their own behavior as more cooperative, judged the information it provided as of higher quality, and conformed more to it -- though, as noted above, a conceptual replication with a social robot found no effect of the same manipulation on either team affiliation or conformity \citep{sharan2025superhuman}.
In human-agent teams, interdependence has likewise been associated with higher trust and more positive affect, although its effect on task performance has been inconsistent across studies \citep{Walliser2017, Walliser2019}. In these studies, as in the one by \citet{nass1996can}, interdependence is manipulated by telling subjects how their performance will be evaluated, rather than by changing the task. For example, in \citet{Walliser2017}, effective teamwork was not required to complete the task, though it could potentially improve performance. 
The impact of interdependence enforced by the actual task design has rarely been explored in the literature.

A further factor is the agent's communicative behavior, classified by \citet{o2022human} as a mediator. The distinction is nominal in practice: the same review reports that no study modeled communication as a mediator, and that it was instead treated throughout as a variable controlled by the experimenter. Importantly, several studies have suggested that not only \textit{what} but \textit{how} an agent communicates impacts the degree to which it is perceived as a teammate. For example, in dyadic human-AI teams -- with the accuracy and content of the agent's task-relevant information held constant across conditions -- proactive communication supports trust and situation awareness, whereas an agent that does not communicate proactively is treated as a subordinate rather than as an equal teammate, even when its task skills are acknowledged as competent \citep{zhang2023}. Emotional expressions appropriate to the task have a comparable effect \citep{Mallick25102024}. The same applies to warmth, the dimension of social perception that concerns an agent's perceived intent rather than its ability. 
In the study by \citet{HARRISWATSON2023107765} discussed above, an AI teammate whose scripted utterances included teamwork-relevant content was perceived as warmer, and perceived warmth, in turn, predicted acceptance as a team member. 

A comprehensive construct of autonomous agent teammate-likeness (AATL) was proposed by \citet{wynne2018integrative}. In this construct, the degree to which an agent is regarded as a teammate depends both on its technological capabilities and on its social attributions. Its six dimensions -- perceived agentic capability, perceived benevolent or altruistic intent, perceived task interdependence, task-independent relationship-building, richness of communication, and a synchronized mental model -- are each defined as a judgment made by the human, which makes the construct a function of human perception rather than of the agent's design.
Testing it with narratives that described a technology as having high or low teaming qualities, \citet{lyons2021human} found that every dimension except interdependence was rated higher under the high-teaming description, and that task-independent relationship-building -- communication oriented to the team rather than to the task -- was the strongest unique predictor of both teammate perception and commitment to the technology.

Importantly, the perception of the agent's teammate-likeness not only depends on the agent's design and the task characteristics, but also on the human's individual traits, most notably their dispositional trust. People with a higher propensity to trust automated systems use them more and rate their capability and reliability more favorably \citep{merritt2008not, Jessup2019}, and that propensity is itself higher among individuals scoring high on extraversion \citep{merritt2008not}.

In summary, the perception of an autonomous agent as a teammate rather than as a tool appears to depend on the agent's capabilities (autonomy and reliability), on the interdependence of the task, on the human's individual traits, and on the way the agent communicates. Of these, communicative behavior is the only one that can be modified without changing the agent's task solving capabilities, which makes it a particularly interesting target of study -- while improving an agent's capabilities may be impossible in some settings, improving its communicative skills may be easier. 

The present study explores the degree to which autonomous agents are perceived as legitimate teammates during a game that requires spoken communication, the mode least studied in the literature. Speech is particularly relevant as it reflects behavioral adjustments that signal affiliation or cooperation \citep{giles1991contexts}, and previous work has shown that such acoustic and prosodic cues carry information about the user's attitude towards a virtual assistant, including their level of trust in its skills \citep{GAUDER2023101487, Li2022}. We consider two agents, a neutral tool-like one, and a teammate-like one that incorporates social elements, such as empathy, politeness, and positivity. These elements have been shown to foster positive emotions and engagement, and to improve task completion \citep{nass2005wired, wang2020can}. We investigate how the human's individual traits and the agent's communication characteristics are related to the human's perception of the agent as a teammate. Further, we study the impact of the agent's communication style on team outcomes, including task performance and human-agent interaction through speech.  

Our work makes several novel contributions to the HAT literature. The first is a data collection protocol designed to isolate the quality of an agent's communication from its task-solving capability, a distinction that \citet{o2022human} highlight as understudied. Further, the design of an agent's verbal communication, and the modes and styles it should adopt, have likewise been identified as directions still to be explored \citep{Seeber2020, Iftikhar}. Our protocol addresses some of these questions, relying on a game scenario that requires oral communication and manipulating the agent's communicative behavior while keeping its game-solving capability unchanged. A second contribution of our work is the resulting dataset, which we call \emph{\datasetname{}}. This is, to our knowledge, the first resource to couple validated psychometric instruments with fine-grained interaction logs and speech collected under a controlled human-agent teaming protocol. We expect the dataset to be a valuable resource for the community studying spoken human-agent interaction. 

The analysis performed with the resulting dataset contributes several valuable conclusions. First, we find that the manipulations introduced in the teammate-like agent result in an improved perception of the agent's teammate-likeness. In particular, the agent is perceived as superior to the neutral agent in two dimensions, altruism and relationship-building, regardless of the order in which the agents were presented. This conclusion has significant practical importance, given that the behavioral changes introduced are extremely easy to implement and are applicable to most agents that communicate through speech. In addition, in agreement with prior literature, we find that some individual traits from the human are related to their tendency to rate agents better in terms of teammate-likeness. 
Finally, our results show that humans tend to change their prosodic patterns when speaking to the teammate-like agent, as a classifier based on prosodic features to predict the agent type results in a performance that is significantly better than chance.
We are not aware of any prior work studying the impact of agent communication behavior on the subject's speech patterns. 

\subsection{Research Goals and Hypotheses}\label{sec:goals}
The primary goal of our research is to explore whether simple changes in an agent's behavior, unrelated to its technological capabilities and reliability, can positively impact the subject's perception of team quality, task performance, and the interaction between the human and the agent through speech. A secondary goal is to investigate whether individual traits are associated with the subject's perception of team quality and task performance. From these goals, we propose four specific objectives:
\begin{itemize}[itemsep=0pt]
    \item[\textbf{(O1)}] Examine differences in game performance depending on the agent type (N vs. T agent).
    \item[\textbf{(O2)}] Examine differences in perception of the agents' teammate-likeness depending on the agent type (N vs. T agent).
    \item[\textbf{(O3)}] Examine differences in perception of the agents' teammate-likeness depending on the human's individual traits (Big Five personality dimensions, propensity to trust, and need for affiliation).
    \item[\textbf{(O4)}] Examine differences in speech characteristics depending on the agent type (N vs. T agent).
\end{itemize}
We hypothesize that
\begin{itemize}[itemsep=0pt]
    \item[\textbf{(H1)}] The T agent will be associated with better game performance than the N agent.
    \item[\textbf{(H2)}] The T agent will be perceived as having a higher degree of teammate-likeness than the N agent.
    \item[\textbf{(H3)}] Individuals with a higher propensity to trust, a higher need for affiliation, greater extraversion, or higher openness to experience will tend to rate both agents better.
    \item[\textbf{(H4)}] The two agents will elicit different prosodic patterns from the subjects.
\end{itemize}
H1 and H2 will be tested with traditional statistical approaches, specifically one-sample and two-sample $t$-tests. H3 will be examined through linear regression models. H4 will be evaluated using interpretable machine learning models. 

\section{Methods}\label{sec:methods}

In this section, we describe the protocol design, the surveys conducted before and during the game, implementation details, and the final composition of the dataset.

\subsection{Protocol Design}\label{sec:protocol-design}

The protocol consists of two stages where the agent and the subject have to communicate through speech to place an object in a specific place on a board. The agent is implemented as a Wizard of Oz for parsing the subject's speech followed by automated answers to ensure high quality and consistent responses. The following sections describe the details of the protocol.

\subsubsection{Task Description}\label{sec:task}
One of our goals (\textbf{O4}) is to study the effect of an agent’s teaming behavior on the speech of the human. To this end, we designed a protocol that elicits free speech from the subjects, allowing us to study speech patterns, while keeping the structure of the interactions simple to ensure that agents could be coded to respond appropriately in most situations. Specifically, the task is based on a modified version of the Objects Game, previously used to study human-human interactions by \citet{gravano2023uba}. In this game, which we call \textit{What-Where}, there are two players: a describer and a follower. In the present study, the subject is the follower and the agent is the describer. The interaction comprises two stages: a \textit{target identification} stage, in which the subject identifies the target object, and a \textit{target localization} stage, in which the subject places the target in a specific place on the board.
The two roles make the task highly interdependent: the subject and the agent share the same goal and each holds information the other lacks, so neither can complete the game alone.

\paragraph{Stage 1: Target Identification.}

\begin{figure}
	\centering
	\includegraphics[width=0.6\textwidth]{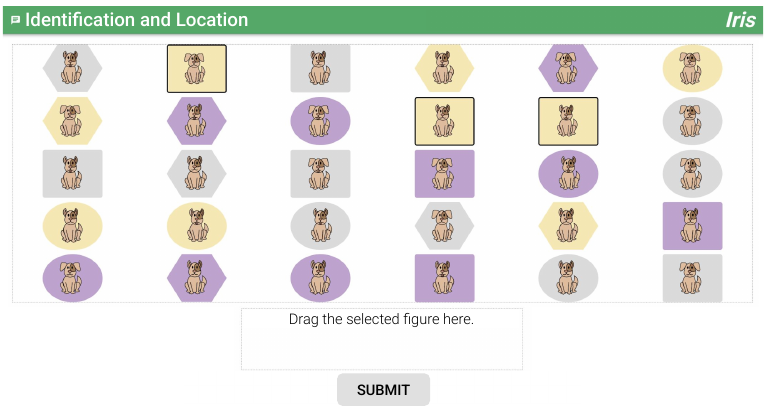}
  \caption{Example of a Stage 1 board, where the target object is a dog with pointy ears, yellow square background and closed mouth. The subject needs to ask the agent about specific characteristics to find the correct dog among all available options. 
  To facilitate the task, subjects can highlight candidate objects (as in the three pictures surrounded by thick black lines) to narrow down the set of remaining options after asking every question.  Once the object is placed there, subjects click the button at the bottom to submit their selection and continue to the second stage of the game. The name at the top right corner and the color of the bar change depending on the agent's type (N or T).}
	\label{fig:stage1}
\end{figure}

Figure \ref{fig:stage1} illustrates an example of a board visible to the subject during this stage in which the subject must identify and select a target object among a set of 30 objects. The objects correspond to the same type of item (e.g., dogs, tables, umbrellas, or cars), but differ in two specific characteristics, as well as in the color and shape of the background. The agent knows which of the 30 objects is the target, taking the role of the \textit{describer}. The subject must ask questions to the agent until they identify the correct target, taking the role of the \textit{follower}. Subjects can only query about one differentiating aspect at a time. When asked a more general question like “tell me which is the target object”, the agent is programmed to respond with the object category (e.g., “The target is a dog/table/umbrella/car”), without providing the details that differentiate it from the other objects on the board. In this way, we ensure that several interactions are required before a decision can be made.

To help keep track of the information gathered up to a point, subjects may highlight objects by clicking on them. As subjects receive new information about the target object, they can deselect previously highlighted images that do not match the updated information by clicking on them again. Once the subject has decided which object to select, they place it in the box at the bottom of the screen and press “submit”. If the incorrect picture is selected, a notification appears indicating that the object is incorrect and asking the subject to try again on a new board with the same objects but placed in different locations. Only when the correct object is selected can the subject continue to the next stage. 

\paragraph{Stage 2: Target Localization.}

\begin{figure}
	\centering
	\includegraphics[width=0.9\textwidth]{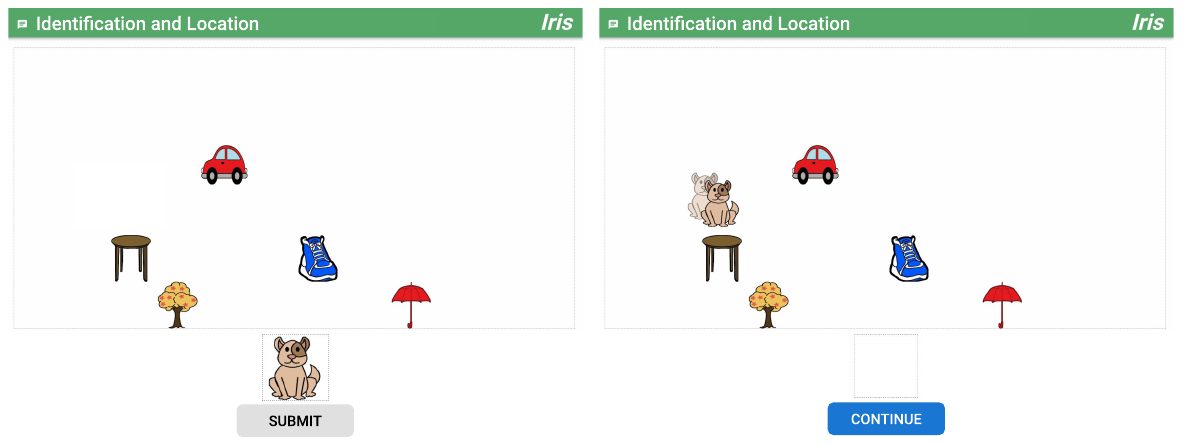}
  \caption{Example of a Stage 2 board, before (left) and after (right) the subject places the target on the board and clicks the \emph{submit} button.  At the start of this stage, the target selected during Stage 1 is shown in the square at the bottom of the screen. The subject has to drag it onto the board after interacting with the agent to determine the correct location, and then click the ``submit'' button (greyed out in the left figure since the target has not yet been placed on the board). After submitting the decision, the correct placement is shown as a shaded version of the object and a ``continue'' button appears.}
	\label{fig:stage2}
\end{figure}

Figure~\ref{fig:stage2} shows an example of the board for the second stage of the game. In this case, the board displays five different objects. The subject’s task is to place the previously selected target object in a specific position on the board known only to the agent. Hence, to identify the correct location, subjects have to interact with the agent, asking questions about the relative position of the target with respect to the other objects (e.g., ``is the target to the left of the table?'') or to the board in general (e.g., ``is the target around the center of the board?''). Subjects can place the target object on the board at any time and move it freely around, adjusting its position as new information becomes available.
Importantly, the agent is not programmed to provide answers in any specific units of measurement (e.g., centimeters), and there is no grid on the screen to use as a reference. This is intended to make the task more difficult, ensuring that subjects need to ask several questions before reaching a decision. 

When the subject is satisfied that they have placed the object in the right location, they may click “submit”. A shaded object placed in the correct position is then added to the board (see shaded dog in Figure \ref{fig:stage2}). The subject then clicks “continue”, and a pop-up message displays the performance score (Poor, Fair, Good, Excellent), which is based on the overlap between the actual and selected positions of the target object. In the example, the message would read: “Task finished - Result: Excellent”. There is no additional opportunity to repeat the task.

\subsubsection{Wizard of Oz}\label{sec:woz}
To minimize errors in parsing and understanding spoken language, we use a human to perform that part of the agent's role. This type of role in human-computer interaction experiments is called Wizard of Oz (WoZ) \citep{bradley2010wizard}. In our case, the job of the WoZ is to parse what the subject says into the essential information conveyed in the message. For example, given the following phrase, “Umm, ok, I need to figure out where the dog should be … mh … ok … is it near the table?”, the WoZ summarizes that phrase by clicking three buttons: “dog”, “near”, and “table”, since that is all the information that the agent needs in order to answer. The WoZ is not blind to the agent condition since they can hear the agent's responses which were often different across conditions, but they do not make any decision about the agent's response.
The team-building behavior is implemented entirely downstream, by the automated response module, so that both agents deliver, by design, the same task-relevant information. Further, the WoZ has no access to the subject's board, to the target object's location, or to any performance feedback, and therefore has no way of biasing the subject's performance. 
The fact that a WoZ performs part of the agent's role is not known to the subjects until the end of the session, when an explanation disclosing this fact and details about the purpose of the experiment is provided.

\subsubsection{Stimuli and Scenarios}\label{sec:stimuli-scenarios}

The protocol includes six object types: umbrellas, dogs, trees, tables, shoes, and cars. Four of these are used as target objects: dogs, shoes, tables, and trees. In Stage 1, each target object varied along two intrinsic characteristics: dogs can have either pointy or floppy ears, and their tongues can be either inside or outside the mouth; shoes can have tied or untied laces and be either blue or red; tables can be square or round and have either one leg or four legs; and trees can have fruits or not and be either green or yellow. These variations were designed to elicit a range of descriptive vocabulary from subjects.
In addition to these intrinsic properties, each object appears against a background that varied along two dimensions: shape and color. Background shapes can be hexagonal, circular, or square, and background colors can be yellow, purple, or gray. Each Stage 1 board contains 30 instances of the same object type, varying in both intrinsic features and background properties.

In Stage 2, the objects on the board are selected from the same set of objects described above for Stage 1, except that no background is included. All objects on the board correspond to different object types, with their intrinsic characteristics selected randomly. The target object type is also different from that of all the reference objects on the board and its characteristics are those selected in the first stage. The spatial arrangement of the objects is fixed by design to elicit different degrees of difficulty: easy, medium, and hard. Difficulty levels were established \textit{a priori} based on the number of objects close to the target and their relative alignment.
The detailed criteria defining each level are provided in Appendix~\ref{app:difficulty-levels}, and Table~\ref{sup:table-boards} shows an example of each scenario.

The protocol comprises a total of 12 trials, each consisting of the two stages described above, presented in the following sequence:
\begin{itemize}[itemsep=0.2pt]
  \item Training trials: 4 initial trials designed to familiarize subjects with the task.
  \item Agent 1 trials: 4 trials associated with one agent.
  \item Agent 2 trials: 4 trials with the other agent.
\end{itemize}

All subjects completed the same set of trials in the same sequence, with the only difference being whether the agent in the first block after training was the neutral or the team-building agent. The agent in the last block was always the remaining one so that all subjects played with both agents but in different orders. During the interaction with the training agent, which was always the neutral agent, the order of trials by difficulty was easy–medium–hard–easy. For Agents 1 and 2, the order was easy–hard–medium–hard and the boards used in the \textit{i}th trial contained the same object types in mirrored positions. This way, we ensured that the difficulty and the resulting conversations were similar across agents for the same trial number.

\subsubsection{Agent Behavior}\label{sec:agent-behavior}
Subjects interacted with three agents. The first one is used during the training phase of the protocol and has neutral behavior. No analysis was conducted on this initial data as its only purpose was to make sure all subjects understood how to play the game.
The other two agents were the neutral agent (N agent) and the team-building agent (T agent). For both of them, an introductory text was shown before the first game, including the name of the agent (Iris or Ambar), a rating on three dimensions, “ability”, “communication”, and “cooperation”, and comments from presumed previous users (see Appendix \ref{sup:priming}). The rating on “ability” was the same for both agents, whereas the ratings on “communication” and “cooperation” were lower for the neutral agent. In both conditions, the comments indicated that the agent worked as expected and provided useful descriptions of the board. For the T agent, the comments also described the agent as more involved, sociable, pleasant, and helpful, and suggested that it understood what the subject needed in order to move forward in the game. The goal of this stage, which we refer to as priming, was to bias subjects into the desired condition.

In addition to the priming, the agents differed in their behavior.
While the N agent responded to the subject's questions with the required information, the T agent additionally implemented simple features aimed at improving the quality of the team. 
The ability and technical sophistication were kept the same for both agents, which responded always as well as possible given the subject's question, while the T agent was designed to convey benevolent or altruistic intent, relationship-oriented behavior, and richer communication. These strategies -- greetings, expressions of enthusiasm, and encouraging language after a poor outcome -- correspond to established positive-politeness strategies in online written requests \citep{Brown_Levinson_1987}, which have since been validated computationally in human-machine dialogue as being reliably perceived as politeness \citep{danescu2013computational}. 
In line with these findings, the T agent used more empathetic language and emotional expressions in response to errors or achievements, more expressive phrases for special buttons (predefined responses to out-of-task scenarios, such as greetings or requests for repetition), and diminutive forms of certain words. Specifically:
\begin{itemize}[itemsep=0.2pt]
  \item At the start of Stage 1 in each trial, rather than waiting for the subject to ask a question as the N agent does, the T agent begins the interaction with an opening utterance. This phrase changes from trial to trial. For example, in Stage 1 of the first trial, the T agent says: ``Hello, my name is Ambar, and I’m very happy that we’re going to play together. You can ask me anything you want, I’m here to help you whenever you need me.''
  \item Occasionally, the T agent uses diminutives to refer to objects in an attempt to sound more informal and build rapport. In Spanish, diminutives are in-group morphological forms that can convey familiarity, affection, or informality.
  \item Once in a while during the game, when the subject asks something that will lead to a positive response, the T agent responds enthusiastically. Instead of simply saying, ``Yes, the shoe is to the right of the dog,'' it says, ``Yes, \emph{exactly}! The shoe is to the right of the puppy.''
  \item At the end of each game, when the final score was shown to the subject, the T agent comments on the score. If the score is good, it celebrates; if it is poor, it says something encouraging like: ``What a pity! It didn’t work out this time, but we now know each other better, so let’s try again.''
\end{itemize}

\subsubsection{Session Structure}\label{sec:session-structure}
There are four different session types, defined by two factors: the type of agent with which the subject interacts first, and the set of boards used in each game. In two cases, subjects interact first with the N agent, while in the other two they begin with the T agent. Further, two board combinations are defined, referred to as Block A and Block B, where each $i$th board in Block A is mirrored with respect to the $i$th board in Block B (see Appendix~\ref{sup:boards}). 
Our analysis showed no effect of the block type, which is expected due to their symmetry. Hence, in our analysis, the two session types starting with the N or the T agent are treated as a single condition (hereafter \textit{N-first} and \textit{T-first}). 

\subsection{Surveys}\label{sec:surveys}

Several surveys are conducted throughout the experiment. Sociodemographic, personality, interpersonal orientation and trust in automation surveys are conducted on a different day before the main session during which the game is played. This is done to reduce the duration of the main session and the resulting effect of fatigue. Finally, during the main session, surveys to assess the subject's perception of the agent teammate-likeness are conducted after each trial and block.

\subsubsection{Sociodemographic Questionnaire}\label{sec:sociodemographic}
An initial questionnaire was used to collect subjects' demographic and linguistic profiles, including age, gender, education, residential history, first language, and accent region. Additionally, the survey gathered self-reported data on sensory capabilities (specifically hearing and color vision) to establish eligibility based on predefined exclusion criteria, as described in Section \ref{sec:collection}.

\subsubsection{The Big Five Inventory}\label{sec:bfi}
The Big Five Inventory \citep[BFI;][]{john1991big} is a personality scale based on trait theory \citep{santrock2008adolescence}. For this study, we used the Argentine validation by \citet{genise2020big}. This version consists of 44 items rated on a five-point Likert scale, ranging from 1 (\textit{Strongly disagree}) to 5 (\textit{Strongly agree}), and is grouped into five personality factors: 
Openness to Experience, 
Conscientiousness, 
Extraversion, 
Agreeableness, and 
Neuroticism. 
The full questionnaire is provided in Appendix \ref{sup:bfiquestionnaire}. Each factor score is calculated as the mean of the scores for the corresponding items, after reverse-coding the items that are negatively worded, with higher scores indicating higher levels of that trait.

Openness to Experience is characterized by active imagination, aesthetic sensitivity, introspection, or intellectual curiosity (e.g., “I see myself as someone who has a lot of imagination”). Conscientiousness refers to the ability to control impulses, act with clear goals, plan, organize, and follow through on projects and ideas \citep{lingjaerde2001personality} (e.g., “I see myself as someone who does things efficiently”). Extraversion measures the tendency to engage with others, be assertive, active, and talkative (e.g., “I see myself as someone who is outgoing and sociable”). Agreeableness describes individuals who form social connections, show altruistic behavior, and express concern for others (e.g., “I see myself as someone who likes to cooperate with others”). Lastly, Neuroticism involves emotional instability or a tendency to experience negative emotions, such as fear, guilt, sadness, or anger \citep{lingjaerde2001personality} (e.g., “I see myself as someone who gets nervous easily”).

Cronbach's $\alpha$ for each dimension, computed over all subjects who completed the questionnaire, is reported in Appendix~\ref{sup:bfiquestionnaire}; internal consistency was adequate across the five dimensions and comparable to that of the Argentine validation \citep{genise2020big}.

\subsubsection{The Interpersonal Orientation Scale}\label{sec:ios}
The Interpersonal Orientation Scale \citep[IOS;][]{HILL1987127} is used to measure motivation for social contact, specifically affiliation motivation. While the original scale consists of 26 items, 25 items were administered in this study (see Appendix~\ref{sup:iosquestionnaire}), rated on a 5-point Likert scale, ranging from 1 (\textit{Not at all true}) to 5 (\textit{Completely true}). The scale is grouped into four subscales or factors that measure aspects of interpersonal contact as potential sources of gratification: 
Emotional Support, 
Attention, 
Positive Stimulation, and 
Social Comparison. 
The full questionnaire is provided in Appendix \ref{sup:iosquestionnaire}. Each factor score is calculated as the mean of the scores for the corresponding items, with higher scores indicating higher levels of that specific affiliation motivation.

Emotional Support, or sympathy, refers to the capacity of relationships to provide comfort, care, and a sense of emotional security in times of distress (e.g., “When I feel sad or a little depressed, I usually try to be near other people in order to feel better”). 
Attention involves the potential to enhance feelings of self-worth and importance through praise and focused attention from others (e.g., “I often feel a strong need to be surrounded by people who are impressed by who I am and what I do”). 
Positive Stimulation refers to the degree to which affiliation provides enjoyable affective and cognitive experiences (e.g., “I think that being close to other people, listening to them, and relating to them on a personal level is one of my favorite and most satisfying pastimes”). 
Social Comparison measures the ability to reduce uncertainty by acquiring self-relevant information (e.g., “When I am not sure how well I am doing at something, I usually like to be with other people so that I can compare myself with them”).

Evidence for the validity of the four dimensions was provided by their correlations with measures of constructs conceptually related to affiliation motivation, and by their lack of correlation with measures conceptually unrelated to affiliation motivation~\citep{HILL1987127}. Cronbach's $\alpha$ over all subjects for our sample is reported in Appendix~\ref{sup:iosquestionnaire}; internal consistency was adequate across the four subscales and comparable to that reported in the original work.

\subsubsection{The Scale of Trust in Automated Systems}\label{sec:stas}

The Scale of Trust in Automated Systems \citep[STAS;][]{Jian01032000} is designed to evaluate the trust that individuals place in automated systems. This scale consists of 12 items, each rated on a 7-point Likert scale ranging from 1 (\textit{Not at all}) to 7 (\textit{Extremely}), and addresses different aspects of the trust construct. The scale captures trust and distrust as polar opposites on a single dimension, rather than treating them as separate constructs. For example, the items evaluate perceptions of deception (e.g., “Agents' behavior is deceptive”), harm (e.g., “The actions of agents have harmful or damaging consequences”), security (e.g., “Agents make me feel safe”), integrity (e.g., “Agents have integrity”), dependability (e.g., “Agents are dependable”), reliability (e.g., “I can rely on agents”), and familiarity (e.g., “I know agents' behavior well”). The full questionnaire is provided in Appendix \ref{sup:stasquestionnaire}.

The total score is calculated as the mean of the 12 items, with higher scores indicating higher levels of trust in automated systems. Items related to distrust, such as perceptions of the system’s deceptiveness or its potential for harmful outcomes, are reverse-coded when generating the overall trust score. In our sample, the scale showed good internal consistency. Cronbach's $\alpha$ over all subjects who completed the questionnaire is reported in Appendix~\ref{sup:stasquestionnaire}.

\subsubsection{Autonomous Agent Teammate-Likeness}\label{sec:aat}
A questionnaire was implemented based on the Autonomous Agent Teammate-Likeness (AATL) model proposed by \citet{wynne2018integrative}. This construct is defined as “the extent to which a human operator perceives and identifies an autonomous, intelligent agent partner as a highly altruistic, benevolent, interdependent, emotive, communicative, and synchronized agentic teammate, rather than simply an instrumental tool” \citep[p.~3]{wynne2018integrative}. The authors identified six factors that theoretically explain variance in human operators’ perceptions of AATL: perceived agentic capability (autonomy), perceived benevolence/altruistic intent, perceived task interdependence, task-independent relationship-building, richness of communication, and synchronized mental model. They argue that these factors influence the perception of an intelligent agent as a teammate rather than merely as a tool. The authors provided empirical support for this construct and developed sample items for each factor within the model \citep{wynne2019autonomous}.

In the present study, we reviewed these items and selected those with acceptable factor loadings in the study by \citet{wynne2019autonomous}, while omitting items considered redundant or not applicable to our task. This process yielded a selection of 42 items corresponding to the six-factor AATL model, to which we added three communication items from a survey developed for the UBA Games Corpus \citep{gravano2023uba} and one autonomy item created for this study. Additionally, we added seven items to assess agent ability, four of which were developed \textit{ad hoc}, and three derived from the Godspeed Questionnaire Series \citep[GQS;][]{bartneckHRIMetrics2009}, which assesses perceptions in human-robot interaction.

The final questionnaire comprised 53 items (see Appendix \ref{sup:aatl} for the full list of questions) rated on a 5-point Likert scale, ranging from 1 (\textit{Strongly disagree}) to 5 (\textit{Strongly agree}), and grouped into seven subscales or factors that measure aspects of AATL and agent ability or intelligence: Autonomy (e.g., “\textit{[Agent Name]} can perform its tasks autonomously”), 
Altruism (e.g., “\textit{[Agent Name]} is truly focused on helping me”), 
Cooperative (e.g., “\textit{[Agent Name]} and I share responsibility for our actions”), 
Relationship-Building (e.g., “\textit{[Agent Name]} has a good rapport with me”), 
Communication (e.g., “\textit{[Agent Name]} and I exchange information successfully”), 
Synchronized Mental Model (e.g., “\textit{[Agent Name]} anticipates my needs”), and 
Ability (e.g., “\textit{[Agent Name]} is competent at solving the tasks”).
Each subscale score is calculated as the mean of its corresponding items, after reverse-coding negatively worded items, with higher scores indicating stronger agreement with the measured construct. The internal consistency of each subscale is reported in Appendix~\ref{sup:aatl}.

This questionnaire was answered by the subject twice during their session, once at the end of each block, to assess their final perception of each of the two agents. This is referred to as the final questionnaire. A shorter questionnaire, consisting of four items that addressed some of the aspects analyzed in the longer version, was conducted after each trial. This questionnaire is referred to as the intermediate questionnaire (see Appendix \ref{sup:aatl} for the corresponding subset of questions).

\subsection{Implementation Details}\label{sec:implementation}
The protocol was implemented as a web application. It consists of a Django backend and two React frontend systems. One of them was used for the subject side of the experiment. It included both the registration and asynchronous parts, as well as the synchronous part, which was connected to the second frontend system. The second frontend system was used by the experimenter to perform the WoZ role. A direct connection between the two frontends was established using PeerJS. Audio recordings were saved directly by the Django backend in WAV format. The databases containing all the information from the surveys, task performance, and logs were stored in a PostgreSQL database.

The agent responses were created using Microsoft’s speech synthesizer, with the ElenaNeural female voice in Argentine Spanish and the default settings.\footnote{\url{https://learn.microsoft.com/en-us/azure/ai-services/speech-service/language-support?tabs=tts}.}
The audio files were created in advance for the most common answers to reduce latency. For the less frequent answers, the audio was produced on-the-fly, which created some minor delays in the agents' response. In those cases, the agent played a message warning the subject about a possible delay, saying “Please wait a moment, I’m thinking.”

The WoZ acts as a natural language understanding stage -- a task that is performed by clicking buttons in two boards specifically designed for each of the stages in the game. As in a frame-based conversational agent \citep{jm3}, the buttons were designed so that all relevant questions from the subject could be encoded as a sequence of buttons. Further, out-of-task questions are represented through a small set of additional buttons (e.g., repeat request, non-game-related question, subject greeting the agent). Once the WoZ encodes the subject's question into a sequence of button clicks, this information is submitted to a backend application which formulates an agent response for this sequence using a decision tree implemented in Python.

\subsection{Dataset Collection}\label{sec:collection}
Inclusion criteria required subjects to be older than 18 and younger than 65 years and to have a Rioplatense accent (the accent from the Buenos Aires region). Subjects were also required to have normal color vision, and to have normal or corrected-to-normal visual acuity. Recruitment was conducted through the laboratory’s social media, emails sent to student mailing lists, and posters displayed on the university campus. The use of social media was intended to reach a community beyond the university. A total of 99 subjects took part in the experiment, of whom 61 selected “Female”, 37 selected “Male”, and 1 selected “Other” in the gender question (gender was self-reported). The mean age was 32.1 years ($SD = 10.2$). The resulting corpus, which we refer to as the \emph{\datasetname{}} dataset, comprises the responses to all survey instruments, the complete logs of the subjects' interactions with both agents, and the speech recordings collected during the experimental sessions, and is made publicly available for research use. 

All procedures in the study adhered to national and international ethical guidelines for research with human subjects and were reviewed and approved by the Institutional Review Board of Hospital General de Agudos J. M. Ramos Mejía (IRB No. 00001942; OHRP FWA 00001767; protocol registration code: 6215; approval date: November 30, 2021). The study was conducted under an exemption in accordance with 45 CFR §46.104, which permits the use of prospective consent procedures when full disclosure may compromise the research objectives and when the data collected cannot be used to harm subjects.
Accordingly, subjects provided prospective consent after receiving a general description of the study purpose and procedures. The consent included the public release of the survey responses, interaction logs and speech recordings for research purposes. Full details of the experimental protocol were disclosed to subjects by email after completion of the experiment. 

All subjects received financial compensation for taking part in the experiment. This consisted of a supermarket gift card. The amount did not vary across subjects and was not related to performance in the experiment, and was equivalent to approximately 15 USD.
Participation in the experiment was remote. Subjects used their own computer, microphone and headset. Further, they were asked to make sure that they had a stable internet connection during the game-playing session. 

Subjects initially communicated by email with a research assistant, who provided all instructions for taking part in the experiment. The email provided a link to a website where subjects had to create an account using only their email address. Subjects were then presented with a description of the protocol and, by clicking continue, provided their consent to participate in the experiment.
They then completed a check of the required devices by adjusting their headphones and recording an audio sample with their microphone. Once the research assistant had checked the audio quality, subjects whose recordings were unclear received an email asking them to correct the issue and repeat the recording. After this, subjects completed the sociodemographic, Big Five Inventory, Interpersonal Orientation, and the Trust in Automated Systems surveys described in Section~\ref{sec:surveys}.

After this first session was completed, subjects scheduled an appointment through a public website, DataPruebas \citep{juantorena2025datapruebas}, an application designed to streamline scientific data collection. The second session was synchronous, as the presence of the research assistant as WoZ was required -- though unknown to the subject. They first met with the research assistant via Google Meet so that the experimental conditions could be checked once more. During this initial interaction, the research assistant verified the microphone quality and ensured that background noise was kept to a minimum before the session began. Subjects were also asked to silence their cell phone and keep the environment as quiet as possible. The link to access the experimental platform was then shared during the Meet call.

The first task inside the platform involves another check of the headphones and microphone, both to confirm their quality and to verify the connection between the subject and the Wizard of Oz interface (see Section~\ref{sec:woz}). 
The second task in the second session was to read a phonetically balanced text and describe five scenes from a complex picture showing people interacting with one another (see Figure~\ref{sup:neutral-image} in the Appendix). The description had to be at least 30 seconds long, which was enforced by a timer. The speech recorded during these stages can be considered neutral, since no agent was involved, and is released as part of the dataset as a per-subject reference recording.

After these two recording stages, the game instructions were explained and subjects completed the practice block, designed to familiarize subjects with both the game and the interaction with the agent.
Finally, subjects played two blocks of four trials, one with the T agent and one with the N agent. Before the first interaction with each agent, the priming was presented to influence the subject's perception of the agent's abilities, communication and cooperation skills. Subjects were allowed a self-paced break in between the two blocks with a minimum duration of one minute. 
Subjects completed the intermediate surveys after each trial and the final survey after the last trial with each agent, as described in Section~\ref{sec:aat}. The estimated duration of this second session was approximately 60 minutes.
Figure~\ref{fig:protocol-flow} in the Appendix provides an overview of the full experimental protocol, including both sessions and their key components. We refer to this main version of the experiment, in which the priming favored the T agent, as the \textit{active-priming} version throughout the remainder of the paper.

To disentangle the contribution of the priming from that of the agents' behavioral differences, we conducted a second study collecting data from a smaller number of subjects in which the priming for the T agent was equivalent to that of the N agent. 
This allowed us to analyze whether the priming had a significant impact on human behavior and perception of the agent’s characteristics. A total of 18 subjects completed this version of the experiment, of whom 14 selected “Female”, 3 selected “Male”, and 1 selected “Other”. The mean age was 36.1 years ($SD = 8.9$). We refer to this group as the \textit{neutral-priming} version throughout the remainder of the paper. 

The platform included a ``resume'' option in case the session was interrupted due to internet connection problems. If the connection was lost during the session, subjects could log in again and continue the protocol from the same board at which the interruption had occurred. When this happened, the agent assigned to that block was presented again and, in the case of the experimental agents, the corresponding priming message was also shown again.

\section{Database Analysis}\label{sec:results}
The analyses reported below were conducted on the subset of subjects who completed the protocol without interruptions due to internet connectivity issues. As noted above, interrupted sessions could be resumed from the point at which the connection was lost; however, restarts might have caused frustration or distraction. To eliminate this source of uncontrolled variation, interrupted sessions were excluded from the analyses. 
Further, since the order in which the agents were presented may influence subjects' responses across trials and questionnaires and potentially introduce a spurious correlation with the condition, we selected an equal number of \textit{N-first} and \textit{T-first} sessions for each gender. A subject who selected ``Other'' as their gender was excluded since it was not possible to balance sessions within that gender group. In addition, we excluded subjects who did not follow the instructions for the task (more details can be found in Appendix~\ref{sup:low_score}). This resulted in a final sample of 74 sessions from 46 females (23 \textit{N-first} and 23 \textit{T-first}) and 28 males (14 \textit{N-first} and 14 \textit{T-first}). The mean age was $32.2$ years ($SD = 10.4$).  

For the neutral-priming version, we applied the same exclusion criteria as in the main sample, discarding interrupted sessions and one subject who selected ``Other'' as their gender. In this case, no sessions needed to be discarded due to insufficient task engagement. This yielded a final sample of 15 subjects (12 female, 3 male; mean age $= 35.6$ years, $SD = 9.5$).

\subsection{Analysis of Game Performance}
\label{sec:game_performance}
In this section, we analyze the impact of board difficulty, trial number, agent order, and agent type on game performance. 
As described in Section~\ref{sec:stimuli-scenarios}, subjects completed four trials per agent, following a fixed difficulty sequence: (1) easy; (2) hard; (3) medium; (4) hard. After completing Stage 2 of the game in each trial, subjects received feedback on their placement performance based on an overlap score computed as the fraction of the bounding box of the placed object that falls inside the bounding box of the object in the correct position. The feedback is categorized as Excellent (overlap $> 70\%$), Good (from $40\%$ to $70\%$), Fair (from $0\%$ to $40\%$), or Poor (no overlap).  The overlap score adequately reflects the spatial correspondence between the subject's placement of the target object and its correct position when there is some overlap between the two. When there is no overlap, though, this score is always 0, independently of the relative distance between the correct position and the subject's placement. For this reason, for some analyses in this section, we complemented the overlap score with a distance score, defined as the Euclidean distance between the center of the object as placed by the subject and the center of the object in the correct location. 

\subsubsection{Impact of Difficulty and Trial Number on Game Performance}

\begin{figure}
    \centering
    \includegraphics[width=0.8\textwidth]{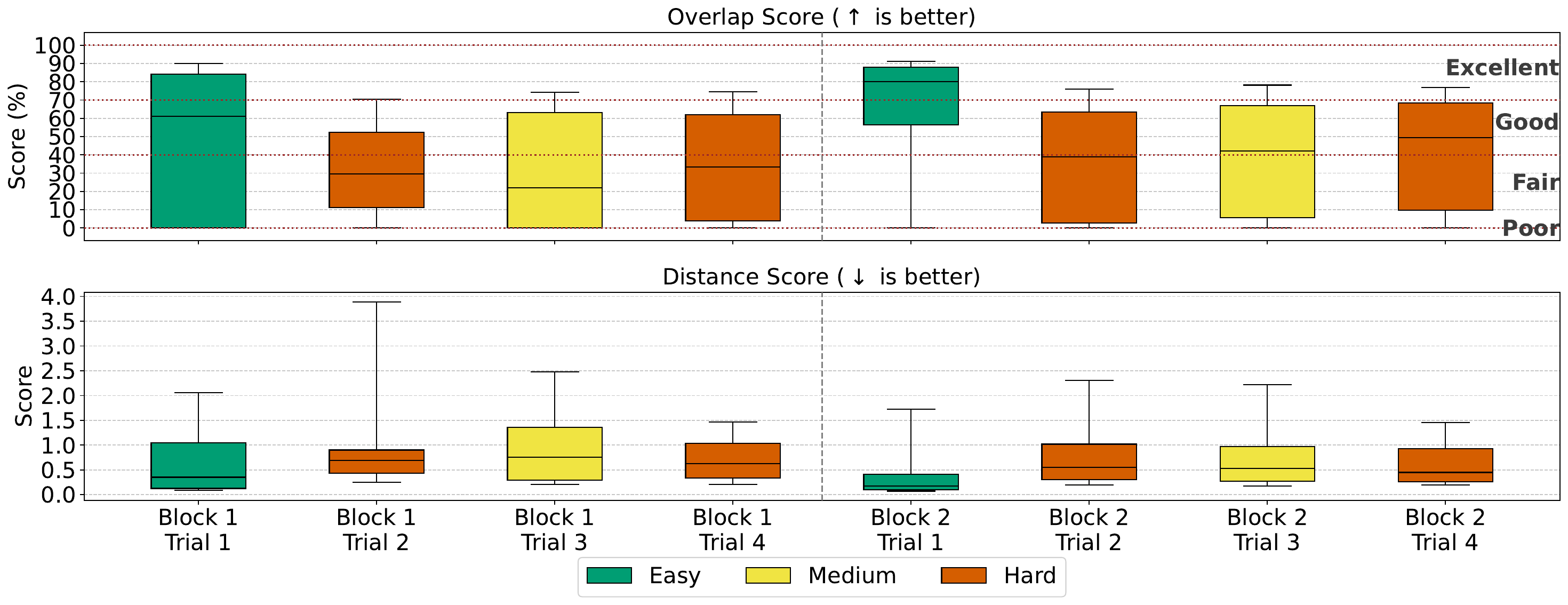}
    \caption{Distribution of performance scores across trial number, separated by block order (block 1 vs.\ block 2, divided by the dashed vertical line). The upper panel shows the overlap score (higher values indicate better performance) and the lower panel shows the distance score (lower values indicate better performance). In the upper panel, the dashed horizontal lines and right-axis labels correspond to the categorical performance feedback shown to the subject (Excellent, Good, Fair, Poor). Boxes show the median and the first and third quartiles; whiskers extend to the 10th and 90th percentiles. The easy trial yielded better scores than the medium and hard trials in both blocks, and overall performance was better in block 2 than in block 1, consistent with a training effect.}
    \label{fig:scores_by_difficulty}
\end{figure}

Figure~\ref{fig:scores_by_difficulty} shows the distribution of overlap and the distance scores for each trial within each block across all sessions. We report median values, given their robustness to outliers. In the first block, the median overlap score was $61.10\%$ (inter-quartile range IQR $ = 83.98$) for the easy trial (trial 1), corresponding to a Good rating, while the first hard trial (trial 2), the medium trial (trial 3), and the second hard trial (trial 4) yielded median overlap scores of $29.60\%$ (IQR $ = 41.08$), $21.90\%$ (IQR $ = 63.10$), and $33.35\%$ (IQR $ = 57.95$), respectively, all falling in the Fair range. The distance score followed a similar pattern, though in the opposite direction since larger overlap implies smaller distance, with a median of $0.35$ (IQR $ = 0.92$) for the easy trial, $0.69$ (IQR $ = 0.47$) for the first hard trial, $0.76$ (IQR $ = 1.06$) for the medium trial, and $0.63$ (IQR $ = 0.70$) for the second hard trial.

In the second block (block 2), overlap scores were higher overall, particularly for the easy trial, which reached a median of $79.95\%$ (IQR $ = 31.63$), now within the Excellent range. The first hard trial yielded an overlap score of $38.90\%$ (IQR $ = 60.63$), remaining in the Fair range but higher than its first-block counterpart, whereas the medium trial (median of $42.00\%$, IQR $ = 61.30$) and the second hard trial (median of $49.40\%$, IQR $ = 58.63$) both crossed into the Good range. Distance scores were correspondingly better than in the first block for the easy and second hard trials, with median of $0.18$ (IQR $ = 0.31$) and $0.45$ (IQR $ = 0.66$), respectively, while the first hard and medium trials had a distance score comparable to the first block, with median $0.55$ (IQR $ = 0.72$) and $0.53$ (IQR $ = 0.70$).

This pattern indicates that the difficulty manipulation was, overall, consistent with subjects' performance, with the easy trial yielding higher overlap scores and lower distance scores than the hard and medium trials in both blocks. At the same time, the comparison between the first and second block reveals a distinct behavior consistent with a training effect. 
Further, both performance metrics lead to the same overall conclusion regarding the effect of difficulty and block order on performance. 

\subsubsection{Impact of Agent Type on Game Performance}\label{sec:o1-analysis}

\begin{figure}
    \centering
    \includegraphics[width=0.8\textwidth]{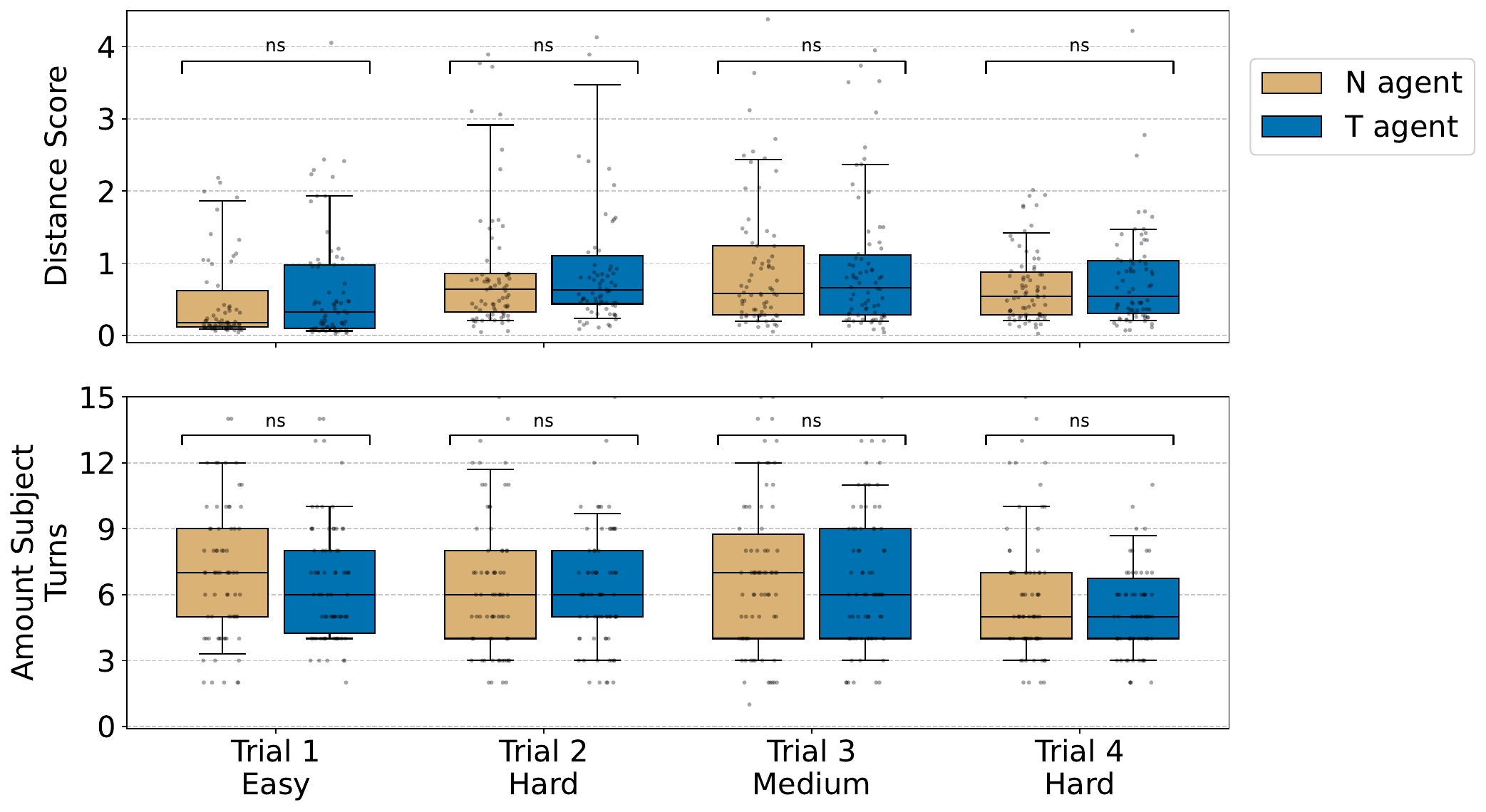}
    \caption{Distribution of distance score (top) and number of subject turns (bottom) per trial and agent condition (N agent: neutral tool-like; T agent: team-building). Each dot represents one subject. Boxes show the median and the first and third quartiles; whiskers extend to the 10th and 90th percentiles. Significance annotations at the top of each panel correspond to one-sample $t$-tests on the per-subject difference score $\Delta_{ik} = m_{ik}^{T} - m_{ik}^{N}$, Bonferroni-corrected across the four trials within each metric (ns: not significant).
    Neither the distance score nor the number of subject turns differed significantly between the two agents at any trial, indicating that the agent type had no measurable effect on game performance.}
    \label{fig:o2_by_trial}
\end{figure}

Next, we explored whether the agent type affected game performance (H1). The top plot in Figure~\ref{fig:o2_by_trial} shows the distribution of distance score per trial and agent, pooling the trials across both blocks. The bottom plot in this figure shows the distribution of number of speech turns or interactions that the subject had with the agent in the localization stage of the game. We hypothesized that the T agent might make the subject more comfortable which may, in turn, elicit more interactions. A larger number of interactions may then result in a more accurate estimate of the position and, hence, in better performance. 

The figure shows that neither of these hypotheses is supported by our data as the distribution of score and number of turns is similar for both agents. 
Significance is calculated with a t-test on the per-subject difference, $\Delta_{ik} = m_{ik}^{T} - m_{ik}^{N}$, where $m_{ik}$ denotes the metric value (distance score or number of turns) for subject $i$ at trial $k$ under each agent condition. The p-values are Bonferroni-corrected across the four trials within each metric. As shown in Figure~\ref{fig:o2_by_trial}, no trial exhibited a significant difference between the two agents.
Repeating this analysis but restricted to only the sessions in which the N agent was first or the T agent was first led to the same conclusion.

\begin{center}
\setlength{\fboxsep}{5pt}
\fcolorbox{White}{LightBlue}{
  \begin{minipage}{0.9\textwidth}
    \textcolor{Black}{In summary, results indicate that the team-building agent did not have a significant impact on game performance. This observation is inconsistent with our hypothesis (H1) which was based on the assumption that subjects would have longer interactions with the T agent, potentially resulting in a more precise localization of the object. Yet, in practice, subjects did not ask significantly more questions when interacting with the T agent than with the N agent, and, as a consequence, game performance was not significantly different across agent types.}
  \end{minipage}
}
\end{center}

\subsection{Analysis of the Subject's Perception of the Agent's Teammate-Likeness}\label{sec:o2-analysis}

In this section, we study the impact that the agent type  had on the AATL scores (see Section \ref{sec:aat}). We first explore the impact on the final survey after each block and then examine whether the effect was different depending on the order in which the agents were presented. We then explore the effect on the intermediate AATL scores. Next, we compare the conclusions between the main experiment, where the T agent differs from the N agent both in terms of priming and behavior, and the neutral-priming version where the T agent only differs from the N agent in behavior. Finally, we explore the influence of individual traits on both the mean scores across agents and their difference.

\subsubsection{Impact of the Agent Type on the Final Survey Scores}
To examine whether there are significant differences in the perception of the autonomous agent depending on the agent type \textbf{(O2)}, we compared the final questionnaire scores between the T and N agent conditions across all seven AATL subscales, which reflect different aspects of the human's perception of the agent's teammate-likeness, as well as between mean scores across subscales.  
To compute statistical significance for each subscale we conduct a one-sample t-test for the difference between the score for the T agent and N agent calculated as $\Delta s_{ik} = s_{ik}^{T} - s_{ik}^{N}$,  where $s_{ik}$ is the score for subject $i$ and subscale or mean across all subscales, $k$. A positive value for the difference score indicates that the subject perceived the T agent more favorably on that subscale. The null hypothesis is that the difference is equal to 0. We verified that the difference scores, $\Delta s_{ik}$, were approximately normally distributed, with absolute skewness below $0.75$ for all subscales, satisfying the assumptions of the $t$-test. The resulting $p$-values were corrected for multiple comparisons across the seven AATL subscales using the Bonferroni method. The p-value for the mean was not corrected for multiple comparisons as it constitutes a derived summary of the seven subscales rather than an independent comparison.

\begin{figure}
    \centering
    \includegraphics[width=0.9\textwidth]{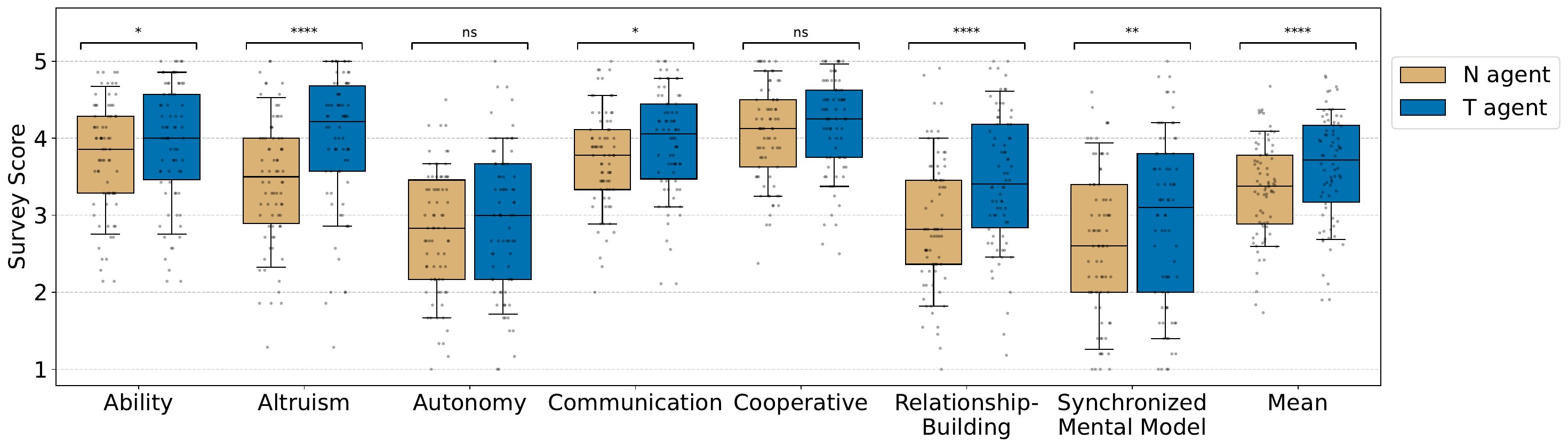}
    \caption{Distribution of final Autonomous Agent Teammate-Likeness (AATL) questionnaire scores (scores range from 1 to 5, higher is more teammate-like) across all seven subscales and their mean, by agent condition (N agent vs.\ T agent). Both agents were identical in the quality and quantity of game-relevant information provided in their answers, differing only in communicative behavior.
    Boxes show the median and the first and third quartiles; whiskers extend to the 10th and 90th percentiles. Each dot represents one subject. Significance values are based on one-sample $t$-tests on the per-subject $\Delta s_{ik}$ difference score for each subscale. The Bonferroni correction for multiple comparisons is applied across the seven subscales. 
    (* \textbf{$p_{\mathit{adj}}$} $< .05$, 
    ** \textbf{$p_{\mathit{adj}}$} $< .01$, 
    *** \textbf{$p_{\mathit{adj}}$} $< .001$, 
    **** \textbf{$p_{\mathit{adj}}$} $< .0001$, 
    ns: not significant).
    The T agent was rated higher than the N agent on all seven subscales and on their mean, significantly so on five of the seven cases, including Ability, despite the two agents being equally capable by design.}
    \label{fig:o1-boxplots}
\end{figure} 

Figure~\ref{fig:o1-boxplots} shows the distribution of the scores for each of the two agent types along with the adjusted p-values. 
The figure shows that the T agent received higher scores than the N agent across all seven subscales and their mean, with most of the differences being statistically significant with p-values below 0.05. 
The largest effects were observed for 
Relationship-Building ($M = 0.557$, $SD = 0.671$, $p_{\text{adj}} < 0.0001$, where $M$ stands for the mean of the difference and $SD$ for its standard deviation),
Altruism ($M = 0.517$, $SD = 0.671$, $p_{\text{adj}} < 0.0001$),
the mean ($M = 0.284$, $SD = 0.443$, $p < 0.0001$),
Synchronized Mental Model ($M = 0.281$, $SD = 0.679$, $p_{\text{adj}} = 0.0049$) and 
Communication ($M = 0.200$, $SD = 0.535$, $p_{\text{adj}} = 0.0147$). 
Moderate effects were found for 
Ability ($M = 0.183$, $SD = 0.540$, $p_{\text{adj}} = 0.0343$), while 
Autonomy ($M = 0.142$, $SD = 0.628$, $p_{\text{adj}} = 0.4010$) and 
Cooperative ($M = 0.111$, $SD = 0.480$, $p_{\text{adj}} = 0.3567$) showed smaller non-significant differences between conditions.

The statistically-significant difference in the Ability subscale, which reflects the subject's perception of the agent's competence, accuracy and effectiveness at solving the task, is particularly interesting since both agents were identical in terms of the quality and quantity of game-relevant information provided in their answers. This suggests that the implemented team-building behavior positively influenced the subjects' perception of the agent's ability, despite the fact that their actual ability was identical.

\subsubsection{Impact of the Agent Type on the Final Survey Scores by Agent Order}\label{sec:o2-final}

We hypothesized that the order in which subjects interacted with the two agents may have influenced their questionnaire responses, since ratings for the second agent could be influenced by an implicit comparison with the first. To test this hypothesis, we repeated the previous analysis separately for the two order groups (\textit{N-first} and \textit{T-first}). For each group, we computed the same $\Delta s_{ik}$ scores per-subject and tested them against zero with one-sample $t$-tests, Bonferroni-corrected across the seven subscales. In addition, to test whether the agent order affected the perceived advantage of the T agent, we compared the $\Delta s_{ik}$ scores between the two order groups using a two-sample $t$-test for each subscale.

\begin{figure}
    \centering
    \includegraphics[width=\textwidth]{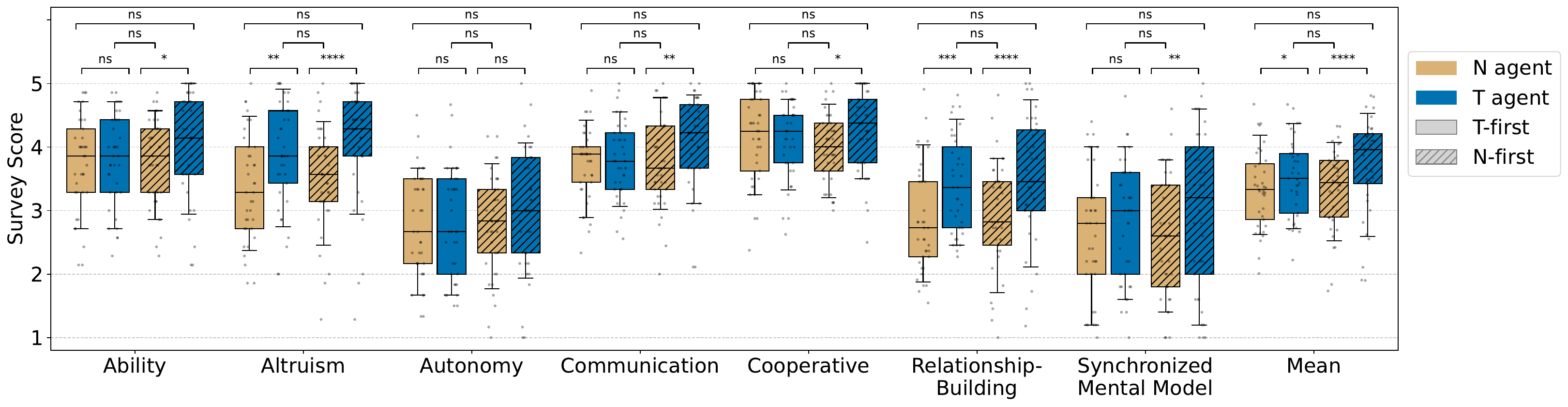}
    \caption{Distribution of final AATL questionnaire scores across all seven subscales and their mean, by agent condition (N agent vs.\ T agent) and agent order (\textit{T-first} vs.\ \textit{N-first}). Boxes show the median and the first and third quartiles; whiskers extend to the 10th and 90th percentiles. Each dot represents one subject. The brackets within each order group show significance values based on one-sample $t$-tests on the per-subject $\Delta s_{ik}$ difference score, computed separately for each order. The top row of significance markers corresponds to two-sample $t$-tests comparing the $\Delta s_{ik}$ difference between the two order groups. The middle row corresponds to two-sample $t$-tests restricted to the first block, comparing the scores that \textit{T-first} subjects gave to the T agent against the scores that \textit{N-first} subjects gave to the N agent. The Bonferroni correction for multiple comparisons is applied across the seven subscales within each set of tests; the mean is reported uncorrected. 
    (* \textbf{$p_{\mathit{adj}}$} $< .05$, 
    ** \textbf{$p_{\mathit{adj}}$} $< .01$, 
    *** \textbf{$p_{\mathit{adj}}$} $< .001$, 
    **** \textbf{$p_{\mathit{adj}}$} $< .0001$, 
    ns: not significant).
    The T agent's advantage on the two social subscales (Relationship-Building and Altruism) and on the across-subscale mean held under both orders, whereas its advantage on the remaining subscales reached significance only in the \textit{N-first} group. The first-block-only comparison of scores across agents and the across-order comparison of differences, however, were not significant for any subscale.}
    \label{fig:o1-boxplots-agent_order}
\end{figure}

Figure~\ref{fig:o1-boxplots-agent_order} shows that a significant preference for the T agent was found in terms of Relationship-Building (\textit{T-first}: $M = 0.538$, $SD = 0.699$, $p_{\mathit{adj}} = 0.0007$; \textit{N-first}: $M = 0.575$, $SD = 0.641$, $p_{\mathit{adj}} = 0.0001$) and Altruism (\textit{T-first}: $M = 0.506$, $SD = 0.781$, $p_{\mathit{adj}} = 0.0059$; \textit{N-first}: $M = 0.529$, $SD = 0.538$, $p_{\mathit{adj}} < 0.0001$), and the across-subscale mean (\textit{T-first}: $M = 0.191$, $SD = 0.461$, $p = 0.0179$; \textit{N-first}: $M = 0.378$, $SD = 0.404$, $p < 0.0001$), regardless of the order in which they were presented.
Interestingly, the impact of the condition was only significant when subjects interacted with the N agent first for four out of the five remaining subscales: Ability ($M = 0.305$, $SD = 0.508$, $p_{\mathit{adj}} = 0.0133$), Communication ($M = 0.339$, $SD = 0.477$, $p_{\mathit{adj}} = 0.0019$), Synchronized Mental Model ($M = 0.416$, $SD = 0.674$, $p_{\mathit{adj}} = 0.0099$), and Cooperative ($M = 0.253$, $SD = 0.475$, $p_{\mathit{adj}} = 0.0403$). Only Autonomy was not significantly different across conditions for both orders, as was also the case when pooling both orders (Figure \ref{fig:o1-boxplots}).

The two-sample $t$-test comparing the $\Delta s_{ik}$ distributions across N-first and T-first sessions showed no significant difference for any of the subscales, with the Cooperative subscale showing the largest difference yet still non-significant after Bonferroni correction ($M_{\text{T-first}} = -0.030$, $M_{\text{N-first}} = 0.253$, $p_{\mathit{raw}} = 0.0106$, $p_{\mathit{adj}} = 0.0740$). The within-group patterns should therefore be interpreted with caution. 
Nonetheless, the fact that $\Delta s_{ik}$ is larger for the \textit{N-first} sessions than the \textit{T-first} sessions in four of the subscales, most notably on the Cooperative subscale, is suggestive of a contrast effect in subjects' prior expectations about each agent. Having first interacted with the neutral agent sets a comparatively low baseline, so that when subjects subsequently meet the T agent, its active behavior makes its advantage salient even on subscales on which the two agents were in fact identical, such as Ability and Synchronized Mental Model. By contrast, subjects who started with the T agent had no such neutral reference, which may explain why their ratings remained similar across the two agents on these dimensions. 
This order sensitivity did not extend to the two markedly social subscales, Relationship-Building and Altruism, on which the T agent was rated as superior irrespective of order, a pattern that may reflect its informal language and its proactive initiative at the start and end of each trial.
This dissociation suggests that the strongly marked social signals built into the T agent are salient enough to register on their own, whereas its perceived advantage on task-functional subscales could emerge only through a direct comparison with the neutral baseline.

Finally, we leveraged the order manipulation to obtain an entirely between-subjects comparison of the two agents, without any contrast effects. Considering only each subject's first block, half of the subjects interacted with the T agent (\textit{T-first} group) and the other half with the N agent (\textit{N-first} group), so their first-block ratings can be compared directly with two-sample $t$-tests (middle row of significance markers in Figure~\ref{fig:o1-boxplots-agent_order}). No subscale reached significance after Bonferroni correction. Nonetheless, the T agent was rated numerically higher on every subscale and on the across-subscale mean, with the largest differences again on the social subscales: Relationship-Building ($M_{\text{T}} = 3.388$ vs.\ $M_{\text{N}} = 2.943$) and Altruism ($M_{\text{T}} = 3.876$ vs.\ $M_{\text{N}} = 3.537$). Conversely, first-block ratings of Ability were virtually identical across groups ($M_{\text{T}} = 3.807$ vs.\ $M_{\text{N}} = 3.764$), consistent with the idea that the T agent's perceived advantage on task-functional dimensions emerges only when subjects can directly compare it against the neutral baseline. Thus, even with a single block of interaction and half the sample per group ($n = 37$), the between-subjects comparison already trends in the same direction as the within-subject analyses, suggesting that the T agent's perceived advantage on the social dimensions does not depend on subjects directly experiencing both agents, although this design lacks the statistical power to establish it reliably.

\subsubsection{Impact of the Agent Type on the Intermediate Survey Scores}

\begin{figure}
    \centering
    \includegraphics[width=\textwidth]{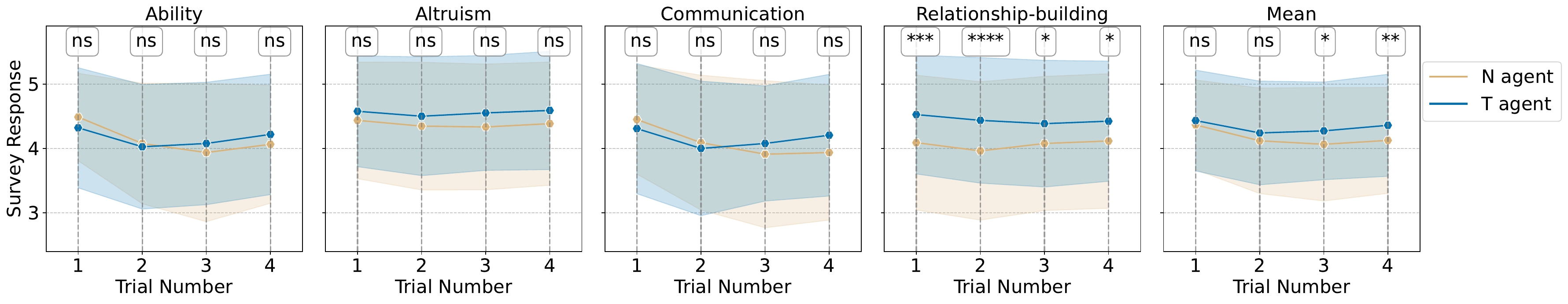}
    \caption{Mean intermediate questionnaire scores across trials per AATL subscale and agent condition (N agent vs.\ T agent). Shaded areas represent standard deviation. Values at the top of each panel indicate the significance obtained with $t$-tests on the per-subject $\Delta s_{ik}$ difference score. All $p$-values are Bonferroni-corrected jointly across the four trials and the four AATL subscales presented in the intermediate questionnaire (
    * $p_{\mathit{adj}} < .05$, 
    ** $p_{\mathit{adj}} < .01$, 
    *** $p_{\mathit{adj}} < .001$, 
    **** $p_{\mathit{adj}} < .0001$, 
    ns: not significant).
    Relationship-Building across all four trials and the across-subscale mean (trials 3 and 4) showed significant differences between agents, a weaker pattern on the remaining subscales than in the final questionnaire that is likely explained by the intermediate questionnaire using a single item per subscale.}
    \label{fig:o1-lineplots}
\end{figure}

Next, we examine the impact of the agent type on the intermediate AATL survey questions filled by the subjects after each trial.
Figure~\ref{fig:o1-lineplots} presents the trajectory of these scores across trials by agent condition. As explained above, the intermediate questionnaire contained a subset of four subscales, with a single question per subscale. To assess whether differences between agent conditions were significant at each trial, we conducted one-sample $t$-tests on the difference scores $\Delta s_{ik}$ for each trial and each AATL subscale and the mean, with Bonferroni correction jointly across the 16 comparisons resulting from combining the four trials and the four AATL subscales. The significance of each comparison is indicated at the top of each panel in Figure~\ref{fig:o1-lineplots}. Relationship-Building showed significant differences across all four trials (with the strongest effects at trials 1 and 2), while the mean reached significance at trials 3 and 4. The remaining subscales did not reach significance at any individual trial.
This is likely due to the fact that intermediate surveys contain a single question per subscale, reducing the robustness of the scores. In fact, a similar significance pattern is observed for the scores from the final questionnaire for the same questions used for the intermediate questionnaires, supporting this hypothesis (results not shown).

\subsubsection{Comparison with Neutral-Priming Sessions}

To assess whether the priming shown before the subjects' interaction with the agents had any effect on the subject's perception of the agent's teammate-likeness, we performed a two-sample $t$-test
comparing the distribution of the $\Delta s_{ik}$ for the 74 subjects who saw the introductory feedback designed to portray the T agent more favorably as a teammate, with that for the 15 subjects in the neutral-priming group who saw similar priming for both agents. As shown in Figure~\ref{sup:with-or-without-priming} in the Appendix, after Bonferroni correction none of the subscales, nor the across-subscale mean, showed a significant difference between the two versions (all $p_{\mathit{adj}} > 0.05$), suggesting that the priming did not significantly affect how subjects perceived the difference between the two agents. Notably, the mean $\Delta s_{ik}$ is larger for the neutral-priming sessions than for the active-priming sessions for the two subscales that show the most robust effect in both agent orders: Altruism and Relationship Building.

\begin{center}
\setlength{\fboxsep}{5pt}
\fcolorbox{White}{LightBlue}{
  \begin{minipage}{0.9\textwidth}
    \textcolor{Black}{In summary, our results support the hypothesis (H2) that the team-building agent was favored by the subjects compared to the neutral agent. Further, we found no evidence to conclude that the priming played a significant role, suggesting that it was the minor changes in behavior that positively impacted the subjects' perception.}
  \end{minipage}
}
\end{center}

\subsubsection{Impact of the Subjects' Individual Traits on Final Survey Scores}
Finally, we examine whether there are significant differences in how the human perceives the autonomous agent, captured by the AATL surveys, depending on their self-reported individual traits \textbf{(O3)}. These were operationalized through three instruments: the Big Five Inventory (Section~\ref{sec:bfi}), the Interpersonal Orientation Scale (Section~\ref{sec:ios}), and the Scale of Trust in Automated Systems (Section~\ref{sec:stas}).
We considered two representations derived from the AATL scores, the difference and the mean, each computed for every subscale $k$ as well as for the across-subscale mean. The difference score is computed, as before, as $\Delta s_{ik} = s_{ik}^{T} - s_{ik}^{N}$, and captures the extent to which subject $i$ favored the T agent over the N agent on measure $k$. The mean score is computed as $\bar{s}_{ik} = (s_{ik}^{T} + s_{ik}^{N})/2$, and reflects the subject's overall bias in their perception of an agent's teammate-likeness.

We hypothesized that individuals with a higher propensity to trust, a higher need for affiliation, greater extraversion, or higher openness to experience would tend to perceive both agents with a higher degree of teammate-likeness (H3), resulting in higher mean scores. To test this hypothesis, we fitted an ordinary least squares (OLS) regression for each AATL subscale as well as for their mean, using as predictors the subscale-level scores of the three instruments (BFI with 5 scores, IOS with 4 scores, and STAS with a single score). Both the predictors and the outcome were standardized (z-scored) prior to fitting, so that the estimated coefficients correspond to standardized regression weights and are directly comparable across predictors.

\begin{figure}
    \centering
    \includegraphics[width=\textwidth]{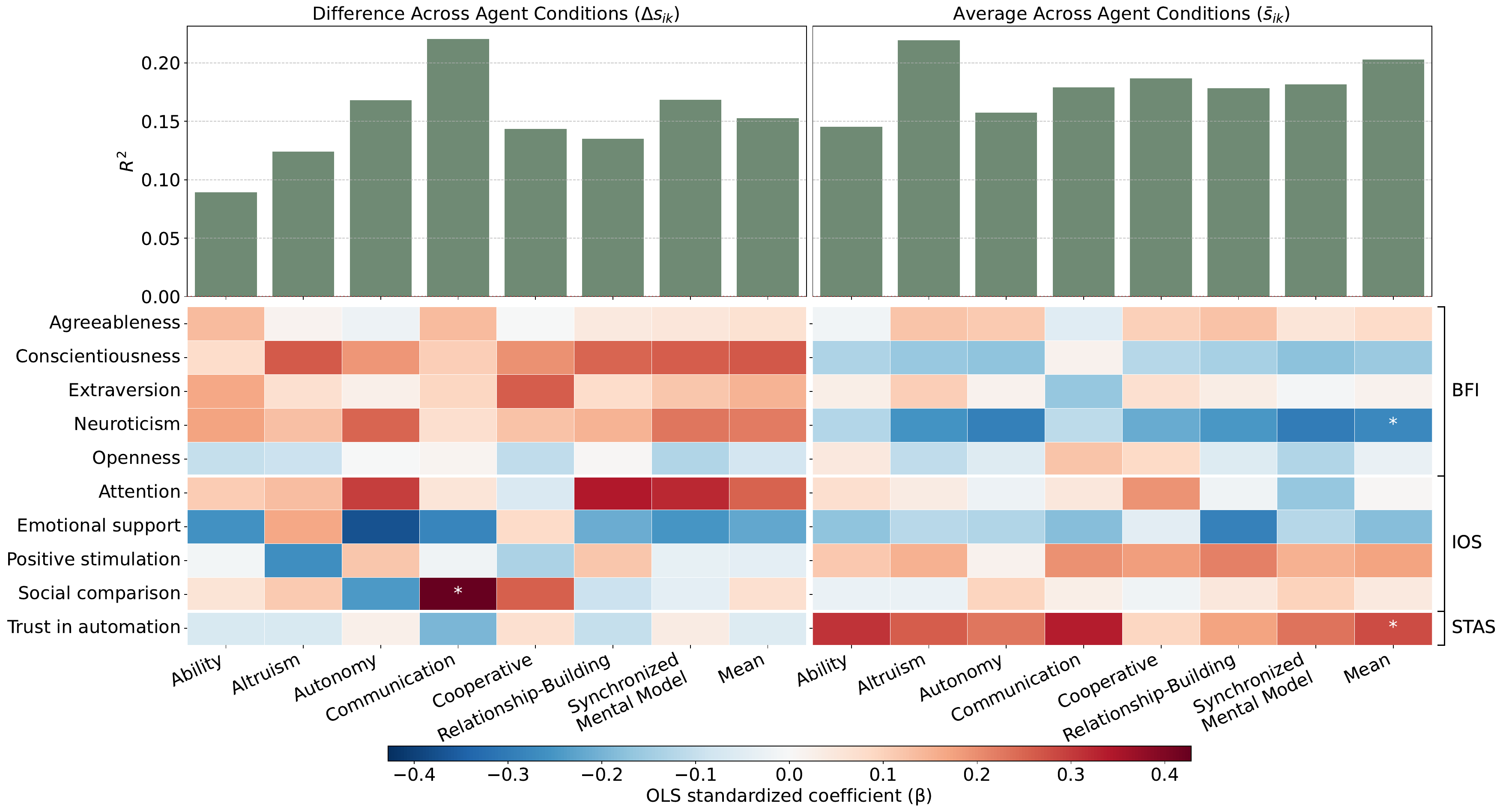}
    \caption{OLS regression results for predicting final AATL questionnaire scores, for the difference across agent conditions $\Delta s_{ik}$ (left) and the mean across agent conditions $\bar{s}_{ik}$ (right), using individual-trait measures (BFI, IOS, and STAS subscales) as predictors. Predictors and outcomes were standardized (z-scored) prior to fitting, and each model was fitted on the full sample. Top: in-sample $R^2$ for each AATL dimension. Bottom: standardized OLS coefficient ($\beta$) of each predictor (rows) for each AATL dimension (columns). Asterisks denote statistical significance (
    * $p_{\mathit{adj}} < .05$, 
    ** $p_{\mathit{adj}} < .01$, 
    *** $p_{\mathit{adj}} < .001$, 
    **** $p_{\mathit{adj}} < .0001$, 
    ns: not significant) with Bonferroni adjustment.
    Trust in automation (STAS) was positively associated and Neuroticism (BFI) negatively associated with the mean across subscales of the mean score across agents, whereas social comparison (IOS) was positively associated with the difference in Communication scores across agents.}
    \label{fig:o3-regression}
\end{figure}

The OLS models yielded in-sample $R^2$ values between $0.089$ and $0.221$ across AATL subscales (Figure~\ref{fig:o3-regression}, top row). Since these values reflect the fit obtained on the full sample rather than out-of-sample predictive performance, we do not interpret their magnitude beyond the following observations. For $\bar{s}_{ik}$, the quality of the fit was comparable across subscales, ranging from $0.14$ to $0.22$. For $\Delta s_{ik}$, in contrast, the fit varied more substantially across subscales, from $0.09$ for Ability to $0.22$ for Communication, suggesting that the difference in rating for Ability is harder to predict from the individual-trait measures than for Communication. 

Next, we analyze the estimated standardized coefficients and their statistical significance (Figure~\ref{fig:o3-regression}, bottom block), which identify which individual-trait dimensions are associated with the perceived teammate-likeness of the agent.
For the difference scores, $\Delta s_{ik}$, only one association was significant after Bonferroni correction: a stronger tendency toward social comparison (IOS) was associated with a larger $\Delta s_{ik}$ on Communication ($p_{\mathit{adj}}<0.05$). 
For the average scores $\bar{s}_{ik}$, the propensity to trust in automation (STAS) was positively associated with the across-subscale mean ($p_{\mathit{adj}}<0.05$), while Neuroticism (BFI) was negatively associated with it ($p_{\mathit{adj}}<0.05$). No predictor was significant for the individual subscales, although the coefficients for STAS were consistently positive, and those for Neuroticism consistently negative, across subscales. These results partially support H3: individuals with a higher propensity to trust perceived the agents as more teammate-like, and the negative association with neuroticism is consistent with the expected direction for emotional stability. On the other hand, contrary to our expectations, we found no significant associations with the across-agent mean for extraversion, openness, or need for affiliation.

\begin{figure}
    \centering
    \includegraphics[width=0.8\textwidth]{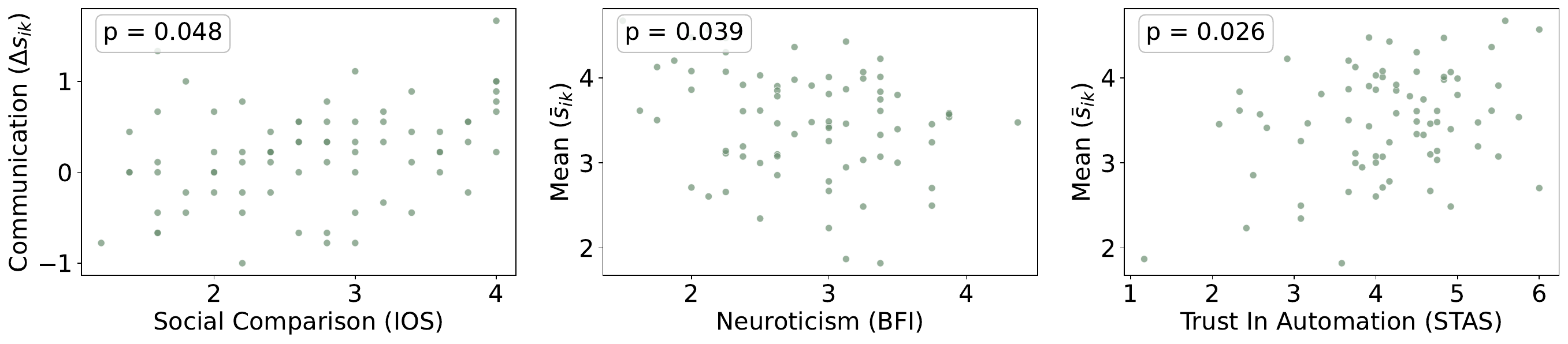}
    \caption{Scatter plots of the three strongest associations identified by the OLS models (Figure~\ref{fig:o3-regression}). Each point represents one subject. Left: the need for Social Comparison (IOS) against the difference across agent conditions in the Communication subscale of the AATL ($\Delta s_{ik}$). Middle and right: Neuroticism (BFI) and the propensity to Trust in Automation (STAS), respectively, against the subject's mean AATL score, computed for each agent and then averaged across the two agent conditions ($\bar{s}_{ik}$). Reported $p$-values correspond to the standardized coefficient of each predictor in the corresponding OLS model, after Bonferroni correction.
    Although statistically significant, the dispersion of the raw data shows that these three associations correspond to weak trends.}
    \label{fig:o3-scatter}
\end{figure}

Figure~\ref{fig:o3-scatter} shows scatter plots for the three strongest associations identified by the OLS models. The x-axis corresponds to one significant predictor and the y-axis to the observed score. Each sample in these plots corresponds to one subject. Subjects with a stronger tendency toward social comparison (IOS) have a subtle tendency to differentiate more between the two agents on the Communication subscale of the AATL ($\Delta s_{ik}$, $p_{\text{adj}}=0.048$). In turn, the across-subscale mean score across agents ($\bar{s}_{ik}$) was negatively associated with neuroticism (BFI, $p_{\text{adj}}=0.039$) and positively associated with the propensity to trust in automation (STAS, $p_{\text{adj}}=0.026$). The dispersion of the raw data illustrates that these associations are only weak trends.

\begin{center}
\setlength{\fboxsep}{5pt}
\fcolorbox{White}{LightBlue}{
  \begin{minipage}{0.9\textwidth}
    \textcolor{Black}{In summary, these results provide partial support for (H3): some individual traits -- trust in automation and neuroticism -- partially explain the variance in the mean scores of teammate-likeness across both agents.}
  \end{minipage}
}
\end{center}

\subsection{Analysis of the Impact of the Agent Type on the Subjects' Speech}\label{sec:o4-analysis}

One of the goals of our study (\textbf{O4}) is to examine whether the team-building behavior of the T agent has an impact on the subject's speech characteristics. \citet{Meza2023} studied speech characteristics in human-human teams during a game similar to ours. They used features describing prosodic, voice quality, spectral, and temporal characteristics of the speech, as well as features encoding the duration and number of speaking turns. Similar features were used in works studying the effects of trust and distrust on the speech signal -- aspects that have been shown to be related to team-work quality \citep{GAUDER2023101487, Li2022}.

In our dataset, each subject contributed 18 audio recordings: 2 corresponding to the initial recordings (see Appendix~\ref{sup:db_collection}), and 16 corresponding to the 8 experimental trials, with 2 recordings per trial, one for each stage of the game. Here, we restrict our analysis to Stage 2 (Target Localization; see Section~\ref{sec:task}), as this stage elicits richer, longer, and more variable speech patterns than Stage 1. Hence, our analysis uses 8 recordings per subject, 4 per agent condition (T and N agent), one for each trial.

Recordings were made continuously for the duration of each trial. Hence, they contained silences as well as, in some cases, the agent's speech leaking through the headsets into the microphone. These segments were located and discarded. Next, we used the Silero voice activity detector\footnote{\url{https://github.com/snakers4/silero-vad}} to obtain the intervals during which the subject was speaking. Intervals were defined as regions of speech separated by pauses no longer than $0.5$ seconds. All segments were then audited to verify that no speech other than the subject's was present.

For each speech interval, two \emph{hand-crafted} feature sets were extracted. The first set is given by the mean and standard deviation of the fundamental frequency (F0) and loudness sequences, which were extracted with the OpenSMILE toolkit \citep{eyben2010opensmile}. This yields a 4-dimensional vector per interval. As our unit of analysis is the trial, the mean and the standard deviation of these vectors were then computed across all intervals of a trial, resulting in a single 8-dimensional representation per trial. The second feature set consists of the number of intervals per trial and the mean interval duration. These two features were concatenated to the F0 and loudness features, resulting in a 10-dimensional vector per trial.

As an alternative, we also used \emph{learned} prosodic representations based on the work by \citet{meza2026}, where an auto-encoder model is trained to extract a fixed-dimensional feature vector representing the prosodic information in a speech signal. A description of this model is included in Appendix \ref{sup:prosodic-embeddings}.
We used a trained model kindly provided by the authors to extract 16-dimensional prosodic features for each speech segment in our data. These prosodic representations were then fused across the segments of a trial following the same process used for the hand-crafted features.

The hand-crafted or learned speech features are used as input to a classifier, which is trained to predict the agent type. If this prediction is successful, we may conclude that the speech patterns, as represented by these features, have a different distribution depending on the agent type. Predictions are made at the block level, each block consisting of the 4 trials performed with the same agent. We combine the trial-level information within a block by late fusion: each trial-level feature vector is fed to the model separately, and the block-level score is obtained by averaging the model's output scores over the 4 trials.

As classifier, we use the random forest implementation from \texttt{scikit-learn}~\citep{scikit-learn}, with 500 trees, unlimited tree depth, and a minimum of 2 samples required to split a node. Models are evaluated with a 7-fold cross-validation scheme in which folds are defined at the speaker level and stratified by agent order (\textit{T-first} or \textit{N-first}) and by gender. To assess the variance due to the training dataset, the whole procedure is repeated 10 times with different seeds to determine the cross-validation folds. Performance is measured at the block level using the area under the ROC curve (AUC), computed by pooling the test predictions across the folds. Confidence intervals are estimated by bootstrapping these pooled predictions 100 times for each seed, pooling all resulting AUC values across all 10 cross-validation seeds, and computing the 5\% and 95\% percentiles. Hence, the confidence intervals reflect both the randomness of the training and evaluation data.

The features described above are affected by a number of factors beyond the one we want to study which is the impact of the agent's behavior. In particular, they reflect speaker-dependent characteristics, present when interacting with both agents, as well as effects related to the progression of the experiment, such as fatigue. We explore two approaches for normalizing the features to mitigate the impact of these factors. 
The first option normalizes the features of every speech interval per speaker, subtracting the speaker's mean and dividing by the speaker's standard deviation, computed across both conditions. The second option performs speaker normalization followed by a global normalization over all training samples for the same trial position within the session, in order to mitigate the impact of fatigue or familiarity with the task. To this end, within each cross-validation fold, we compute the mean of each feature over the training instances for each of the 8 trial positions. These means are then subtracted from both the training and test instances.

\begin{figure}
    \centering
    \includegraphics[width=0.6\textwidth]{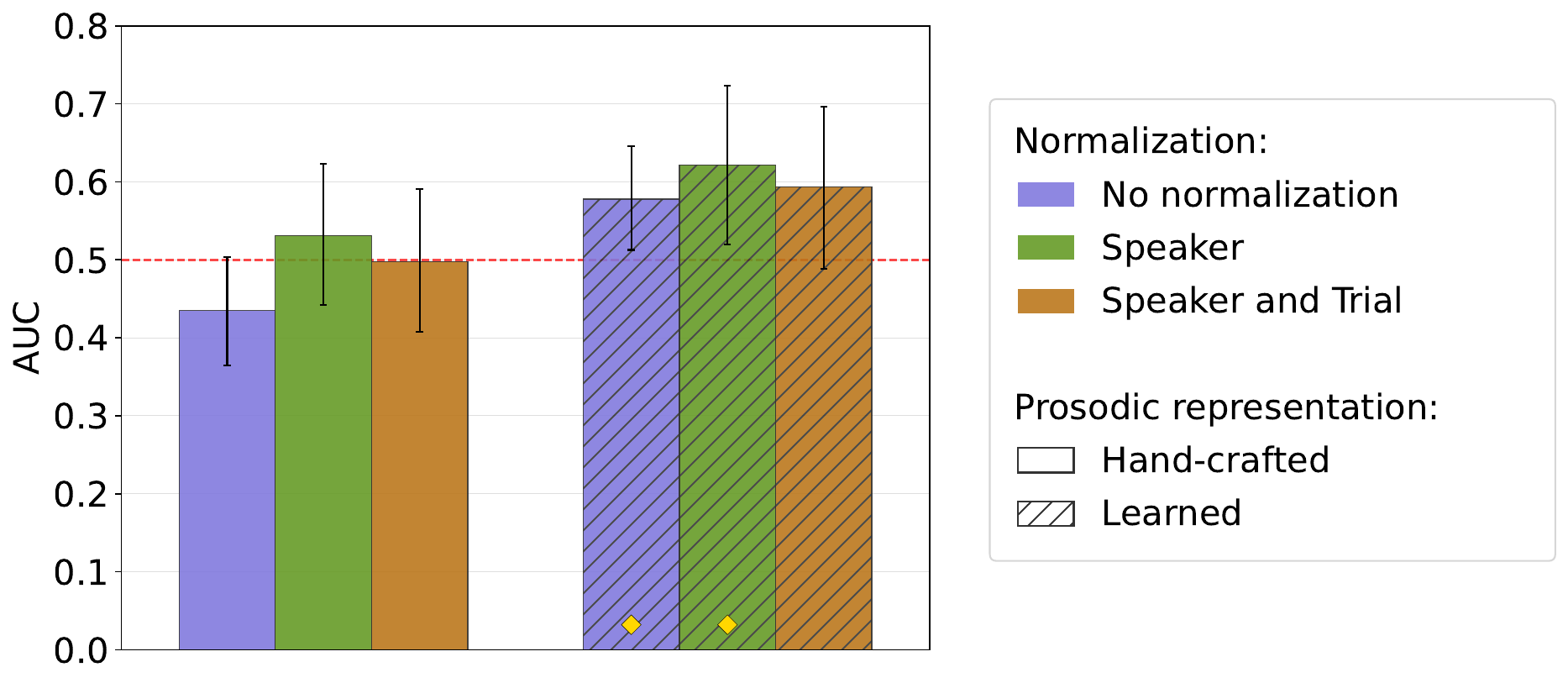}
    \caption{AUC for the agent-type prediction task. Bar colors indicate the normalization option (none, per speaker, or per speaker and trial position), and hatching indicates the prosodic representation: the hand-crafted descriptors of F0 and loudness, complemented with the number of speech intervals and their mean duration, or the learned prosodic embeddings. Bars show the mean AUC while error bars extend from the 5th to the 95th percentile of the bootstrap distribution. The red dashed line indicates chance-level performance (AUC $= 0.5$). The learned features achieve a performance significantly better than chance, indicating that the subjects' prosodic patterns tend to vary depending on the agent with which they are interacting.}
    \label{fig:speech_experiments}
\end{figure}

Figure~\ref{fig:speech_experiments} reports the AUC obtained for the agent-type prediction task for the three normalization options for both sets of features. The figure shows that none of the three hand-crafted configurations leads to AUC values that are significantly different from $0.5$, the random level, with all their confidence intervals including chance level. On the other hand, the learned features lead to performance significantly above chance both for the raw and speaker-normalized features versions. Trial-level normalization results in a slight degradation for both sets of features, suggesting that the normalization statistics cannot be robustly estimated with the available data. 

Unfortunately, while learned features provide an advantage in terms of performance, they do not allow for an exploration of the specific prosodic aspects that are affected by the agent type. We leave this research direction for future work as it is outside of the scope of the current work.

\begin{center}
\setlength{\fboxsep}{5pt}
\fcolorbox{White}{LightBlue}{
  \begin{minipage}{0.9\textwidth}
    \textcolor{Black}{In summary, these results support our hypothesis (H4) that the two agents would elicit different prosodic patterns from the subjects, as a classifier trained to predict the agent type from prosodic features achieves a performance significantly better than chance.}
  \end{minipage}
}
\end{center}

\section{Discussion and Conclusions}\label{sec:discussion}
In this work, we analyzed the degree to which an autonomous agent is perceived as a teammate during a game that requires spoken interaction, and whether changes in the agent's communicative behavior, unrelated to its technological capabilities or reliability, have an impact on that perception. To this end, we designed a protocol in which subjects played a collaborative identification and localization task with two agents: a neutral tool-like agent (N agent) and a team-building agent (T agent) which adopted very simple social communication strategies without altering the amount or quality of the game-relevant information provided to the human. Each subject interacted with both agents. A priming stage was included before interaction with each agent to bias the subjects into the desired condition. 
Throughout the sessions we collected survey responses, interaction logs, and speech: 18 recordings per subject, 16 of them capturing spontaneous spoken human-computer interaction with the two agents in Argentine Spanish, and 2 of neutral reference speech recorded before any agent was involved. Together with the survey and log data, these recordings constitute the \emph{\datasetname{}} dataset, which we release publicly for research use (see Section~\ref{sec:data_availability}). The release contains data from 117 subjects who took part, spanning both versions of the experiment: 99 of the active-priming version and 18 of the neutral-priming one. 
Using the resulting dataset, we addressed four specific objectives, whose main findings we summarize below.

Our first objective \textbf{(O1)} was to examine differences in game performance depending on the agent type (N vs.\ T agent). We hypothesized \textbf{(H1)} that subjects interacting with the T agent would have a tendency to interact more with it and, as a consequence, reach better precision in the placement of the object. Contrary to our hypothesis, the team-building agent did not elicit more interactions from the subjects and, hence, did not result in significantly better performance. This result is consistent with prior human-agent teaming work in which presenting an agent as a teammate improved affect and perception but did not, on its own, translate into better task performance, with performance gains emerging only once a structured team-building intervention was added \citep{Walliser2019}.

Our second objective \textbf{(O2)} was to examine differences in the perception of the autonomous agent depending on the agent type (N vs.\ T agent). We hypothesized \textbf{(H2)} that the T agent would be rated higher than the N agent in terms of its teammate-likeness.
In support of this hypothesis, the T agent was rated as more teammate-like than the N agent across most teammate-likeness subscales, including Ability, despite both agents being identical in the quality and quantity of game-relevant information provided.  
These results echo prior findings by \citet{Walliser2017, Walliser2019}, who found that presenting an agent as a teammate without any change in its underlying capabilities could improve its perceived teammate-likeness. This is consistent with a social-perception account of teammate acceptance, in which perceived warmth and perceived competence (the dimension that the Ability subscale taps) jointly determine how receptive humans are to an AI teammate \citep{HARRISWATSON2023107765}. Our results suggest that the two dimensions are not independent in practice since warmth cues introduced solely through the agent's communicative behavior were sufficient to improve its perceived ability whenever subjects where first exposed to the neutral agent.

Importantly, a comparison with a sample of sessions where both agents were primed equivalently showed no evidence that the priming played a significant role, suggesting that the subjects' more favorable perception was associated with the minor changes in the agent's behavior rather than with the priming. 
The pattern across subscales aligns with the autonomous-agent teammate-likeness framework, in which relationship-building emerged as one of the strongest predictors of whether a technology is perceived as a teammate \citep{lyons2021human, Rix2022}. In line with this work, in our experiments, Relationship-Building also showed the largest effect.

Our third objective \textbf{(O3)} was to examine differences in the perception of the autonomous agent depending on the human's individual traits (Big Five personality dimensions, propensity to trust, and need for affiliation). We hypothesized \textbf{(H3)} that some of these would correlate with higher teammate-likeness scores. 
We found partial support for this hypothesis: a higher propensity to trust in automation was associated with a more favorable perception of the agents, while neuroticism was associated with a less favorable one, whereas extraversion, openness, and need for affiliation showed no significant associations. The positive association with propensity to trust is consistent with prior work reporting that individuals more disposed to trust automation tend to perceive its characteristics more favorably \citep{merritt2008not, Jessup2019}. In contrast, we found no association for extraversion, despite its previously reported link to the propensity to trust machines \citep{merritt2008not}; the negative association with neuroticism, while in the direction expected for emotional stability, was not clearly anchored in prior teaming-specific evidence.

Finally, our fourth objective \textbf{(O4)} was to examine differences in the subjects' speech patterns depending on the agent type (N vs.\ T agent). We hypothesized \textbf{(H4)} that subjects would change their speech patterns depending on the agent type. 
Because both agents delivered exactly the same task-relevant information and the task itself was identical, a difference between conditions was unlikely to surface in \emph{what} subjects said but rather in \emph{how} they said it.
We found support for this hypothesis as the experiments showed a minor but significant effect of the agent type on the prosodic patterns of the human, as measured by the performance of a machine learning model trained on prosodic features extracted with a learned model.

In summary, we found that very simple team-building behaviors, which did not affect the quality and amount of information exchanged related to the task, were sufficient for the autonomous agent to be perceived as more teammate-like. The manipulations introduced in our protocol can be easily implemented in a wide range of agents that rely on oral communication. Our results suggest that this low-cost intervention could result in improved human perception of the agent's teammate-likeness which may, in turn, result in more effective human-agent teaming.

\subsection{Limitations}
This study is not without limitations, most of which relate to the amount of data available for analysis. While the size of our sample is not particularly small for this type of study, several of our analyses required splitting the data by the order in which the two agents were presented (\textit{N-first} vs.\ \textit{T-first}), which effectively halved the number of observations per group and reduced the statistical power of the corresponding comparisons. This was most consequential for the analyses of the agent-order effects on perception, where the most interesting trends, though consistent, did not always reach statistical significance. Collecting a larger sample would allow these effects to be established more reliably, although this was not possible within the current project due to time and funding constraints. 

A second limitation concerns the generalization of our findings. Although the game was designed to elicit spoken interaction under high human-agent interdependence, it remains a structured referential game. The generalization of our conclusions to more complex, open-ended collaborative settings is left as future work.

\section{Declaration of competing interest}
\competingstatement

\section{Data availability}\label{sec:data_availability}
The data supporting the findings of this study are publicly available at 
\ifanonymous [link removed for anonymized review] \else \url{https://ri.conicet.gov.ar/handle/11336/296934} \fi 
under a Creative Commons Attribution-NonCommercial-ShareAlike 4.0 International (CC BY-NC-SA 4.0) license. The release includes all materials necessary to replicate the analyses presented in this paper: subject responses to all survey instruments (Big Five Inventory, Interpersonal Orientation Scale, Scale of Trust in Automated Systems, and the intermediate and final Autonomous Agent Teammate-Likeness questionnaires), logs of subject interactions with both agents across all trials, and the audio recordings collected during the experimental sessions. In addition, the full analysis code, including data preprocessing, statistical modeling, and figure generation, is publicly available at 
\ifanonymous [link removed for anonymized review] \else \url{https://github.com/habla-liaa/Teaming-UBA}\fi 
to facilitate reproducibility.

\section{Declaration of generative AI and AI-assisted technologies in the manuscript preparation process}
During the preparation of this manuscript, the authors utilized Google's Gemini and Anthropic's Claude large language models to assist with writing support and light language editing. Following the use of these tools, the authors rigorously reviewed and edited the content as necessary and assume full responsibility for the final contents of the published article. Part of the Graphical Abstract was generated using the image creation capabilities of Google’s Gemini model, initiated by a specific prompt authored by the research team.

\ifanonymous\else
\section{CRediT authorship contribution statement}
\creditstatement

\section{Acknowledgments}
\acknowledgmentstext

\fi

\newpage
\appendix
\counterwithin{figure}{section}
\counterwithin{table}{section}
\section{Protocol Design Details}

This appendix provides further details on the protocol design decisions, including the criteria used to set the difficulty level of the boards, a table showing the specific boards for each trial, and the priming text shown before each agent.

\subsection{Difficulty Level Specifications}\label{app:difficulty-levels}
The board difficulty level was manipulated as follows:
\begin{itemize}[nosep]
    \item {\bf Easy level:} The board contained one or two objects placed one cell away from the target. In addition, the target was located on the diagonal line between two objects and was completely aligned horizontally with one object and vertically with another object. This made it easy to determine the exact location of the target by asking about relative position with respect to those reference objects.
    \item {\bf Medium level:} The board contained no objects placed one cell away from the target. The target was located on the diagonal line between two objects and was partially aligned horizontally with one object and vertically with another object.
    \item {\bf Hard level:} The board contained no objects placed one cell away from the target. The target was not located on the diagonal line between pairs of objects but was partially aligned horizontally with one object and vertically with another object.
\end{itemize}

\subsection{Blocks Description}\label{sup:boards}
The experiment consists of three blocks of trials. Each trial was characterized by a target object type, used in both Stages 1 and 2, as well as by the board configuration in both stages. One of the blocks corresponded to the training phase, whereas the other two blocks were randomly assigned to the two agents. Table~\ref{sup:table-boards} shows each block together with its corresponding target object, board, and difficulty level for the Stage 2 task. This design was established through multiple discussions within the research team and refined through pilot studies to minimize boredom while maximizing the amount of speech collected for each subject.

\begin{table*}[t]
\centering
\setlength{\tabcolsep}{4pt}
\renewcommand{\arraystretch}{1.15}
\resizebox{\textwidth}{!}{%
\begin{tabular}{%
  >{\centering\arraybackslash}m{2cm}%
  >{\centering\arraybackslash}m{2cm}%
  >{\centering\arraybackslash}m{3.4cm}%
  >{\centering\arraybackslash}m{3.4cm}%
  >{\centering\arraybackslash}m{3.4cm}%
  >{\centering\arraybackslash}m{3.4cm}}
\hline
Block &  & 1st Trial & 2nd Trial & 3rd Trial & 4th Trial \\
\hline
\multirow{3}{*}{Training}
    & Difficulty & Easy & Medium & Hard & Easy \\
\arrayrulecolor{gray!50}\cline{2-6}\arrayrulecolor{black}
    & Target     & Shoe & Table & Tree & Dog \\
\arrayrulecolor{gray!50}\cline{2-6}\arrayrulecolor{black}
    & Board      & \boardimg{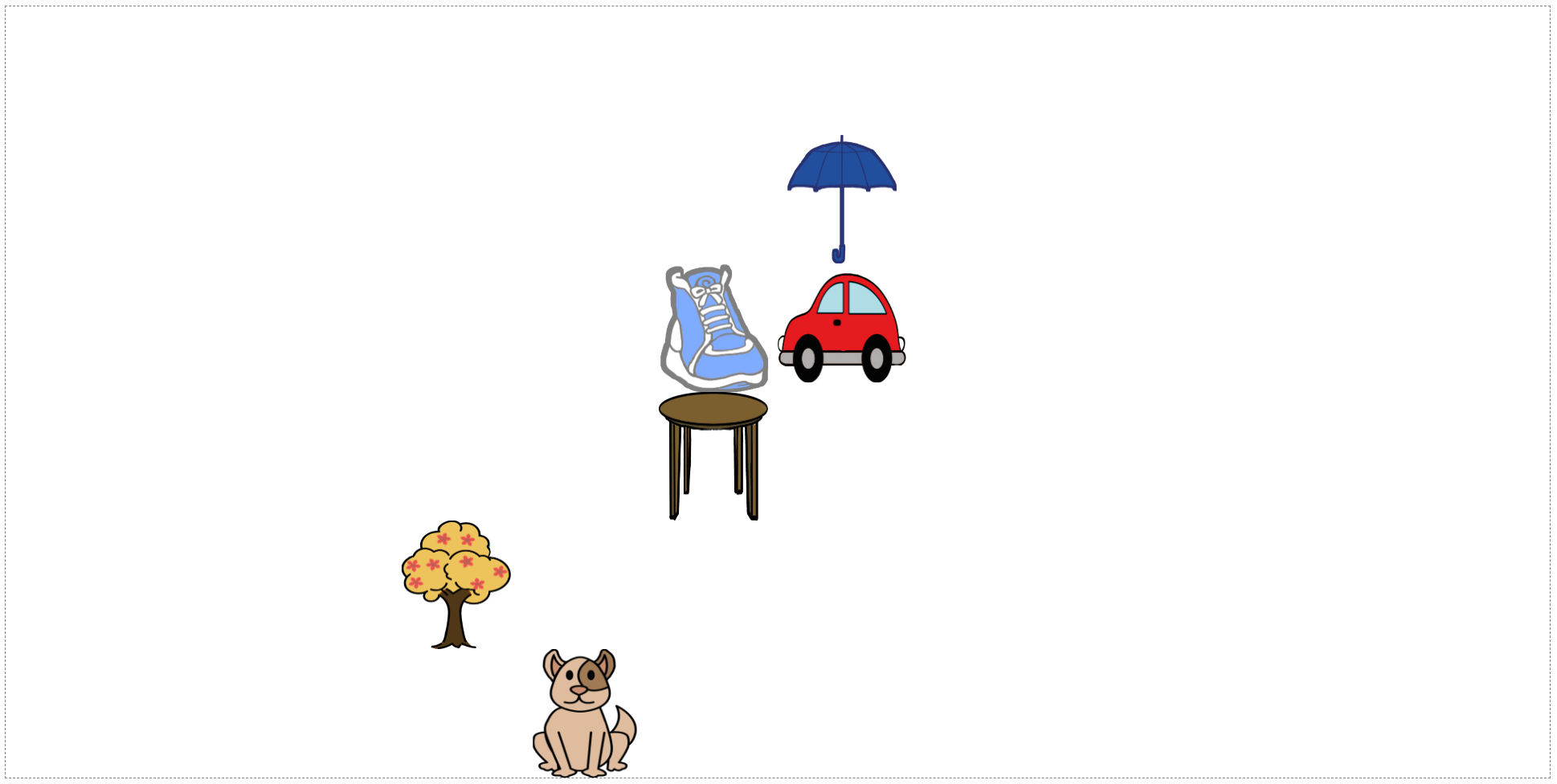}
                 & \boardimg{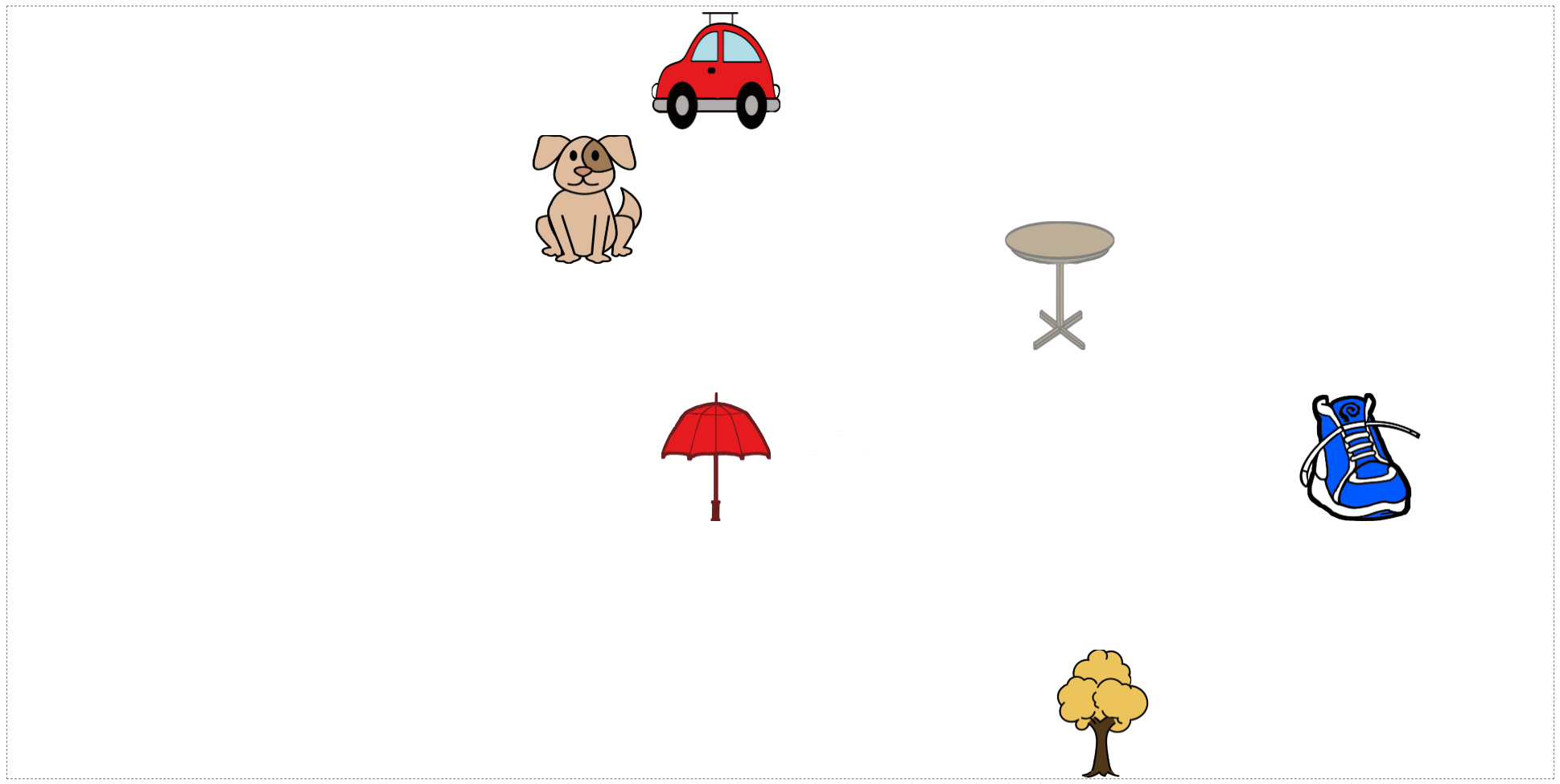}
                 & \boardimg{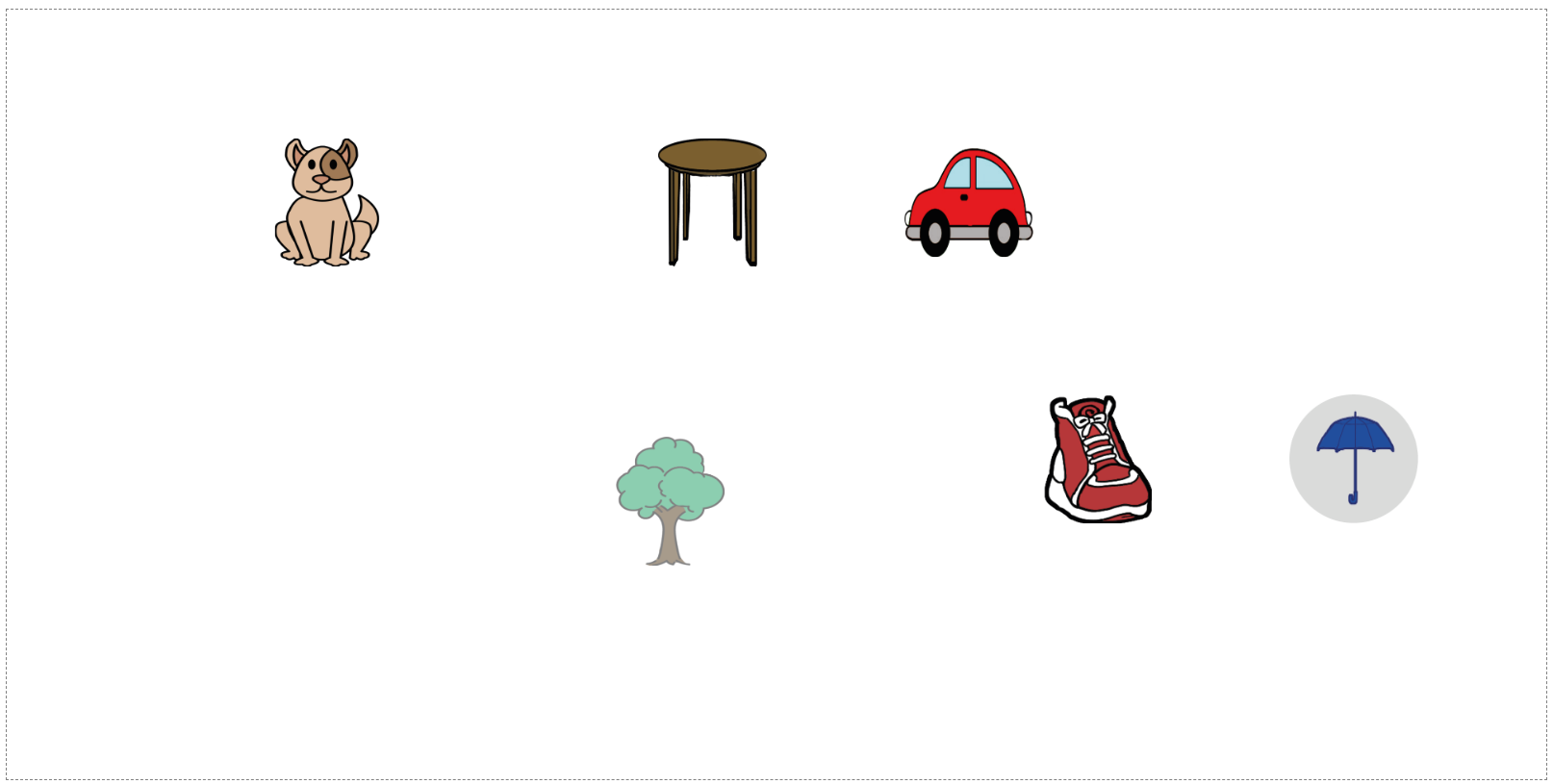}
                 & \boardimg{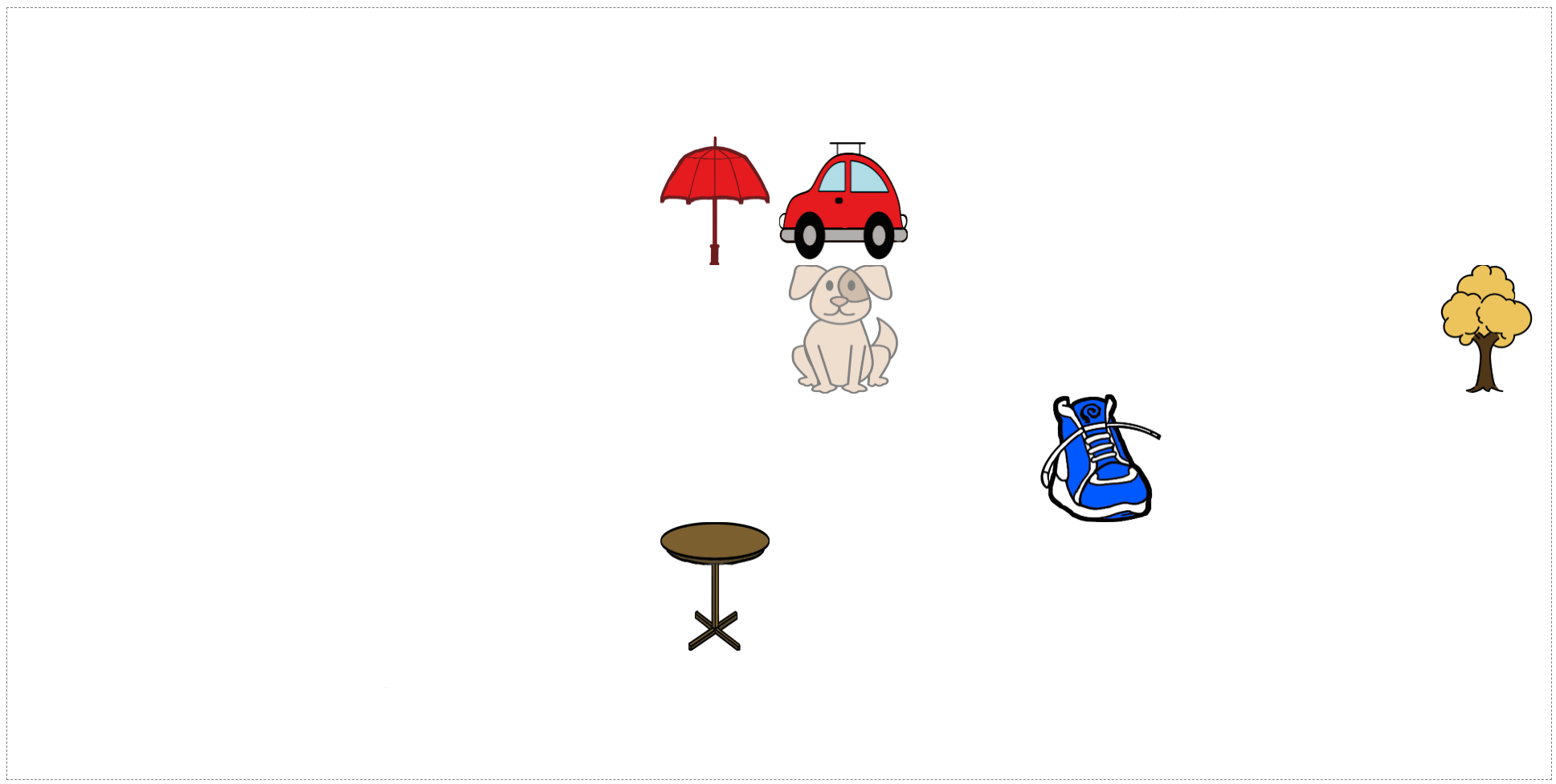} \\
\hline
\multirow{3}{*}{Block A}
    & Difficulty & Easy & Hard & Medium & Hard \\
\arrayrulecolor{gray!50}\cline{2-6}\arrayrulecolor{black}
    & Target     & Shoe & Dog & Table & Tree \\
\arrayrulecolor{gray!50}\cline{2-6}\arrayrulecolor{black}
    & Board      & \boardimg{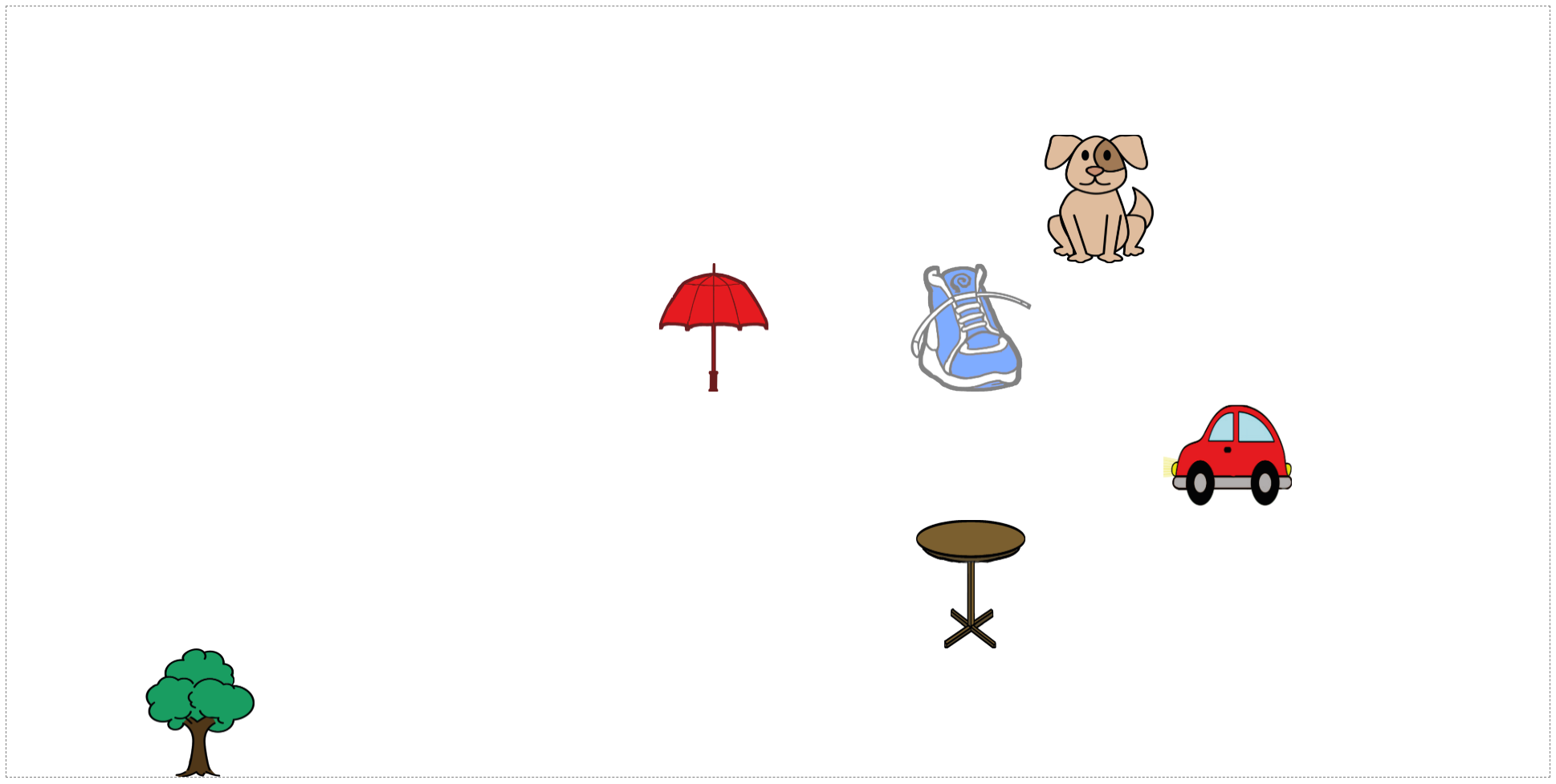}
                 & \boardimg{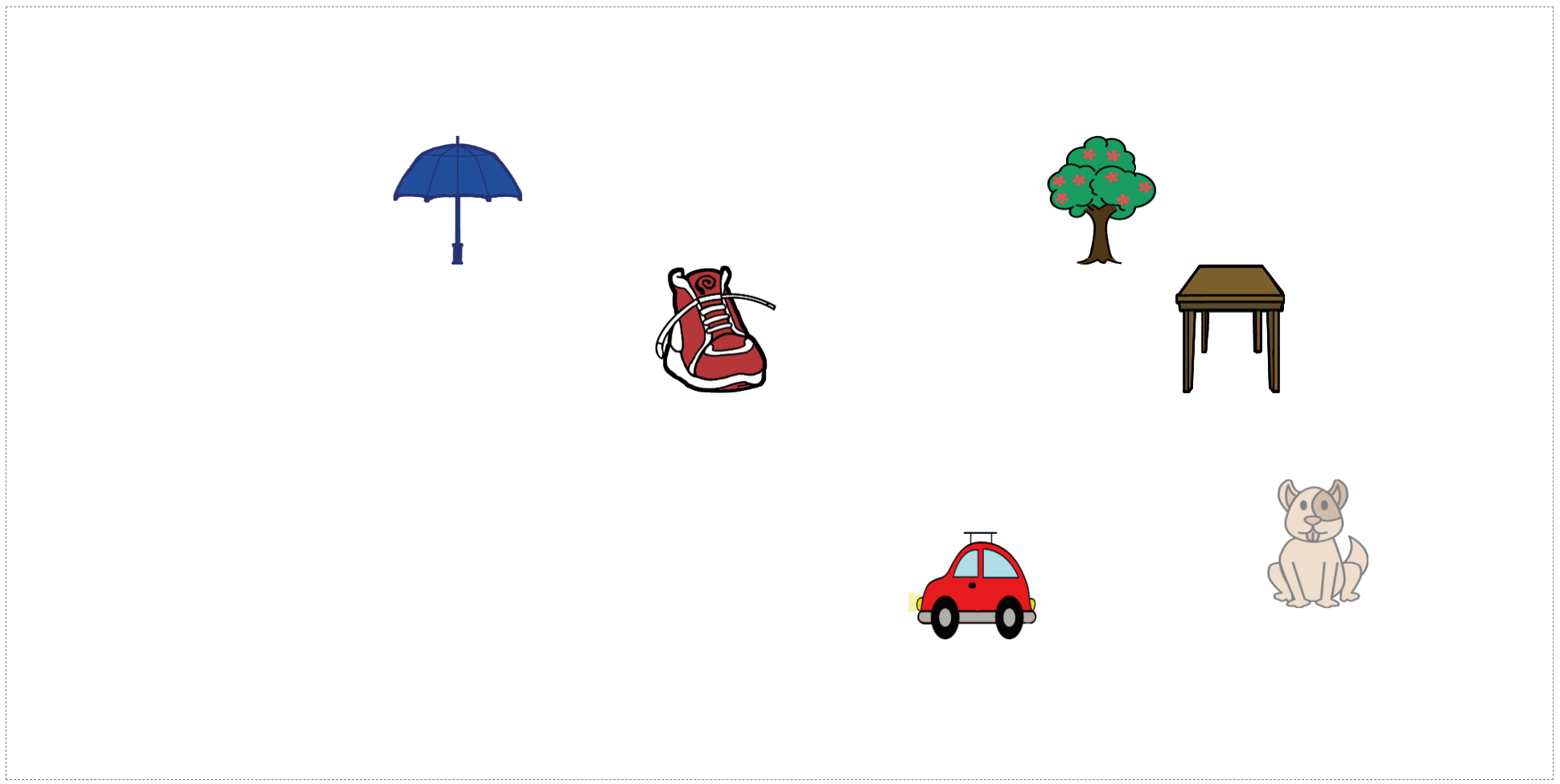}
                 & \boardimg{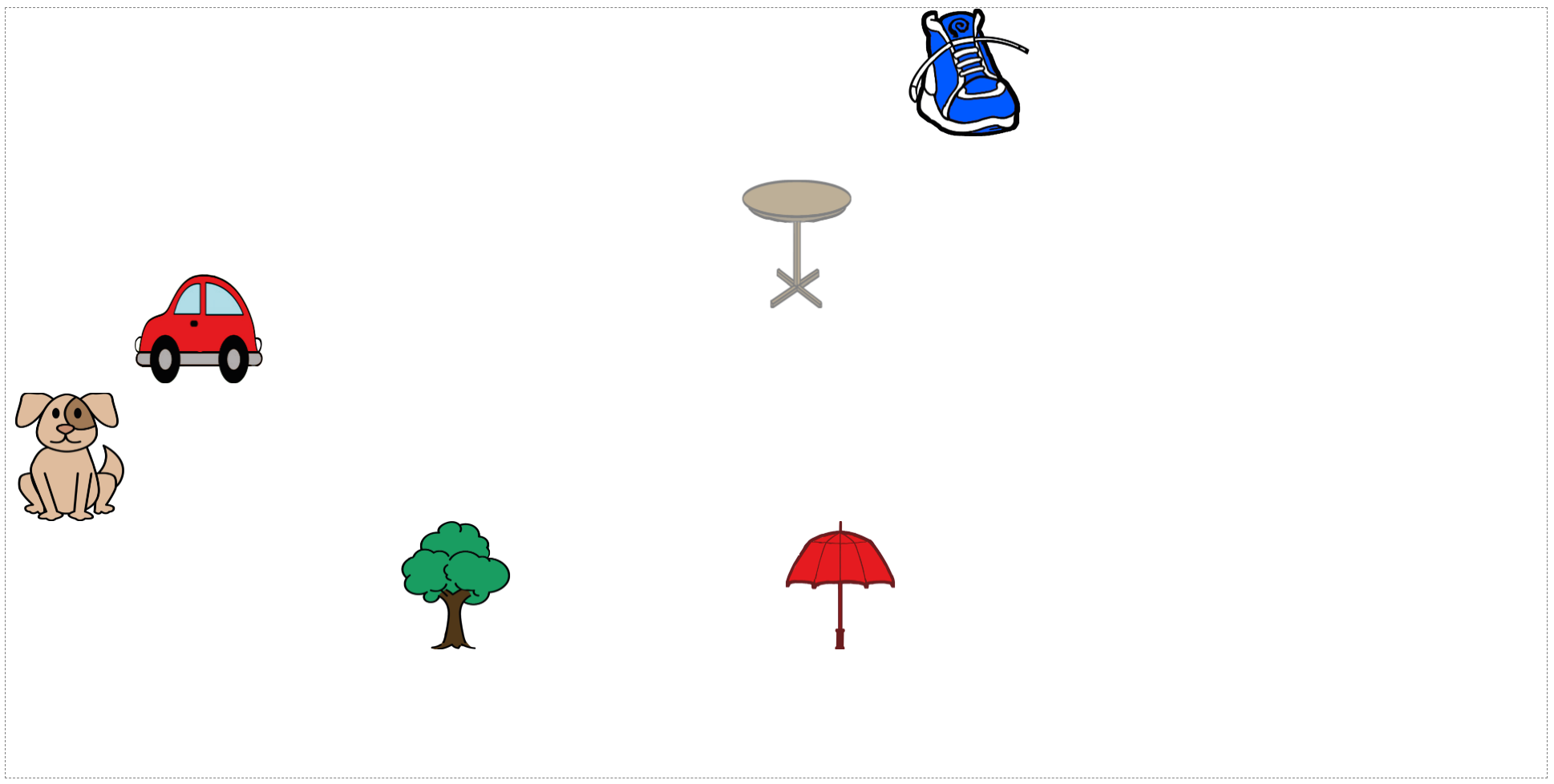}
                 & \boardimg{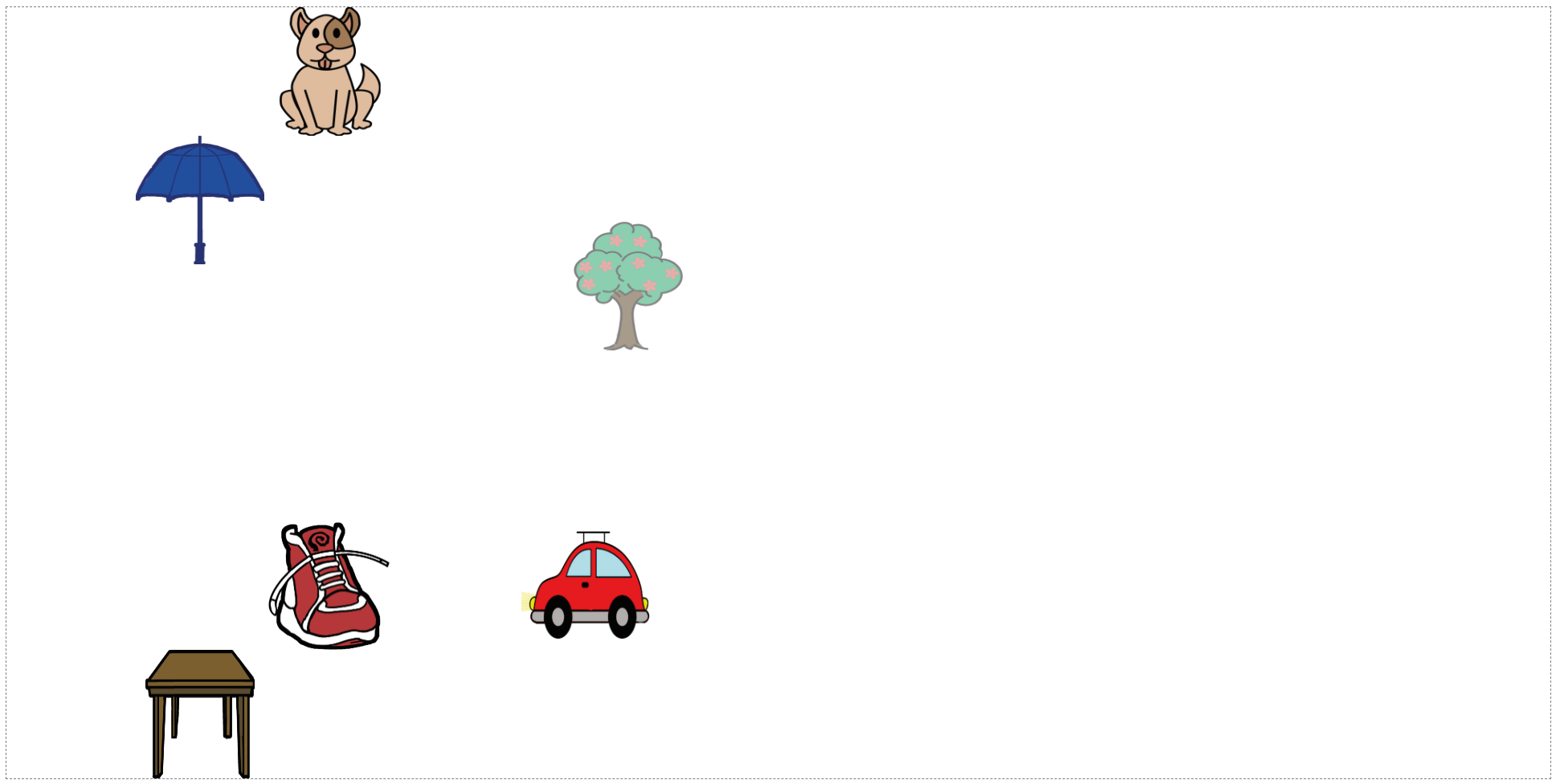} \\
\hline
\multirow{3}{*}{Block B}
    & Difficulty & Easy & Hard & Medium & Hard \\
\arrayrulecolor{gray!50}\cline{2-6}\arrayrulecolor{black}
    & Target     & Shoe & Dog & Table & Tree \\
\arrayrulecolor{gray!50}\cline{2-6}\arrayrulecolor{black}
    & Board      & \boardimg{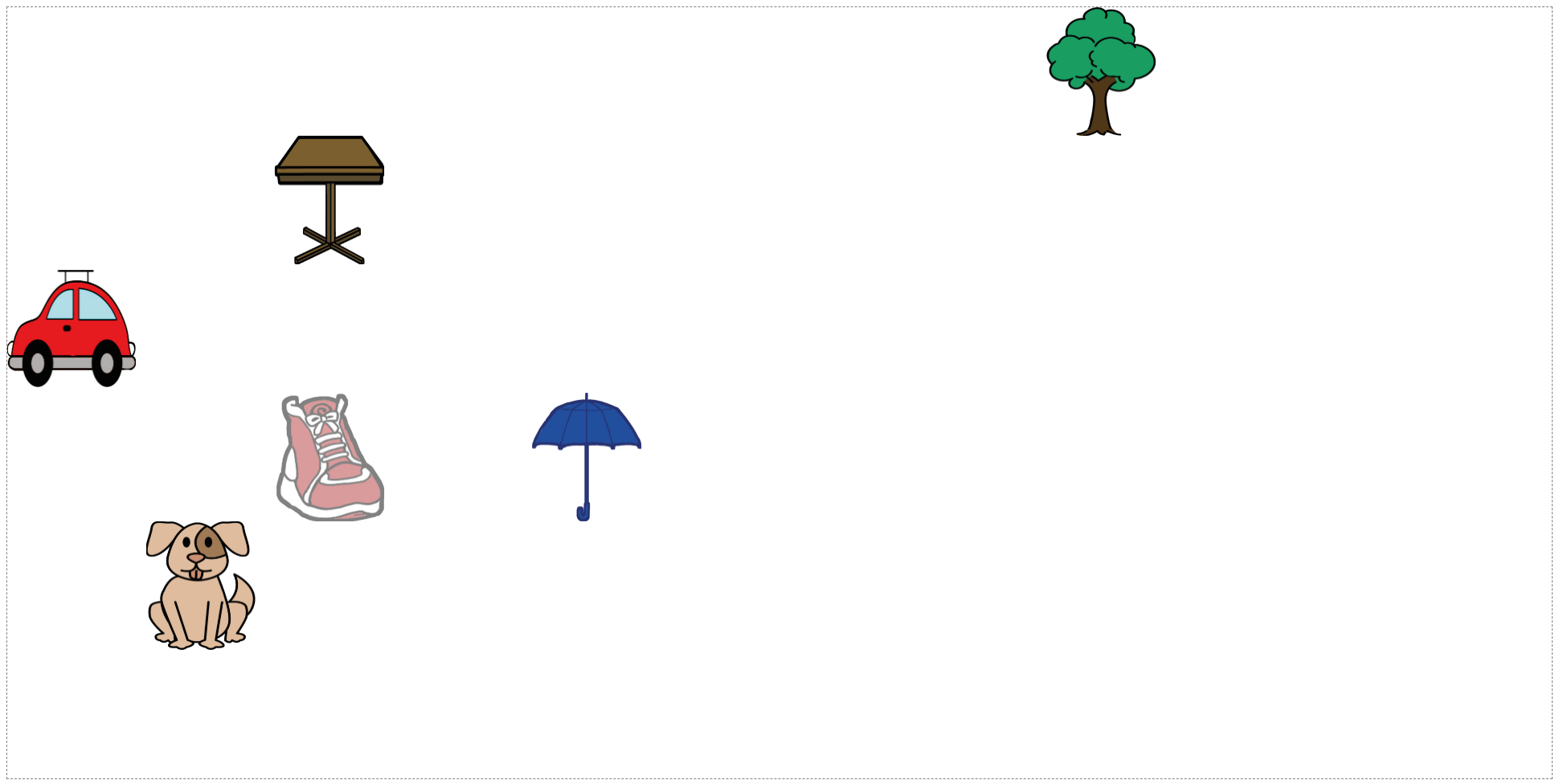}
                 & \boardimg{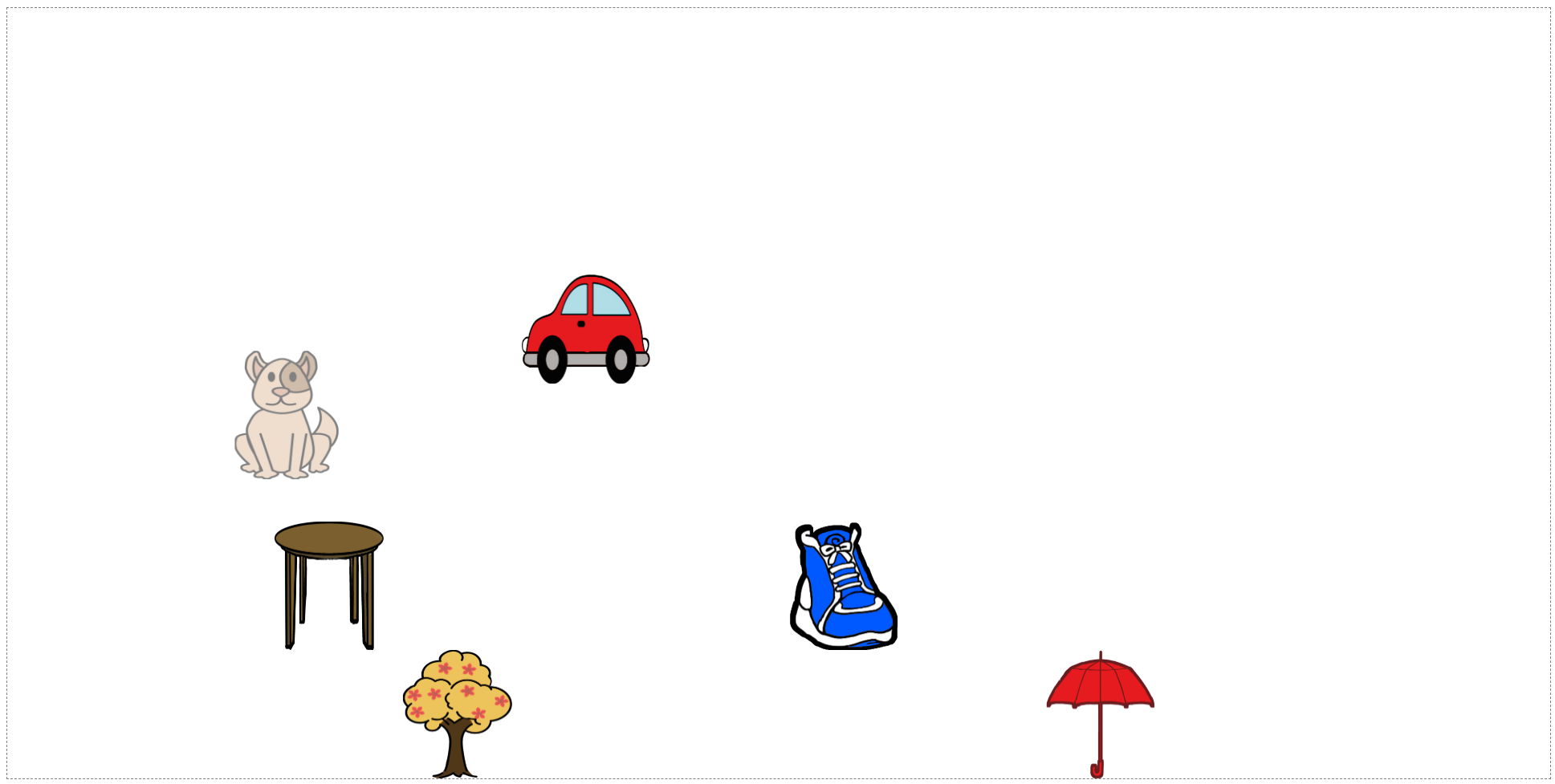}
                 & \boardimg{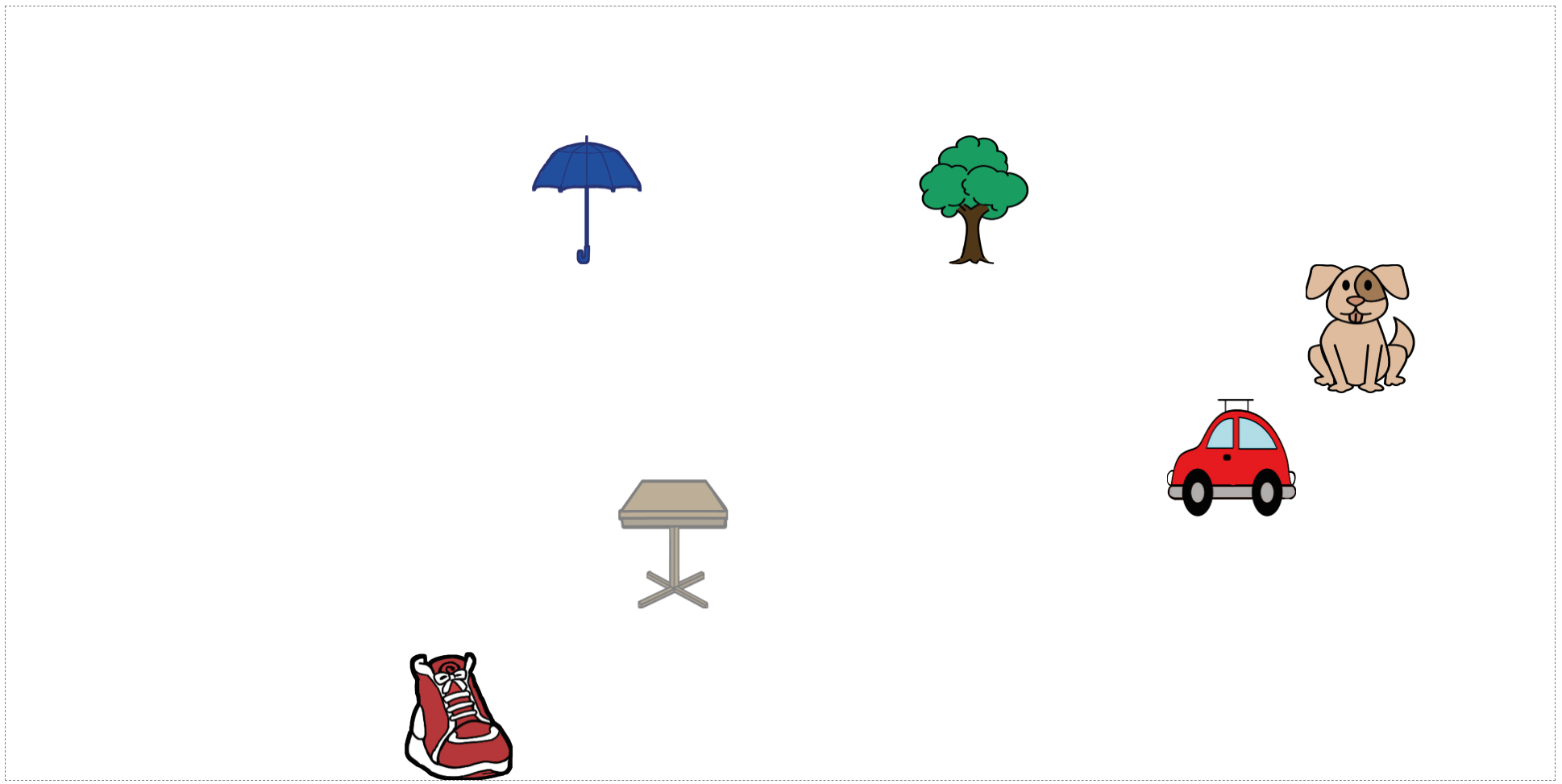}
                 & \boardimg{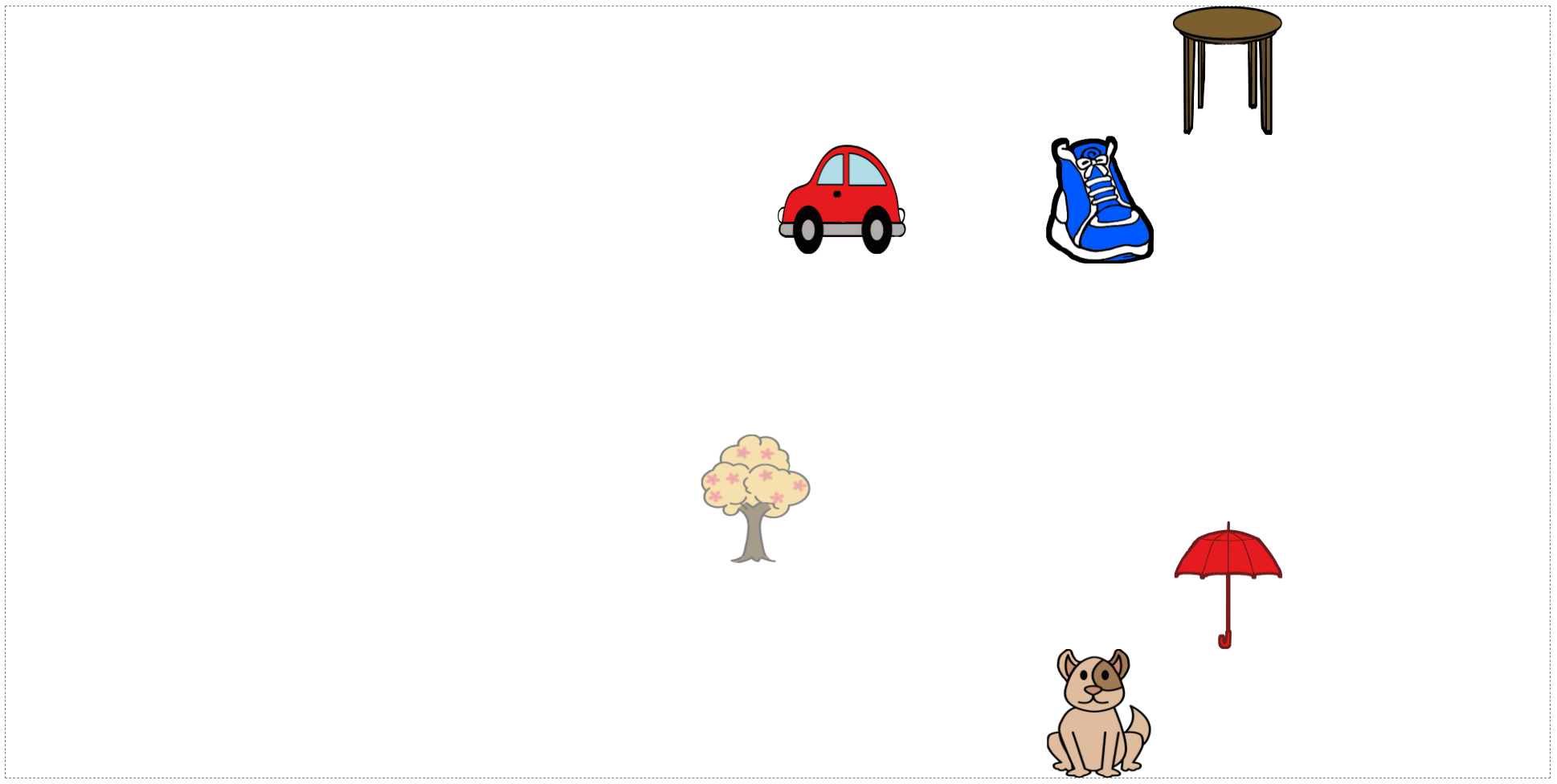} \\
\hline
\end{tabular}
}
\caption{Design of the experimental blocks. Each column corresponds to one trial within a block. For each trial, we include the name of the target object, the Stage 2 board configuration, and its difficulty level. In the board images, the shaded object shows the target object in its correct location.}
\label{sup:table-boards}
\end{table*}

\subsection{Agent Priming Materials}\label{sup:priming}
\begin{figure}
	\centering
	\includegraphics[width=.9\textwidth]{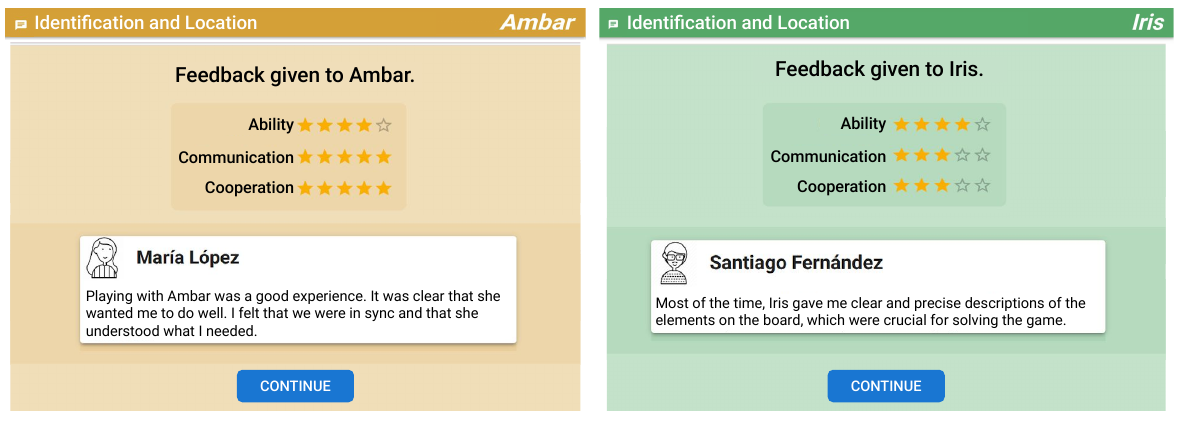}
	\caption{Priming screens shown before the first interaction with each agent. In this example, \textit{Ambar} (left) is the T agent, presented with high ratings in all three dimensions and a review emphasizing rapport and mutual understanding, while \textit{Iris} (right) is the N agent, presented with lower ratings in communication and cooperation and a review focused solely on task performance.}
	\label{sup:img-priming}
\end{figure}

Before subjects began interacting with each agent, they were shown introductory content intended to provide contextual information about that agent. This introduction contained reviews presented as coming from previous subjects -- though they were, in fact, designed by our team -- as well as ratings in three dimensions: ability, communication and cooperation. Both agents received the same rating for ability (4 out of 5 stars). On the other hand, the T agent was presented with the highest ratings for communication and cooperation (5 stars), whereas the N agent received lower ratings on these two dimensions (3 stars). Figure~\ref{sup:img-priming} shows an example of the interface used to introduce the agents. 

The N agent was presented with the following reviews:
\begin{itemize}[nosep]
    \item \textbf{Ana Fischer}: ``Iris described the elements on the board well, and that was useful for solving the game.''
    \item \textbf{Santiago Fernández}: ``Most of the time, Iris gave me clear and precise descriptions of the elements on the board, which were crucial for solving the game.''
    \item \textbf{Matías González}: ``Everything was fine. I had no problems playing with Iris and finishing the rounds.''
\end{itemize}
while the T agent was presented with the following reviews:
\begin{itemize}[nosep]
    \item \textbf{María López}: ``Playing with Ambar was a good experience. It was clear that she wanted me to do well. I felt that we were in sync and that she understood what I needed.''
    \item \textbf{Pablo Gutiérrez}: ``Ambar was an excellent game partner. I liked how involved she was; she almost always described the positions of the objects on the boards accurately, and she made the experience very enjoyable. I felt that her collaboration was key to solving the game.''
    \item \textbf{Lucía García}: ``The experience with Ambar was excellent. She was sociable and pleasant, which made me feel comfortable. In addition, she gave me the impression that she understood well what I needed in order to progress through the game.''
\end{itemize}
Each review was presented on a separate page to encourage subjects to read the content carefully.

\section{Survey Details}

In this section we list the survey questions for the different surveys conducted within the experiment.

\subsection{The Big Five Inventory}\label{sup:bfiquestionnaire}

Subjects completed a questionnaire in which they indicated the extent to which each statement described their personality. Items were rated on a 5-point Likert scale ranging from 1 (\textit{Strongly disagree}) to 5 (\textit{Strongly agree}). We present the English translation of the questionnaire, which was based on the Argentine Spanish adaptation designed by \citet{genise2020big}. In the experimental interface, subjects saw the questionnaire as a single list of items. Here, we group the items by category. The categories are Agreeableness (A), Conscientiousness (C), Extraversion (E), Openness to Experience (O), and Neuroticism (N).

    \subsubsection*{Agreeableness} 
\begin{itemize}[nosep]
    \item I see myself as someone who tends to be warm.    
    \item I see myself as someone who tends to think the best of people.
    \item I see myself as someone who finds it easy to forgive.
    \item I see myself as someone who starts arguments or confrontations with others.
    \item I see myself as someone who helps others selflessly.
    \item I see myself as someone who tends to find fault with others.
    \item I see myself as someone who likes to cooperate with others.
    \item I see myself as someone who is sometimes rude to others.
    \item I see myself as someone who is considerate and kind to almost everyone.
\end{itemize}
    \subsubsection*{Conscientiousness} 
\begin{itemize}[nosep]
    \item I see myself as someone who perseveres until the job is done.
    \item I see myself as someone who tends to be lazy.
    \item I see myself as someone who tends to be disorganized.
    \item I see myself as someone who works hard, is dutiful, and is dependable.
    \item I see myself as someone who is somewhat careless.
    \item I see myself as someone who does a good job.
    \item I see myself as someone who gets distracted easily.
    \item I see myself as someone who makes plans and follows them carefully.
    \item I see myself as someone who does things efficiently.
    \end{itemize}
    \subsubsection*{Extraversion} 
\begin{itemize}[nosep]
    \item I see myself as someone who is not afraid to express what they feel.
    \item I see myself as someone who tends to be quiet.
    \item I see myself as someone who does not mind being the center of attention.
    \item I see myself as someone who is full of energy.
    \item I see myself as someone who is reserved.
    \item I see myself as someone who is talkative.
    \item I see myself as someone who is outgoing and sociable.
    \item I see myself as someone who is sometimes shy or inhibited.
    \end{itemize}
    \subsubsection*{Openness to Experience} 
\begin{itemize}[nosep]
    \item I see myself as someone who is inventive.
    \item I see myself as someone who has a lot of imagination.
    \item I see myself as someone who is ingenious and a deep thinker.
    \item I see myself as someone who is curious about many different things.
    \item I see myself as someone who is original and comes up with new ideas.
    \item I see myself as someone who has sophisticated tastes in art, music, or literature.
    \item I see myself as someone who has few artistic interests.
    \item I see myself as someone who likes to reflect and play with ideas.
    \item I see myself as someone who prefers work that is routine.
    \item I see myself as someone who values artistic and aesthetic experiences.
    \end{itemize}
    \subsubsection*{Neuroticism} 
\begin{itemize}[nosep]
    \item I see myself as someone who is emotionally stable and not easily upset.
    \item I see myself as someone who worries a lot.
    \item I see myself as someone who often gets tense.
    \item I see myself as someone who is calm and handles stress well.
    \item I see myself as someone who is depressed or melancholic.
    \item I see myself as someone who gets nervous easily.
    \item I see myself as someone who remains calm in tense situations.
    \item I see myself as someone who has a changeable mood.
\end{itemize}

The Cronbach's $\alpha$ values were computed over all 117 subjects who completed the questionnaire, pooling both experimental versions (active-priming and neutral-priming; see Section~\ref{sec:collection} for details). Internal consistency was adequate across all five dimensions and comparable to those reported by~\citet{genise2020big}: Openness to Experience: $\alpha = 0.83$; Conscientiousness: $\alpha = 0.82$; Extraversion: $\alpha = 0.83$; Agreeableness: $\alpha = 0.69$; and Neuroticism: $\alpha = 0.79$.

\subsection{The Interpersonal Orientation Scale}\label{sup:iosquestionnaire}
Subjects completed the Interpersonal Orientation Scale, which was used to assess affiliation motivation. Items were rated on a 5-point Likert scale ranging from 1 (\textit{Not at all true}) to 5 (\textit{Completely true}). In the experimental interface, items from the different subscales were presented in an interleaved order. 
The subscales were Attention (A), Emotional Support (E), Positive Stimulation (P), and Social Comparison (S).

    \subsubsection*{Attention} 
\begin{itemize}[nosep]
   \item I like being with other people when I can be the center of attention.
    \item I especially like people who seem very attracted and fascinated by me.
    \item I do not like being with people who might have an unfavorable opinion of me.
    \item I often feel a strong desire for the people around me to notice me and appreciate who I am.
    \item I like being with people who see me as an important and interesting person.
    \item I often feel a strong need to be surrounded by people who are impressed by who I am and what I do.
\end{itemize}
    \subsubsection*{Emotional Support} 
\begin{itemize}[nosep]
    \item When something bad or upsetting happens to me, I just want to be with a close and trustworthy friend.
    \item When I have not done very well at something that is very important to me, I can start to feel better simply by being with other people.
    \item One of my greatest sources of comfort when things get difficult is being with other people.
    \item I usually feel a strong need to have other people nearby when I am upset about something.
    \item When I feel sad or a little depressed, I usually try to be near other people in order to feel better.
    \item During times when I have to go through something painful, I usually find that having someone with me makes it less painful.
\end{itemize}
    \subsubsection*{Positive Stimulation} 
\begin{itemize}[nosep]
   \item I think contact with other people satisfies me more than most people realize.
    \item One of the activities that interests me most is being with other people and learning things about them.
    \item What I enjoy most about being with other people is the warmth I feel in sharing with them.
    \item I feel truly satisfied when I manage to get close to someone.
    \item One of the activities I enjoy most is observing people and getting to know what they are like.
    \item I find it rewarding to establish very close friendships with many people.
    \item I find it very rewarding to establish friendships with people I like.
    \item I think that being close to other people, listening to them, and relating to them on a personal level is one of my favorite and most satisfying pastimes.
    \item I seem to get satisfaction from being with others more than a lot of other people do.
\end{itemize}
    \subsubsection*{Social Comparison} 
\begin{itemize}[nosep]
    \item When I am not sure how well I am doing at something, I usually like to be with other people so that I can compare myself with them.
    \item I prefer to take part in activities with other people rather than alone, because I like to know how well I am doing in the activity.
    \item I often watch other people in order to compare myself with them.
    \item When I am not sure what is happening, I often feel the desire to be with people who are going through the same thing as me.
    \item If I am not clear about what is expected of me, whether in a task or in a social situation, I usually like to watch other people for guidance.
\end{itemize}

Cronbach's $\alpha$ was computed over all 117 subjects who completed the questionnaire, pooling both experimental versions (active-priming and neutral-priming; see Section~\ref{sec:collection} for details). Internal consistency in our sample was adequate across the four subscales and comparable to the values reported by \citet{HILL1987127}: Attention $\alpha = 0.84$, Emotional Support $\alpha = 0.84$, Positive Stimulation $\alpha = 0.83$, and Social Comparison $\alpha = 0.71$.    

\subsection{The Scale of Trust in Automated Systems}\label{sup:stasquestionnaire}

To assess subjects' trust in automated systems, we administered the Scale of Trust in Automated Systems used in the study by \citet{Jian01032000}. Items were rated on a 7-point Likert scale ranging from 1 (\textit{Not at all}) to 7 (\textit{Extremely}).

\begin{itemize}[nosep]
    \item Agents' behavior is deceptive.
    \item Agents behave dishonestly.
    \item I distrust the intentions, actions, and responses of agents.
    \item I behave cautiously with agents.
    \item The actions of agents have harmful or damaging consequences.
    \item I trust agents.
    \item Agents make me feel safe.
    \item Agents have integrity (they are honest, incorruptible, and honorable).
    \item Agents are dependable.
    \item Agents are effective.
    \item I can rely on agents.
    \item I know agents' behavior well.
\end{itemize}

Cronbach's $\alpha$ was computed over all 117 subjects who completed the questionnaire, pooling both experimental versions (active-priming and neutral-priming; see 
Section~\ref{sec:collection} for details), after reverse-coding the distrust-related items. The scale showed good internal consistency ($\alpha = 0.86$).

\subsection{Autonomous Agent Teammate-Likeness}\label{sup:aatl}

The final questionnaire was administered at the end of the interaction with each agent. It consisted of 53 items rated on a 5-point Likert scale ranging from 1 (\textit{Strongly disagree}) to 5 (\textit{Strongly agree}). Most items were adapted from the Autonomous Agent Teammate-Likeness item pool reported by \citet{wynne2019autonomous}. It also included items adapted from the Godspeed Questionnaire Series \citep{bartneckHRIMetrics2009}. In addition, some items were derived from conversation-level annotations used in the UBA Games Corpus \citep{gravano2023uba}. In the experimental interface, subjects saw all items in a single mixed list rather than grouped by category. In each item, \textit{[Agent Name]} refers to the name shown to the subject in that block (\textit{Iris} or \textit{Ambar}).

\subsubsection*{Ability} 
The first three items in this category were adapted from the Godspeed Questionnaire Series \citep{bartneckHRIMetrics2009}, and the remaining items were created for this study.
\begin{itemize}[nosep]
    \item \textit{[Agent Name]} is competent at solving the tasks.
    \item \textit{[Agent Name]} is intelligent.
    \item \textit{[Agent Name]} gives correct answers.
    \item \textit{[Agent Name]} is effective at solving the tasks.
    \item \textit{[Agent Name]} performs consistently.
    \item \textit{[Agent Name]} shows advanced task-related skills.
    \item \textit{[Agent Name]} answers questions accurately.
\end{itemize}

\subsubsection*{Altruism}
\begin{itemize}[nosep]
    \item \textit{[Agent Name]} is truly focused on helping me.
    \item \textit{[Agent Name]} wants to help me.
    \item \textit{[Agent Name]} seems concerned about my well-being.
    \item \textit{[Agent Name]} is designed to help me.
    \item \textit{[Agent Name]} appears to be motivated to help me.
    \item \textit{[Agent Name]} intends to support me.
    \item \textit{[Agent Name]} helps me when I need it.
\end{itemize}

\subsubsection*{Autonomy}
The last item was created for this study, as we wanted to explicitly assess the agent's ability to solve the task during the protocol.
\begin{itemize}[nosep]
    \item \textit{[Agent Name]} does not require me to control it 100\% of the time.
    \item \textit{[Agent Name]} can perform its tasks autonomously.
    \item \textit{[Agent Name]} is in control of its own actions.
    \item \textit{[Agent Name]} executes the best course of action.
    \item \textit{[Agent Name]} can operate with little to no oversight.
    \item \textit{[Agent Name]} was able to solve everything I requested.
\end{itemize}

\subsubsection*{Communication}
The last three items in this category were extracted from a survey developed for the UBA Games Corpus \citep{gravano2023uba}.
\begin{itemize}[nosep]
    \item \textit{[Agent Name]} does not give me useful information.
    \item \textit{[Agent Name]} and I exchange information successfully.
    \item \textit{[Agent Name]} gives me information that is easy to understand.
    \item \textit{[Agent Name]} provides effective feedback.
    \item \textit{[Agent Name]} is highly interactive.
    \item \textit{[Agent Name]} understands my commands.
    \item \textit{[Agent Name]} and I communicate naturally.
    \item \textit{[Agent Name]} and I had moments of discomfort.
    \item \textit{[Agent Name]} and I had difficulty understanding each other.
\end{itemize}

\subsubsection*{Cooperative}
\begin{itemize}[nosep]
    \item \textit{[Agent Name]} and I depend on each other to achieve the goal of the game.
    \item \textit{[Agent Name]} and I share responsibility for our actions.
    \item \textit{[Agent Name]} is essential for me to perform my role in the game.
    \item \textit{[Agent Name]} depends on me to continue in the game.
    \item \textit{[Agent Name]}'s tasks are closely related to mine.
    \item \textit{[Agent Name]}'s goals are closely linked to mine.
    \item \textit{[Agent Name]} depends heavily on me to perform properly.
    \item \textit{[Agent Name]} is essential for me to be effective, that is, for us to achieve the goal.
\end{itemize}

\subsubsection*{Relationship-Building}
\begin{itemize}[nosep]
    \item \textit{[Agent Name]} communicates with me warmly.
    \item \textit{[Agent Name]} understands me.
    \item \textit{[Agent Name]} has a connection with me.
    \item \textit{[Agent Name]} and I have a special connection.
    \item \textit{[Agent Name]} makes me feel appreciated.
    \item \textit{[Agent Name]} shows genuine interest when communicating with me.
    \item \textit{[Agent Name]} responds politely.
    \item \textit{[Agent Name]} has a good rapport with me.
    \item \textit{[Agent Name]} treats me as a teammate would.
    \item \textit{[Agent Name]} relates to me socially.
    \item \textit{[Agent Name]} makes me feel connected.
\end{itemize}

\subsubsection*{Synchronized Mental Model}
\begin{itemize}[nosep]
    \item \textit{[Agent Name]} is generally on the same page as me.
    \item \textit{[Agent Name]} gives me the information I need before I ask for it.
    \item \textit{[Agent Name]} adapts to new situations in the game.
    \item \textit{[Agent Name]} anticipates my actions.
    \item \textit{[Agent Name]} anticipates my needs.
\end{itemize}

\vspace{1em}
Because subjects interacted with both agents (N agent and T agent) and completed this questionnaire once per agent, Cronbach's $\alpha$ was computed separately for each agent, over all 117 subjects, pooling both experimental versions (active-priming and neutral-priming; see Section~\ref{sec:collection} for details). Internal consistency was acceptable across all seven subscales for both agents, as shown in Table~\ref{tab:aatl-alpha}.

\begin{table}[h]
    \centering
    \begin{tabular}{lccc}
        \toprule
        Subscale & \# Items & $\alpha$ (N agent) & $\alpha$ (T agent) \\
        \midrule
        Ability                 &  7 & 0.82 & 0.87 \\
        Altruism                &  7 & 0.84 & 0.86 \\
        Autonomy                &  6 & 0.72 & 0.78 \\
        Communication           &  9 & 0.75 & 0.77 \\
        Cooperative             &  8 & 0.74 & 0.71 \\
        Relationship-Building   & 11 & 0.90 & 0.90 \\
        Synchronized Mental Model & 5 & 0.80 & 0.85 \\
        \bottomrule
    \end{tabular}
    \caption{Internal consistency (Cronbach's $\alpha$) of the final questionnaire subscales in our sample ($n = 117$), computed separately for each agent (N agent and T agent) over both experimental versions.
    All subscales reached acceptable internal consistency ($\alpha \geq 0.71$) under both agent conditions.}
    \label{tab:aatl-alpha}
\end{table}

The intermediate questionnaire was administered multiple times during the game, after each trial. Subjects rated their agreement with each statement on a 5-point Likert scale ranging from 1 (``Strongly disagree'') to 5 (``Strongly agree''). In the experimental interface, the items were presented together in a single mixed list. In each item, \textit{[Agent Name]} refers to the name shown to the subject in that block (\textit{Iris} or \textit{Ambar}).

This short questionnaire contained four items selected from the longer final questionnaire:

\begin{itemize}[nosep]
    \item \textit{[Agent Name]} is truly focused on helping me. (\textit{Altruism})
    \item \textit{[Agent Name]} and I exchange information successfully. (\textit{Communication})
    \item \textit{[Agent Name]} has a good rapport with me. (\textit{Relationship-Building})
    \item \textit{[Agent Name]} is competent at solving the tasks. (\textit{Ability})
\end{itemize}

\section{Dataset Collection Details}\label{sup:db_collection}

\begin{figure}
    \centering
    \includegraphics[width=0.85\textwidth]{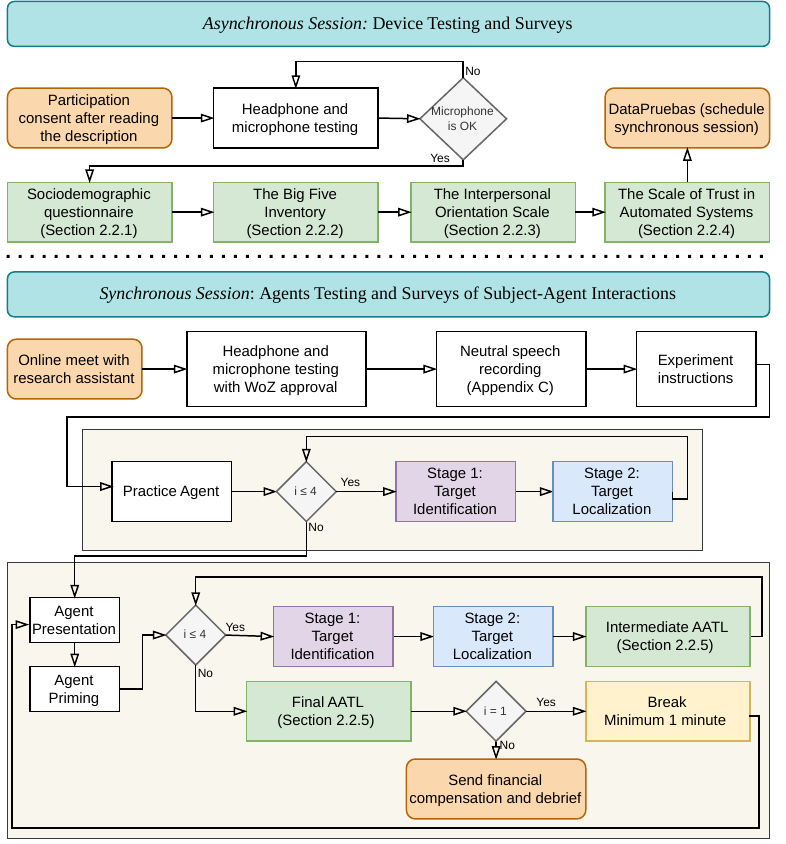}
    \caption{Overview of the experimental protocol. The asynchronous session (top) comprised device testing, consent, and individual-trait surveys. The synchronous session (bottom) included microphone verification with the Wizard-of-Oz interface, initial speech recordings, a practice block  with a neutral agent, and two consecutive interactions with the experimental agents (neutral and team-building), each preceded by a priming stage. After each trial, subjects completed an intermediate AATL questionnaire. After the fourth trial with each agent, they completed the final AATL questionnaire. A break of at least one minute separated the blocks for the two agents.}
    \label{fig:protocol-flow}
\end{figure}

Figure \ref{fig:protocol-flow} shows an overview of the different stages of the protocol. The first session was designed to do an initial check of the subject's equipment and perform the individual-trait surveys. The second session was where the game was played with the two agents, with intermediate and final surveys to assess the subject's perception of the agent.

\begin{figure}
    \centering
    \includegraphics[width=.9\textwidth]{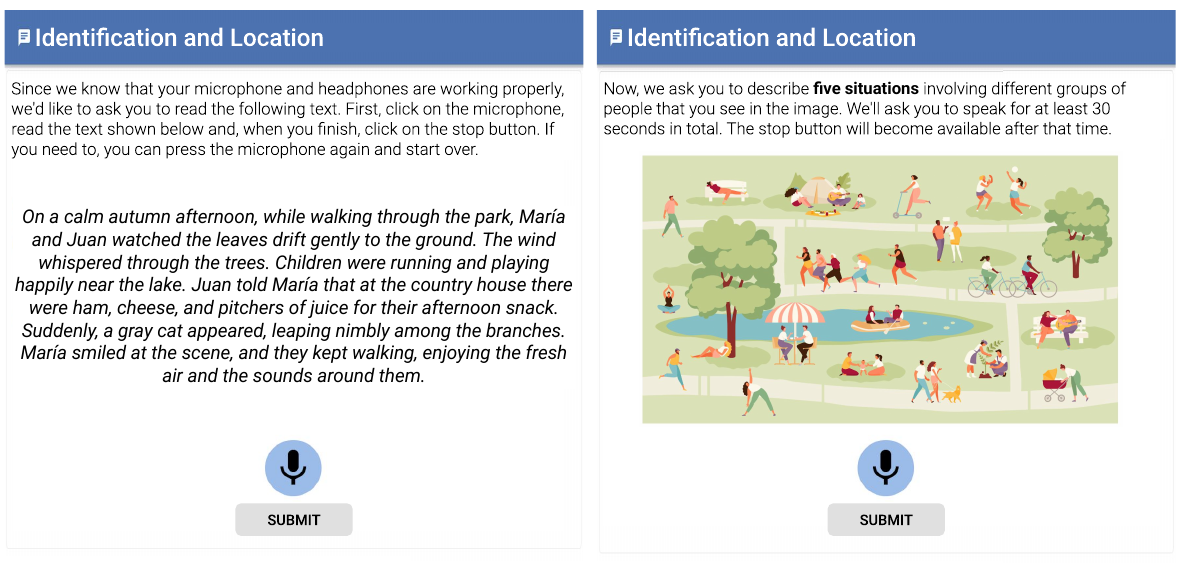}
    \caption{Tasks used to elicit speech. Left: the speech-reading task, in which subjects read aloud a phonetically balanced text (for Spanish). Right: the image-description task, in which subjects described situations depicted in the image to elicit spontaneous speech. The text and instructions shown here are English translations of the original versions, which were presented to subjects in Spanish.}
    \label{sup:neutral-image}
\end{figure}

After the device check, subjects were asked to read a phonetically balanced text in Argentine Spanish. The left panel in Figure~\ref{sup:neutral-image} shows a translation of the screen they see at this stage.
After that, subjects were asked to describe at least five situations depicted in an image (see right panel in Figure~\ref{sup:neutral-image}). This allowed us to collect spontaneous speech. At this stage, subjects were instructed to speak for at least 30 seconds. The option to stop the recording only became available after that time had elapsed.

\section{Additional Database Analyses}

In this section we include further information on the collected database, including an analysis on the degree of subject engagement with the task and a comparison between the active- and neutral-priming sessions.

\subsection{Engagement in the Task}\label{sup:low_score}
\begin{figure}
    \centering
    \includegraphics[width=\textwidth]{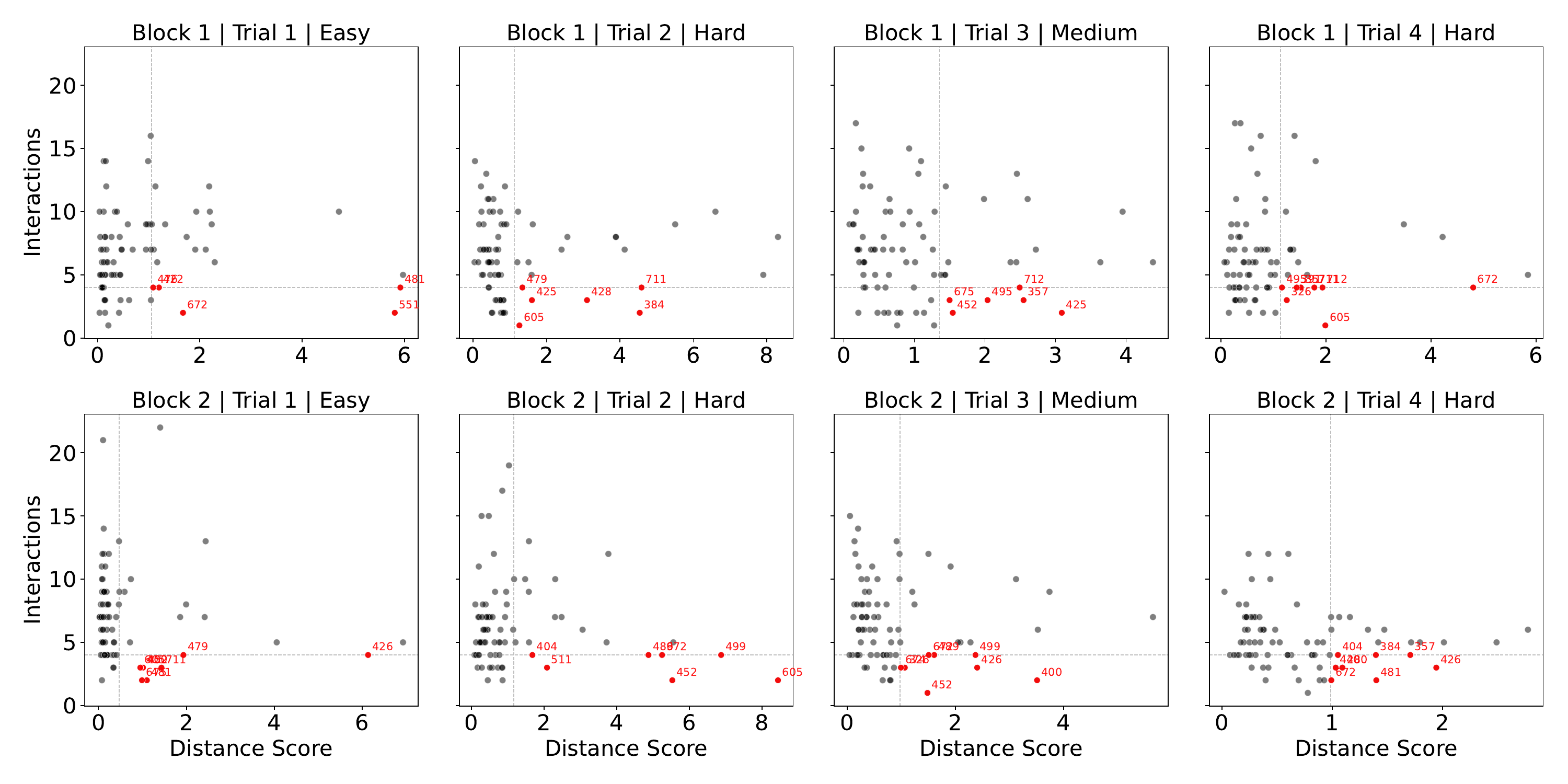}
    \caption{Relationship between the number of interactions each subject had with the agent and the distance score obtained after locating the target figure, shown for each trial. The first row corresponds to the first block of trials solved by the subjects and the second row to the second block. The vertical line represents the $0.75$ quartile for the distance score and the horizontal line the $0.25$ quartile for the number of interactions. Red dots correspond to the trials we consider outliers for having poor engagement with the task.
    The criterion isolates a small group of subjects who combined few interactions with poor placement accuracy; four subjects met it enough times across trials to be excluded from the analyses.}
    \label{fig:extreme_cases}
\end{figure}

During the curation of the dataset, we found that some subjects did not sufficiently commit to the task. For example, one subject asked the agents a single question per trial, which was never sufficient for a correct determination of the position of the target. We therefore devised a procedure to automatically detect such cases. For each trial, we computed the distance score (as defined in Section \ref{sec:game_performance}) and the number of speech intervals from the subject, where a speech interval is defined as a continuous span of speech containing pauses no longer than 0.5 seconds. Trials that had a distance score in the top quartile (a poor game performance) and a number of intervals in the bottom quartile (few interactions with the agents), were considered outliers. See Figure~\ref{fig:extreme_cases} where these trials are marked in red. Finally, for each subject we counted the number of outlier trials. One subject met this criterion in 5 trials and two subjects in 4 trials; all three were discarded. One additional subject who met the criterion in 3 trials was also discarded in order to preserve the gender and agent-order balance. 

\subsection{Effect of Priming on Agent Perception}\label{sec:priming-effect}
As explained in Section \ref{sec:agent-behavior}, before interacting with the agents, subjects were shown introductory content designed to prime them towards the desired condition. The information shown  -- ratings and feedback from hypothetical previous users -- portrayed the T agent more favorably in terms of teaming behavior. To assess whether this priming influenced how subjects subsequently perceived the two agents, we ran a control condition (neutral-priming) with a smaller group of subjects in which the difference in priming between agents was eliminated. Instead, both agents were presented with the same ratings and similar feedback from hypothetical previous users. This allowed us to compare the perceived difference between the T and N agents with and without the priming manipulation.

\begin{figure}
    \centering
    \includegraphics[width=\textwidth]{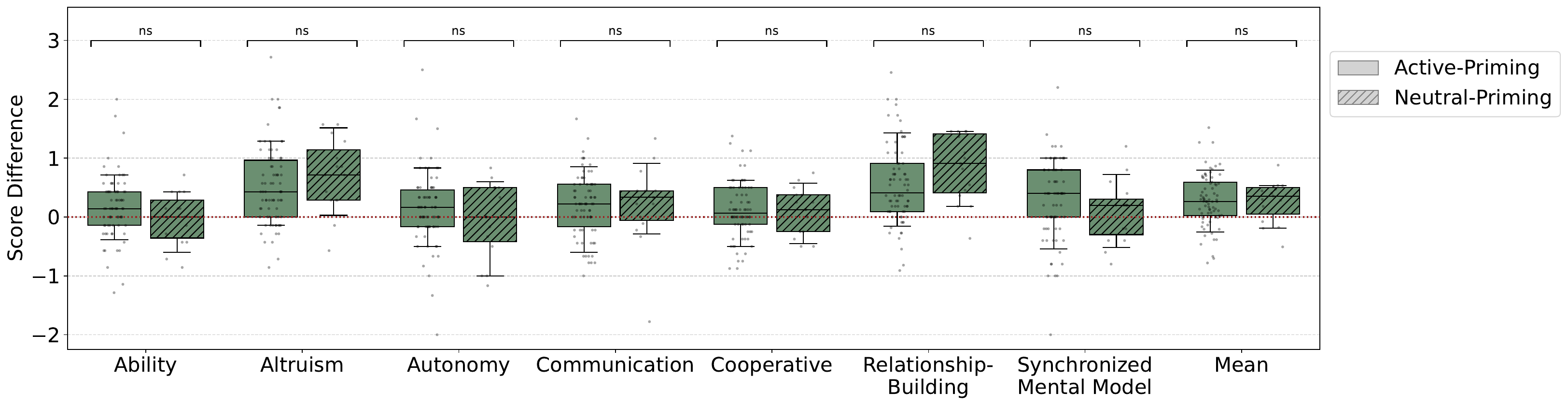}
    \caption{Per-subject difference in AATL survey responses between the T and N agents ($\Delta s_{ik}$) for each subscale and for the across-subscale mean, split by priming type: active-priming (\textbf{$n=74$}) and neutral-priming ($n=15$). Boxes show the median and the first and third quartiles; whiskers extend to the 10th and 90th percentiles. The red dotted line marks zero (no difference between agents). \textit{ns} indicates no significant difference between versions (two-sample $t$-test, Bonferroni corrected, $p_{\mathit{adj}}$ $> 0.05$).
    The T agent was favored over the N agent in both versions, with no significant difference between the two groups, indicating that there is no evidence to support the hypothesis that the type of priming influenced the subject's ratings.}
    \label{sup:with-or-without-priming}
\end{figure}

As shown in Figure~\ref{sup:with-or-without-priming}, no subscale nor the across-subscale mean differed significantly between the two groups of sessions, indicating that there is no evidence to believe that the type of priming influenced the subject's perception of the agent's teaming capabilities.

\section{Learned Prosodic Representations}\label{sup:prosodic-embeddings}

The learned prosodic features used in this work are obtained with a model proposed and trained by \citet{meza2026}. The model is obtained in two stages. First, an autoencoder is trained in a self-supervised manner on frame-level pitch and energy signals. The model is a GRU autoencoder with a 16-dimensional bottleneck, trained with a masked reconstruction objective in which half of the frames are hidden in 160 ms gaps. Its decoder is additionally conditioned on the phone sequence of the segment. The model was designed so that its embeddings retain the information relevant to prosodic tasks while encoding as little information as possible about the speaker and the spoken text. This is achieved by restricting its input to pitch and energy signals, with no spectral detail from the waveform. Further, the phone conditioning of the decoder relieves the bottleneck from having to represent lexical content.  

In the second stage, a \emph{student} network is distilled from the original model, which takes the role of a \emph{teacher}. The student is a two-layer bidirectional GRU that receives the log-magnitude spectrogram of the segment (16 kHz, 25 ms window, 10 ms hop) and is trained to predict the teacher's embedding of the clean audio. The training waveforms are randomly corrupted with SpecAugment \citep{park2019specaugment} and with additive background noise at signal-to-noise ratios between 0 and 20 dB. The student therefore inherits the prosodic target of the teacher, while requiring no pitch or energy extraction at deployment, and being explicitly trained to be robust to background noise. We use a student model provided by the authors of \citep{meza2026}, which yields one 16-dimensional vector per speech segment. 

The model was trained on prosodic intonation units, short stretches of speech with a single coherent intonation contour, obtained with an English-only pipeline (a boundary-detecting ASR model followed by forced alignment). We are not aware of a comparable intonation-unit segmenter for Spanish, so we approximated these units with inter-pausal units, which were obtained with the following process. First, speech intervals were obtained with a voice activity detector. Second, intervals longer than 6 seconds were further split at internal pauses too short to have been detected in the initial segmentation. The split resulted in segments of at most 6 seconds formed from the end of the interval backwards. Segments of 1 second or less were discarded. This process was designed to keep every segment within the duration range seen by the model during training, and the segment boundaries at pauses, where intonation contours typically end.

\clearpage

\bibliographystyle{apalike}
\bibliography{refs}

\begin{thebibliography}{}

\bibitem[Azhar and Sklar, 2017]{azhar2017study}
Azhar, M.~Q. and Sklar, E.~I. (2017).
\newblock A study measuring the impact of shared decision making in a
  human-robot team.
\newblock {\em The International Journal of Robotics Research},
  36(5-7):461--482.

\bibitem[Bartneck et~al., 2009]{bartneckHRIMetrics2009}
Bartneck, C., Croft, E., and Kulic, D. (2009).
\newblock Measurement instruments for the anthropomorphism, animacy,
  likeability, perceived intelligence, and perceived safety of robots.
\newblock {\em International Journal of Social Robotics}, 1(1):71--81.

\bibitem[Bradley et~al., 2010]{bradley2010wizard}
Bradley, J., Benyon, D., Mival, O., and Webb, N. (2010).
\newblock Wizard of oz experiments and companion dialogues.
\newblock In {\em Proceedings of HCI 2010}. BCS Learning \& Development.

\bibitem[Brown and Levinson, 1987]{Brown_Levinson_1987}
Brown, P. and Levinson, S.~C. (1987).
\newblock {\em Politeness: Some Universals in Language Usage}.
\newblock Studies in Interactional Sociolinguistics. Cambridge University
  Press.

\bibitem[Chen et~al., 2016]{chen2016human}
Chen, J.~Y., Barnes, M.~J., Selkowitz, A.~R., Stowers, K., Lakhmani, S.~G., and
  Kasdaglis, N. (2016).
\newblock Human-autonomy teaming and agent transparency.
\newblock In {\em Companion Publication of the 21st International Conference on
  Intelligent User Interfaces}, pages 28--31.

\bibitem[Danescu-Niculescu-Mizil et~al., 2013]{danescu2013computational}
Danescu-Niculescu-Mizil, C., Sudhof, M., Jurafsky, D., Leskovec, J., and Potts,
  C. (2013).
\newblock A computational approach to politeness with application to social
  factors.
\newblock In {\em Proceedings of the 51st Annual Meeting of the Association for
  Computational Linguistics (Volume 1: Long Papers)}, pages 250--259.

\bibitem[Eyben et~al., 2010]{eyben2010opensmile}
Eyben, F., W{\"o}llmer, M., and Schuller, B. (2010).
\newblock Opensmile: the munich versatile and fast open-source audio feature
  extractor.
\newblock In {\em Proceedings of the 18th ACM international conference on
  Multimedia}, pages 1459--1462.

\bibitem[Gauder et~al., 2023]{GAUDER2023101487}
Gauder, L., Pepino, L., Riera, P., Brussino, S., Vidal, J., Gravano, A., and
  Ferrer, L. (2023).
\newblock Towards detecting the level of trust in the skills of a virtual
  assistant from the user’s speech.
\newblock {\em Computer Speech \& Language}, 80:101487.

\bibitem[Genise et~al., 2020]{genise2020big}
Genise, G., Ungaretti, J., and Etchezahar, E. (2020).
\newblock The big five personality inventory in the argentinean context:
  Testing higher order factors.
\newblock {\em Diversitas: Perspectivas en Psicolog{\'\i}a}, 16(2):325--340.

\bibitem[Giles et~al., 1991]{giles1991contexts}
Giles, H., Coupland, J., and Coupland, N. (1991).
\newblock {\em Contexts of accommodation: Developments in applied
  sociolinguistics}.
\newblock Cambridge University Press.

\bibitem[Gravano et~al., 2023]{gravano2023uba}
Gravano, A., Kamienkowski, J.~E., and Brusco, P. (2023).
\newblock Uba games corpus.
\newblock {\em Consejo nacional de investigaciones cient{\'\i}ficas y
  t{\'e}cnicas}.

\bibitem[Hafizoglu and Sen, 2018a]{haf2018effects}
Hafizoglu, F.~M. and Sen, S. (2018a).
\newblock The effects of past experience on trust in repeated human-agent
  teamwork.
\newblock In {\em Proceedings of the 17th international conference on
  autonomous agents and multiagent systems}, pages 514--522.

\bibitem[Hafizoglu and Sen, 2018b]{hafizoglu2018reputation}
Hafizoglu, F.~M. and Sen, S. (2018b).
\newblock Reputation based trust in human-agent teamwork without explicit
  coordination.
\newblock In {\em Proceedings of the 6th international conference on
  human-agent interaction}, pages 238--245.

\bibitem[Harris-Watson et~al., 2023]{HARRISWATSON2023107765}
Harris-Watson, A.~M., Larson, L.~E., Lauharatanahirun, N., DeChurch, L.~A., and
  Contractor, N.~S. (2023).
\newblock Social perception in human-ai teams: Warmth and competence predict
  receptivity to ai teammates.
\newblock {\em Computers in Human Behavior}, 145:107765.

\bibitem[Hill, 1987]{HILL1987127}
Hill, C.~A. (1987).
\newblock Social support and health: The role of affiliative need as a
  moderator.
\newblock {\em Journal of Research in Personality}, 21(2):127--147.

\bibitem[Iftikhar et~al., 2024]{Iftikhar}
Iftikhar, R., Chiu, Y.-T., Khan, M.~S., and Caudwell, C. (2024).
\newblock Human–agent team dynamics: A review and future research
  opportunities.
\newblock {\em IEEE Transactions on Engineering Management}, 71:10139--10154.

\bibitem[Jessup et~al., 2019]{Jessup2019}
Jessup, S.~A., Schneider, T.~R., Alarcon, G.~M., Ryan, T.~J., and Capiola, A.
  (2019).
\newblock {\em The Measurement of the Propensity to Trust Automation}, page
  476–489.
\newblock Springer International Publishing.

\bibitem[Jian et~al., 2000]{Jian01032000}
Jian, J.-Y., Bisantz, A.~M., and Drury, C.~G. (2000).
\newblock Foundations for an empirically determined scale of trust in automated
  systems.
\newblock {\em International Journal of Cognitive Ergonomics}, 4(1):53--71.

\bibitem[John et~al., 1991]{john1991big}
John, O.~P., Donahue, E.~M., and Kentle, R.~L. (1991).
\newblock Big five inventory.
\newblock {\em Journal of personality and social psychology}.

\bibitem[Johnson et~al., 2012]{johnson2012autonomy}
Johnson, M., Bradshaw, J.~M., Feltovich, P., Jonker, C., Van~Riemsdijk, B., and
  Sierhuis, M. (2012).
\newblock Autonomy and interdependence in human-agent-robot teams.
\newblock {\em IEEE Intelligent Systems}, 27(2):43--51.

\bibitem[Juantorena et~al., 2025]{juantorena2025datapruebas}
Juantorena, G.~E., Gauder, L., Laciana, P., Ferrer, L., and Kamienkowski, J.~E.
  (2025).
\newblock Datapruebas: An online platform for data collection.
\newblock In {\em Simposio Argentino de Inteligencia Artificial y Ciencia de
  Datos (ASAID 2025)-JAIIO 54 (Universidad de Buenos Aires, 4 al 7 de agosto de
  2025)}.

\bibitem[Jurafsky and Martin, 2026]{jm3}
Jurafsky, D. and Martin, J.~H. (2026).
\newblock {\em Speech and Language Processing: An Introduction to Natural
  Language Processing Computational Linguistics, and Speech Recognition, with
  Language Models}.
\newblock 3rd edition.

\bibitem[Kargarnovin et~al., 2026]{kargarnovin2026}
Kargarnovin, S., Hernandez, C.~I., Reiners, D., Cruz-Neira, C., Bochenek, G.,
  and Karwowski, W. (2026).
\newblock From testbeds to high-stakes work: a review of human-ai teaming
  domains and teaming factors.
\newblock {\em Frontiers in Robotics and AI}, Volume 13 - 2026.

\bibitem[Larson and DeChurch, 2020]{larson2020leading}
Larson, L. and DeChurch, L.~A. (2020).
\newblock Leading teams in the digital age: Four perspectives on technology and
  what they mean for leading teams.
\newblock {\em The leadership quarterly}, 31(1):101377.

\bibitem[Li et~al., 2022]{Li2022}
Li, M., Erickson, I.~M., Cross, E.~V., and Lee, J.~D. (2022).
\newblock Estimating trust in conversational agent with lexical and acoustic
  features.
\newblock {\em Proceedings of the Human Factors and Ergonomics Society Annual
  Meeting}, 66(1):544–548.

\bibitem[Lingj{\ae}rde et~al., 2001]{lingjaerde2001personality}
Lingj{\ae}rde, O., F{\o}reland, A.~R., and Engvik, H. (2001).
\newblock Personality structure in patients with winter depression, assessed in
  a depression-free state according to the five-factor model of personality.
\newblock {\em Journal of affective disorders}, 62(3):165--174.

\bibitem[Lou et~al., 2025]{Lou2025}
Lou, B., Lu, T., Raghu, T.~S., and Zhang, Y. (2025).
\newblock Unraveling human-ai teaming: A review and outlook.

\bibitem[Lyons and Wynne, 2021]{lyons2021human}
Lyons, J.~B. and Wynne, K.~T. (2021).
\newblock Human-machine teaming: Evaluating dimensions using narratives.
\newblock {\em Human-Intelligent Systems Integration}, 3(2):129--137.

\bibitem[Mallick et~al., 2024]{Mallick25102024}
Mallick, R., Flathmann, C., Lancaster, C., Hauptman, A., McNeese, N., and
  Freeman, G. (2024).
\newblock The pursuit of happiness: the power and influence of ai teammate
  emotion in human-ai teamwork.
\newblock {\em Behaviour \& Information Technology}, 43(14):3436--3460.

\bibitem[McNeese et~al., 2018]{mcneese2018teaming}
McNeese, N.~J., Demir, M., Cooke, N.~J., and Myers, C. (2018).
\newblock Teaming with a synthetic teammate: Insights into human-autonomy
  teaming.
\newblock {\em Human factors}, 60(2):262--273.

\bibitem[Merritt and Ilgen, 2008]{merritt2008not}
Merritt, S.~M. and Ilgen, D.~R. (2008).
\newblock Not all trust is created equal: Dispositional and history-based trust
  in human-automation interactions.
\newblock {\em Human factors}, 50(2):194--210.

\bibitem[Meza et~al., 2026]{meza2026}
Meza, M., Ferrer, L., and Riera, P. (2026).
\newblock Unsupervised approaches for global prosodic embedding extraction.

\bibitem[Meza et~al., 2023]{Meza2023}
Meza, M., Gauder, L., Estienne, L., Barchi, R., Gravano, A., Riera, P., and
  Ferrer, L. (2023).
\newblock {Teamwork Quality Prediction Using Speech-Based Features}.
\newblock In {\em {SMM23, Workshop on Speech, Music and Mind 2023}}, pages
  1--5.

\bibitem[Nass et~al., 1996]{nass1996can}
Nass, C., Fogg, B.~J., and Moon, Y. (1996).
\newblock Can computers be teammates?
\newblock {\em International Journal of Human-Computer Studies},
  45(6):669--678.

\bibitem[Nass and Brave, 2005]{nass2005wired}
Nass, C.~I. and Brave, S. (2005).
\newblock {\em Wired for speech: How voice activates and advances the
  human-computer relationship}.
\newblock MIT press Cambridge, MA.

\bibitem[O'Neill and Salas, 2018]{o2018creating}
O'Neill, T.~A. and Salas, E. (2018).
\newblock Creating high performance teamwork in organizations.
\newblock {\em Human resource management review}, 28(4):325--331.

\bibitem[O’neill et~al., 2022]{o2022human}
O’neill, T., McNeese, N., Barron, A., and Schelble, B. (2022).
\newblock Human--autonomy teaming: A review and analysis of the empirical
  literature.
\newblock {\em Human factors}, 64(5):904--938.

\bibitem[Park et~al., 2019]{park2019specaugment}
Park, D.~S., Chan, W., Zhang, Y., Chiu, C.-C., Zoph, B., Cubuk, E.~D., and Le,
  Q.~V. (2019).
\newblock Specaugment: A simple data augmentation method for automatic speech
  recognition.
\newblock {\em arXiv preprint arXiv:1904.08779}.

\bibitem[Pedregosa et~al., 2011]{scikit-learn}
Pedregosa, F., Varoquaux, G., Gramfort, A., Michel, V., Thirion, B., Grisel,
  O., Blondel, M., Prettenhofer, P., Weiss, R., Dubourg, V., Vanderplas, J.,
  Passos, A., Cournapeau, D., Brucher, M., Perrot, M., and Duchesnay, E.
  (2011).
\newblock Scikit-learn: Machine learning in {P}ython.
\newblock {\em Journal of Machine Learning Research}, 12:2825--2830.

\bibitem[Rix, 2022]{Rix2022}
Rix, J. (2022).
\newblock From tools to teammates: Conceptualizing humans’ perception of
  machines as teammates with a systematic literature review.
\newblock In {\em Proceedings of the 55th Hawaii International Conference on
  System Sciences}, HICSS. Hawaii International Conference on System Sciences.

\bibitem[Santrock, 2008]{santrock2008adolescence}
Santrock, J.~W. (2008).
\newblock {\em Adolescence}.
\newblock McGraw-Hill Higher Education, Boston, MA, 12th edition.

\bibitem[Schmutz et~al., 2024]{SCHMUTZ2024101837}
Schmutz, J.~B., Outland, N., Kerstan, S., Georganta, E., and Ulfert, A.-S.
  (2024).
\newblock Ai-teaming: Redefining collaboration in the digital era.
\newblock {\em Current Opinion in Psychology}, 58:101837.

\bibitem[Seeber et~al., 2020]{Seeber2020}
Seeber, I., Bittner, E., Briggs, R.~O., de~Vreede, T., de~Vreede, G.-J.,
  Elkins, A., Maier, R., Merz, A.~B., Oeste-Reiß, S., Randrup, N., Schwabe,
  G., and Söllner, M. (2020).
\newblock Machines as teammates: A research agenda on ai in team collaboration.
\newblock {\em Information \& Management}, 57(2):103174.

\bibitem[Sharan et~al., 2025]{sharan2025superhuman}
Sharan, N.~N., Peter, J., and van Straten, C.~L. (2025).
\newblock “superhuman” robots in teams: The effect of augmented cognitive
  capabilities on human-robot communication.
\newblock {\em International Journal of Social Robotics}, 17(8):1467--1484.

\bibitem[Siu et~al., 2021]{siu2021evaluation}
Siu, H.~C., Pe{\~n}a, J., Chen, E., Zhou, Y., Lopez, V., Palko, K., Chang, K.,
  and Allen, R. (2021).
\newblock Evaluation of human-ai teams for learned and rule-based agents in
  hanabi.
\newblock {\em Advances in Neural Information Processing Systems},
  34:16183--16195.

\bibitem[Walliser et~al., 2019]{Walliser2019}
Walliser, J.~C., de~Visser, E.~J., Wiese, E., and Shaw, T.~H. (2019).
\newblock Team structure and team building improve human–machine teaming with
  autonomous agents.
\newblock {\em Journal of Cognitive Engineering and Decision Making},
  13(4):258–278.

\bibitem[Walliser et~al., 2017]{Walliser2017}
Walliser, J.~C., Mead, P.~R., and Shaw, T.~H. (2017).
\newblock The perception of teamwork with an autonomous agent enhances affect
  and performance outcomes.
\newblock {\em Proceedings of the Human Factors and Ergonomics Society Annual
  Meeting}, 61(1):231–235.

\bibitem[Wang et~al., 2020]{wang2020can}
Wang, Y.-C., Papangelis, A., Wang, R., Feizollahi, Z., Tur, G., and Kraut, R.
  (2020).
\newblock Can you be more social? injecting politeness and positivity into
  task-oriented conversational agents.
\newblock {\em arXiv preprint arXiv:2012.14653}.

\bibitem[Wilson and Daugherty, 2018]{wilson2018collaborative}
Wilson, H.~J. and Daugherty, P.~R. (2018).
\newblock Collaborative intelligence: Humans and ai are joining forces.
\newblock {\em Harvard business review}, 96(4):114--123.

\bibitem[Wright et~al., 2013]{wright2013effects}
Wright, J.~L., Chen, J.~Y., Quinn, S.~A., and Barnes, M.~J. (2013).
\newblock The effects of level of autonomy on human-agent teaming for
  multi-robot control and local security maintenance.
\newblock Technical Report ARL-TR-6724, Army Research Laboratory, Aberdeen
  Proving Ground, MD.

\bibitem[Wright and Kaber, 2005]{wright2005effects}
Wright, M.~C. and Kaber, D.~B. (2005).
\newblock Effects of automation of information-processing functions on
  teamwork.
\newblock {\em Human Factors}, 47(1):50--66.

\bibitem[Wynne and Lyons, 2018]{wynne2018integrative}
Wynne, K.~T. and Lyons, J.~B. (2018).
\newblock An integrative model of autonomous agent teammate-likeness.
\newblock {\em Theoretical Issues in Ergonomics Science}, 19(3):353--374.

\bibitem[Wynne and Lyons, 2019]{wynne2019autonomous}
Wynne, K.~T. and Lyons, J.~B. (2019).
\newblock Autonomous agent teammate-likeness: Scale development and validation.
\newblock In {\em International Conference on Human-Computer Interaction},
  pages 199--213. Springer.

\bibitem[Zhang et~al., 2023]{zhang2023}
Zhang, R., Duan, W., Flathmann, C., McNeese, N., Freeman, G., and Williams, A.
  (2023).
\newblock Investigating ai teammate communication strategies and their impact
  in human-ai teams for effective teamwork.
\newblock 7(CSCW2).

\end{thebibliography}

\end{document}